\documentclass[12pt]{article}
\usepackage{cite}
\usepackage[cmex10]{amsmath}
\usepackage{amsbsy}
\usepackage{tikz}
\usetikzlibrary{shapes, shadows, arrows}
\usepackage{url}
\usepackage[round]{natbib}

\usepackage{graphicx}
\usepackage{float}
\usepackage{stfloats}
\usepackage{wrapfig}

\usepackage{epstopdf}

\usepackage[latin1]{inputenc}

\usepackage{amsthm,dcolumn,graphicx,multicol,fancyhdr}
\usepackage{latexsym}
\usepackage{ifthen}
\usepackage{rotating,amsfonts,color,amssymb,amsmath,amscd,array,enumerate,afterpage}

\usepackage{hyperref}
\usepackage{multirow}

\usepackage{enumitem}
\setlist{parsep=3pt,listparindent=\parindent}

\usepackage[caption=false,font=footnotesize]{subfig}

\usepackage{calligra}

\usepackage{algorithm,algpseudocode}

\newcommand{\blind}{1}

\usepackage[top=1in, bottom=1in, left=1in, right=1in]{geometry}

\theoremstyle{definition}

\def\subfigure{\subfloat}

\newcommand{\babc}{\renewcommand{\labelenumi}{(\alph{enumi})}\begin{enumerate}}
\newcommand{\eabc}{\end{enumerate}}
\newcommand{\biii}{\renewcommand{\labelenumi}{(\roman{enumi})}\begin{enumerate}}
\newcommand{\eiii}{\end{enumerate}}

\newcommand{\beqn}{\begin{eqnarray*}}
\newcommand{\beq}{\begin{eqnarray}}
\newcommand{\eeqn}{\end{eqnarray*}}
\newcommand{\eeq}{\end{eqnarray}}

\DeclareMathOperator* {\argmax}{ arg\,max}

\newcommand{\ckboldon}[1]{#1}

\newcommand{\ckbold}[1]{%
 \ifthenelse{\isundefined{\ckboldon}}{#1}{ \textbf{#1} }
}

\newcommand{\tr}{\mbox{tr}\,}

\begin{document}
\date{}
\def\spacingset#1{\renewcommand{\baselinestretch}%
{#1}\small\normalsize} \spacingset{1}


\if1\blind
{
  \title{\bf Multi-Scale Factor Analysis of High-Dimensional Brain Signals}
  \author{Chee-Ming Ting\footnote{Center for Biomedical Engineering, Universiti Teknologi Malaysia (UTM), 81310 Skudai, Johor, Malaysia;\texttt{cmting@utm.my}}, Hernando Ombao\footnote{Computer, Electrical and Mathematical Sciences and Engineering Division, King Abdullah University of Science and Technology, Thuwal 23955, Saudi Arabia, and also the Department of Statistics, University of California, Irvine CA 92697, USA; \texttt{hombao@uci.edu,hernando.ombao@kaust.edu.sa}} and Sh-Hussain Salleh\footnote{Center for Biomedical Engineering, UTM, 81310 Skudai, Johor, Malaysia;\texttt{hussain@fke.utm.my}}}
  \maketitle
} \fi

\if0\blind
{
  \bigskip
  \bigskip
  \bigskip
  \begin{center}
    {\LARGE\bf Multi-Scale Factor Analysis of High-Dimensional Brain Signals}
\end{center}
  \medskip
} \fi

\bigskip
\begin{abstract}
In this paper, we develop an approach to modeling high-dimensional networks with a large number of 
nodes arranged in a hierarchical and modular structure. We propose a novel multi-scale factor analysis (MSFA) 
model which partitions the massive spatio-temporal data defined over the complex networks into a finite 
set of regional clusters. To achieve further dimension reduction, we represent the signals in each cluster 
by a small number of latent factors. The correlation matrix for all nodes in the network are approximated 
by lower-dimensional sub-structures derived from the cluster-specific factors. To estimate regional 
connectivity between numerous nodes (within each cluster), we apply principal components analysis 
(PCA) to produce factors which are derived as the optimal reconstruction of the observed signals 
under the squared loss. Then, we estimate global connectivity (between clusters or sub-networks) 
based on the factors across regions using the RV-coefficient as the cross-dependence measure. 
This gives a reliable and computationally efficient multi-scale analysis of both regional and global 
dependencies of the large networks. 
The proposed novel approach is applied to estimate brain connectivity networks using functional 
magnetic resonance imaging (fMRI) data. Results on resting-state fMRI reveal interesting modular 
and hierarchical organization of human brain networks during rest.
\end{abstract}

\noindent%
{\it Keywords:}  Multi-dimensional signals; Dimension reduction; Factor analysis; Principal components analysis; fMRI.
\vfill

\newpage
\spacingset{1.45} 
\section{Introduction}

Analysis of complex networks involves characterization and modeling of the coordinated interactions between different entities. A network can be represented as a graph, where nodes correspond to individual units and weights of edges connecting the nodes to the strength of connections. One popular measure for quantifying the connectivity in weighted networks is through statistical dependencies, such as cross-correlations between signals or data measured at each node. Correlation-based network analysis has found applications in many domains, e.g. brain functional networks \citep{Worsley2005,Marrelec2006,Marrelec2005}, gene co-expression networks \citep{Zhang2005} and financial networks \citep{Onnela2003}. The key challenges involved in characterizing and estimating the network covariance matrix are (a.) the high-dimensionality of network data and hence a large number of connectivity parameters due to the large number of nodes in a network; and (b.) the inherent complexity of the connectivity structure between nodes in most real-world networks. The dimension of the data $N$ (refer to the number of nodes in a network) is usually comparable or even larger than the sample size $T$ (the total number of observed time points). In this high-dimensional setting, the traditional sample covariance matrix is no longer a reliable and accurate estimate of the population covariance, due to massive number of correlation coefficients (i.e. $N(N-1)/2$) that have to be estimated relative to sample size. It will lead to low statistical power in detecting the true network connections. For example, in estimating a full-brain network from fMRI data, the number of voxels $N$ can be in order of $10^4$ but the number of scans $T$ is often of order $10^2$. For gene expression data, $N$ can be up to $10^6$ but with $T$  is of order $10^2$.

Another issue is how to quantify the network's community structure or modularity which is an important property of real-world networks \citep{Newman2012,Girvan2002}. Communities (or modules) refer to sub-networks (clusters of nodes that share common properties or$\slash$and roles) where nodes within the same module are densely intra-connected but relatively sparsely inter-connected with nodes at different modules. In brain networks, modules may correspond to groups of regions of interest with the same function \citep{bullmore2009}, i.e., they respond similarly to a stimulus or are highly synchronized during resting state. It has been observed that spatially distant nodes (e.g., on different sites of the brain) may belong to the same module. Moreover, the modular structure are hierarchical, observed over multiple topological scales ranging from individual nodes as a cluster all the way to having all the nodes (or the entire network) as one cluster \citep{shen2010,betzel2016,song2015}. At any particular scale, each of the modules can be divided into smaller sub-modules (further into sub-sub modules) \citep{Meunier2010}. 
In this paper, we address the problem of modeling and inferring correlation between nodes and modules in high-dimensional networks with a multi-scale community structure, by deriving both local and global connectivity measures based on the proposed spatio-temporal model for network data.

Various approaches have been proposed to estimate high-dimensional covariance matrix for large-scale networks. The simplest approach computes pairwise correlation matrix constructing edges between pairs of nodes, however, it ignores potential influence of other nodes in the entire network. More advanced approach is based on regularized estimation, which includes variety of methods such as thresholding \citep{rothman2009}, shrinkage estimation \citep{Fiecas2011,beltrachini2013,schneider2016} and sparse estimation \citep{ryali2012,chen2016} as previously applied to analyze large-dimensional covariance in neuroimaging and genetic data. However, thresholding and imposing sparsity directly on the covariance matrix might not accurately capture the true underlying connectivity structure.

We follow the dimension-reduction approach which characterizes the connectivity structure in high-dimensional data through a small number of latent components. Two common dimension reduction methods via subspace projection are principal components analysis (PCA) and independent component analysis (ICA). These methods aim to find a projection of high-dimensional data to a low-dimensional subspace that contains independent information (thus removing data redundancy) and then extract a reduced set of latent components (or sources) for connectivity analysis. 
As shown in brain connectivity studies using fMRI data, ICA is capable of decomposing a network into spatial sub-networks with similar functions which are temporally correlated \citep{Calhoun2008,Smith2009,Allen2011,Li2011}. PCA has also been shown to be as effective in connectivity analyses. Signal summaries  derived from PCA are more sensitive than using the mean signals in detecting Granger-causality in an ROI-based fMRI analysis \citep{Zhou2009}. It was able to reveal connectivity across different functional networks during rest \citep{Carbonel2011,Leonardi2013}.

In this paper, we employ the factor analysis (FA) model \citep{BaiLi2012} and PCA for estimation of the low-dimensional projection, to produce a consistent estimator for the high-dimensional covariance structure of large networks. We now summarize the rationale for our approach. First, it is directly linked to analysis of large covariance matrix. The FA model is more general than the conventional PCA and ICA, in the sense that it includes a noise component to the common component driven by few latent factors. This model explicitly provides decomposition of a high-dimensional covariance matrix into a low-rank structure (low-dimensional subspace spanned by the factor loadings) plus a sparse noise covariance matrix. Instead of computing connectivity between reduced components obtained by the PCA or ICA, this approach allows analysis of large-dimensional connectivity matrix in the observation space based on lower-dimensional factor space.
Secondly, it enables consistent and efficient estimation of high-dimensional covariance. PCA can be used to extract factors by solving the constrained linear least-squares, a well-defined criteria for the factor model structure, which minimizes the squared error between the original and projected signals based on the lower-dimensional factors. The PCA can consistently estimate the linear factor subspace, as shown in the extensive literature on estimating high-dimensional FA models \citep{StockWatson2002,Bai2003}. The large covariance estimate is only a simple construction based on these low-dimensional PC estimates. The asymptotic theory of the PCA-based FA model and covariance matrix estimators have been well established for large $N$ \citep{Fan2011,Fan2013}, which can provide insights to our proposed estimator. However, these theoretical results are still unclear for other projection methods such as ICA.

Moreover, most of the previous approaches to large network analysis produce only a single-block covariance matrix for the entire graph. The single-block covariance is ineffective for networks with ultra high dimensions mainly due to the difficulty in interpreting the results, and is unable to describe the multi-scale modularity of real networks.
Motivated by the nature of dense connectivity within each network community, with high multi-collinearity (or redundant information) of activities in the same cluster of nodes, our approach applies FA to each cluster where analysis of the whole network correlation matrix is decomposed to analyzes of smaller module-specific matrices.
More precisely, we propose a multi-scale factor analysis (MSFA) model for network data which allows hierarchical partition 
of the model cross-sectional structure to capture modular dependency at different scales in a large network. 
At the lowest level, the nodes (fundamental units) of the entire network domain are partitioned 
into a finite set of regional clusters, according to spatial proximity or functional relevance. 
The high-dimensional dependence between nodes is then characterized by a partitioned correlation matrix with low-dimensional sub-structures derived from the cluster-specific factors. The global dependence between clusters or collections of clusters (networks) can be measured via the correlations between cluster-specific factors. We further use the RV-coefficient \citep{Escoufieri1973} as a single-valued measure to summarize these factor correlation blocks across clusters and networks. The proposed factor-based decomposable network (a.) enables more reliable and efficient estimation (lower estimation error and less computational effort) of large-scale dependency via dimension-reduction; and (b.) can capture the hierarchical, modular dependency through a factor-structured covariance matrix. We develop a two-step estimation procedure. We first apply PCA to estimate model parameters for each cluster, and then use these local estimates to construct the estimators for various global dependence measures. We apply the approach to analyze resting-state brain networks using fMRI data. It permits a multi-scale analysis of regional (within-ROI) and global (between ROIs and between networks) connectivity. While this paper covers applications only to fMRI, our proposed model and estimation procedure is broadly applicable to is broadly applicable to high-dimensional signals over other types of networks. The performance of our method was assessed via simulations and real data.

\vspace{-0.05in}

\section{Multi-Scale Factor Model for Correlation Network Analysis}

In this section, we propose a novel model and related correlation-based measures to characterize the hierarchical, modular dependency structure of a network at multiple topological scales, from the node-level (between fundamental units in a network) to global-level (between clusters of nodes and between larger sub-networks of clusters). The correlation networks are inferred from signals measured from nodes across the entire network.

\vspace{-0.1in}

\subsection{Definitions and Notations}

We consider a hierarchical network structure, as illustrated in Fig. 1. Suppose the entire network space ${\mathcal G} = \left\{V_1, \ldots, V_N \right\}$ consisting of $N$ nodes are partitioned into $R$ disjoint clusters ${\mathcal C}_1, \ldots, {\mathcal C}_R$, which are then grouped (possibly with overlapping nodes) to form $S$ larger sub-networks ${\mathcal W}_1, \ldots, {\mathcal W}_S$, such that $V_i \in {\mathcal C}_r \subset {\mathcal W}_s \subset {\mathcal G}$. 
We define ${\mathcal C}_r = \left\{V_i : i = I_{r1}, \ldots, I_{r n_r} \right\}$ a set of $n_r$ nodes with indexes $I_{r1}, \ldots, I_{r n_r}$ assigned to the cluster $r$, for $r = 1, \ldots, R$, and ${\mathcal W}_s = \left\{{\mathcal C}_r : r = I_{s1}, \ldots, I_{s d_s} \right\}$ as a collection of $d_s$ clusters with indexes $I_{s1}, \ldots, I_{s d_s}$ grouped to the network $s$. For example in brain networks, $V_i$ corresponds to voxels, ${\mathcal C}_r$ to anatomically-parcellated ROIs (spatially-divided clusters of voxels, and ${\mathcal W}_s$ to system networks (collections of ROIs with similar functional relevance). In this paper, we represent the network as a connectivity matrix quantified with statistical dependence, i.e. the covariance between signals associated with the nodes.

We omit the index symbols $I$ for notational brevity. Let ${\mathbf Y}(t) =  [Y_{1}(t), \ldots, Y_{N}(t)]'$, $t$=1, \ldots, $T$ be the $N \times 1$ vector of signals of length $T$ (e.g. fMRI time series) measured from each node in the entire network space, and ${\mathbf Y}^{\mathcal C}_r(t) =  [Y_{r1}(t), \ldots, Y_{r n_r}(t)]'$ be the subset of signals from nodes in cluster $r$, and ${\mathbf Y}^{\mathcal W}_s(t) = [{\mathbf Y}_{s1}(t), \ldots, {\mathbf Y}_{s d_s}(t)]'$ be the signals from all clusters in network $s$, with dimension $D_s = \sum_{r \in \left\{ 1, \ldots, d_s \right\}} n_r$. The total signals for the entire network can be collectively defined by ${\mathbf Y}(t) = [{\mathbf Y}_{1}^{'}(t), \ldots, {\mathbf Y}_{R}^{'}(t)]'$. The cross-sectional dimension of the entire network $N = \sum_{r=1}^{R} n_r$ (total number of nodes) is assumed to be comparable or larger than the sample size $T$. We denote by $\Sigma_{{\mathbf Y}_r {\mathbf Y}_r}$ (with dimension $n_r \times n_r$), $\Sigma_{{\mathbf Y}_s {\mathbf Y}_s}$ ($D_s \times D_s$) and $\Sigma_{{\mathbf Y} {\mathbf Y}}$ ($N \times N$) the covariance matrices of ${\mathbf Y}^{\mathcal C}_r(t)$, ${\mathbf Y}^{\mathcal W}_s(t)$ and ${\mathbf Y}(t)$, which describe the functional (undirected) dependence between nodes within a cluster, within a sub-network and in the whole network respectively. We assume all these covariance matrices to be time-invariant.

\vspace{-0.1in}

\begin{figure}[!t]
	\begin{minipage}[t]{\linewidth}
		\centering
		\includegraphics[width=0.7\linewidth,keepaspectratio]{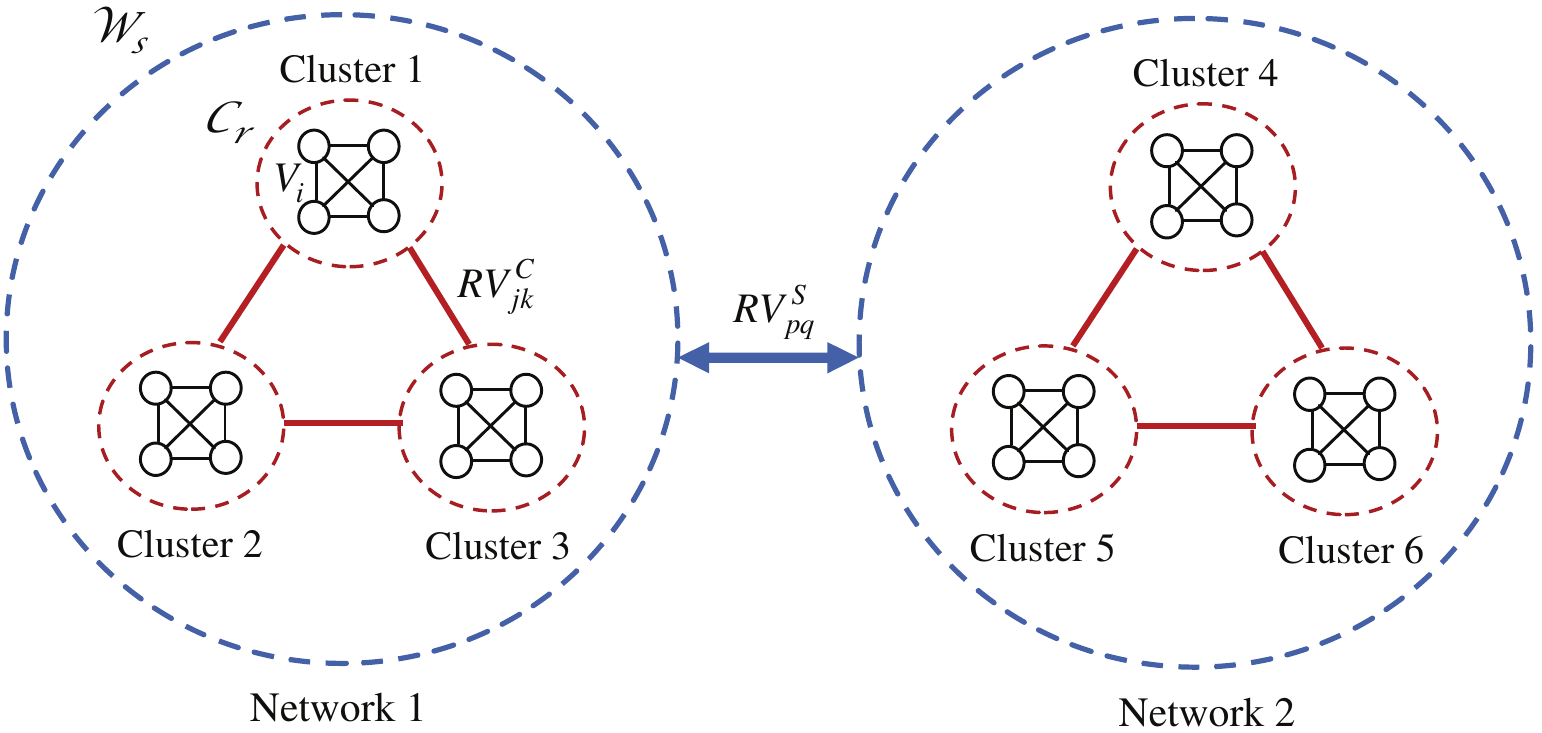}
	\end{minipage}
	\vspace{-0.6 cm}
	\caption{Schematic graphical representation of a network with hierarchical connectivity at the local (nodes $V_i$ within a cluster) and the global (clusters ${\mathcal C}_r$ and networks ${\mathcal W}_s$) levels. The global connectivity is characterized in a lower-dimensional factor space and summarized by single-measure RV coefficients. Red edges: between-cluster connections. Blue edges: between-network connections.}
\label{Fig:Net_Graph}
\vspace{-0.05in}
\end{figure}

\subsection{Modeling Local (Regional) Dependency}

\textit{1) Factor model:} To capture dependence between nodes within a cluster, we first specify the cluster-specific factor model. At each cluster, we assume activity across all nodes can be summarized by a finite number of latent common components, much less than the number of nodes $n_r$. Specifically, the local FA model for signals of each cluster $r$ is
\begin{equation} \label{Eq:ROIfm}
{\mathbf Y}^{\mathcal C}_r(t) = {\mathbf Q}_r {\mathbf f}_{r}(t) + {\mathbf E}_r(t).
\end{equation}
where ${\mathbf f}_r(t) = [f_{r1}(t), \ldots, f_{r m_r}(t)]'$ is a $m_r \times 1$ vector of latent common factors with number of factors $m_r << n_r$. The $m_r \times m_r$ covariance matrix of ${\mathbf f}_r(t)$ is assumed as $\Sigma_{\mathbf{f}_r \mathbf{f}_r} = Cov[\mathbf{f}_r(t)] = {\mbox{diag}}({\sigma}_{f_{r1}}^2, \ldots, {\sigma}_{f_{r n_r}}^2)$, i.e. the factors within cluster $r$ are uncorrelated for any pairs of different factors $j \ne k$, $Cov[f_{rj}(t), f_{rk}(t)] = 0$. The $n_r \times m_r$ factor loading matrix ${\mathbf Q}_r = [{\bf q}_{r1}, \ldots, {\bf q}_{r n_r}]'$ defines the dependence between nodes through the mixing of ${\mathbf f}_r(t)$. It satisfies the condition ${\mathbf Q}'_r {\mathbf Q}_r = \mathbf{I}_{m_r}$, where $\mathbf{I}_{m_r}$ denotes a $m_r \times m_r$ identity matrix. ${\mathbf E}_r(t) = [e_{r1}(t), \ldots, e_{r n_r}(t)]'$ is a $n_r \times 1$ vector of white noise with $E[{\mathbf E}_r(t)] = {\bf 0}$ and $\Sigma_{{\mathbf E}_r {\mathbf E}_r} = Cov[{\mathbf E}_r(t)] = {\mbox{diag}}({\sigma}_{e_{r1}}^2, \ldots, {\sigma}_{e_{r n_r}}^2)$. Our approach enables dimension reduction via (1.) piece-wise partitioning of the entire high-dimensional observation space ${\mathbf Y}(t)$ into a finite number of smaller components ${\mathbf Y}^{\mathcal C}_r(t)$ at each cluster, and (2.) the serial and cross-dependence within each cluster are further summarized by a much lower dimensional factor process ${\mathbf f}_r(t)$ and mixing matrix ${\mathbf Q}_r$.

The components ${\mathbf Q}_r$ and $\mathbf{f}_r(t)$ are not separately identifiable. For any $m_r \times m_r$ invertible matrix ${\mathbf U}_{r r}$, ${\mathbf Q}_r \mathbf{f}_r(t) = {\mathbf Q}_r {\mathbf U}_{r r} {\mathbf U}_{r r}^{-1} \mathbf{f}_r(t) = {\mathbf Q}^*_r \mathbf{f}^*_r(t)$ with ${\mathbf Q}^*_r = {\mathbf Q}_r {\mathbf U}_{r r}$ and $\mathbf{f}^*_r(t) = {\mathbf U}_{r r}^{-1} \mathbf{f}_r(t)$. The model (\ref{Eq:ROIfm}) is observationally equivalent to ${\mathbf Y}_r(t) = {\mathbf Q}^*_r {\mathbf f}^*_{r}(t) + {\mathbf E}_r(t)$. The orthonormality of ${\mathbf Q}_r$ restricts ${\mathbf U}_{r r}$ to be orthonormal (imposing $m_r(m_r + 1)/2$ restrictions), together with the diagonality of $\Sigma_{\mathbf{f}_r \mathbf{f}_r} = E(\mathbf{f}_r(t) \mathbf{f}'_r(t))$ (with $m_r(m_r - 1)/2$ restrictions) restricts ${\mathbf U}_{r r}$ to be a diagonal matrix with diagonal entries of $\pm 1$ (total $m_r^2$ restrictions on ${\mathbf U}_{r r}$). This identifies ${\mathbf Q}_r$ and $\mathbf{f}_r(t)$ up to a sign change.

\textit{2) Determination of clusters:} In this paper, the following are assumed to be fixed and known: hierarchical partitioning of the clusters $\left\{ {\mathcal C}_1, \ldots, {\mathcal C}_R \right\}$ and sub-networks $\left\{ {\mathcal W}_1, \ldots, {\mathcal W}_S \right\}$; the number of clusters $R$; and the number of sub-networks $S$. For fMRI analysis, clustering of brain nodes into ROIs can be defined by prior information on the anatomical parcellation (e.g. according to Anatomical Automatic Labeling (AAL) atlas \citep{Tzourio-Mazoyer2002}), and the brain sub-networks by the functional similarity of the ROIs. When unknown, many algorithms for community detection in networks can be used to automatically identify the modules and their hierarchical (multi-scale) organization. See \citep{fortunato2010} for an extensive review. Among these algorithms, we are particularly interested in spectral clustering which partitions a graph into clusters through the eigenvectors of the connectivity matrix (e.g., a simple adjacency or Laplacian matrix \citep{Von2007} or the recently proposed correlation matrix \citep{shen2010}). These algorithms will be incorporated in the preliminary step in our future work to further refine and generalize the proposed framework.

\vspace{-0.08in}

\subsection{Modeling Global (Inter-Region) Dependency}
To capture dependence between the different clusters and between larger networks of these clusters, we develop the global factor model for the entire network, by concatenating the local factor models in (\ref{Eq:ROIfm}) from all clusters $r = 1, \ldots, R$. The global FA model has a structured form by partitioning the cross-sectional dimension, as defined by
\begin{equation} \label{Eq:ModelAll}
{\mathbf Y}(t) = {\mathbb Q} {\mathbf f}(t) \ + \ {\mathbf E}(t)
\end{equation}
where ${\mathbf E}(t) = [{\mathbf E}_{1}^{'}(t), \ldots, {\mathbf E}_{R}^{'}(t)]'$ is a $N \times 1$ global white noise process with $E[{\mathbf E}(t)] = {\bf 0}$ and $\Sigma_{{\mathbf E} {\mathbf E}} = Cov[{\mathbf E}(t)] = {\mbox{diag}}(\Sigma_{{\mathbf E}_1 {\mathbf E}_1}, \ldots, \Sigma_{{\mathbf E}_R {\mathbf E}_R})$ i.e. the off-diagonal covariance blocks $\Sigma_{{\mathbf E}_j {\mathbf E}_k} = Cov[{\mathbf E}_j(t), {\mathbf E}_k(t)] = {\bf 0}$ for any pair of errors (${\mathbf E}_j(t), {\mathbf E}_k(t)$), $j \ne k$. 
By concatenating factor time series from all clusters in the entire network, we have a $M \times 1$ global factor process ${\mathbf f}(t) = [{\mathbf f}_{1}^{'}(t), \ldots, {\mathbf f}_{R}^{'}(t)]'$ with total dimension $M = \sum_{r=1}^{R} m_r$, which has $E[{\mathbf f}(t)] = {\bf 0}$ and $M \times M$ covariance matrix $\Sigma_{{\mathbf f} {\mathbf f}} = Cov[{\mathbf f}(t)]$. Both processes ${\mathbf E}(t)$ and ${\mathbf f}(t)$ are assumed to be uncorrelated. The global mixing matrix ${\mathbb Q}$ is a $N \times M$ block-diagonal matrix
\[
{\mathbb Q} =
\left(
  \begin{array}{ccc}
    {\mathbf Q}_1 & \ldots & {\bf 0} \\
    {\vdots} & \ddots & \vdots \\
    {\bf 0} & \ldots & {\mathbf Q}_R \\
  \end{array}
\right)
\]
where the diagonal blocks ${\mathbf Q}_r$ explains the mixing between uncorrelated factors in cluster $r$, and the zero off-diagonals indicate that ${\mathbb Q}$ does not capture the dependence between factors of different clusters.

\vspace{-0.1in}

\subsection{Measures of Dependence}
Our aim of assuming factor models is to approximate large covariance matrix with a simpler, lower dimensional structure for efficient network analysis. We now describe the model parameters in (\ref{Eq:ROIfm}) and (\ref{Eq:ModelAll}) which quantify the network dependency at the local and global level.

\textit{1) Local (within-cluster) dependency:} The between-node dependency within cluster $r$ is captured by the covariance matrix $\Sigma_{{\mathbf Y}_r {\mathbf Y}_r}$. The model (\ref{Eq:ROIfm}) implies a decomposition of $\Sigma_{{\mathbf Y}_r {\mathbf Y}_r}$ into a matrix of lower-rank $m_r$ and a diagonal matrix.
\begin{equation} \label{Eq:covROIfm}
\Sigma_{{\mathbf Y}_r {\mathbf Y}_r} = {\mathbf Q}_r \Sigma_{\mathbf{f}_r \mathbf{f}_r} {\mathbf Q}_r' + \Sigma_{{\mathbf E}_r {\mathbf E}_r}
\end{equation}
assuming ${\mathbf f}_r(t)$ to be uncorrelated with ${\mathbf E}_r(t)$.

\textit{2) Global (whole-network) dependency:}
The global model (\ref{Eq:ModelAll}) also implies a low-rank decomposition of the whole-network covariance matrix $\Sigma_{{\mathbf Y} {\mathbf Y}}$, and a block structure
\begin{equation} \label{Eq:covModelAll}
\Sigma_{{\mathbf Y} {\mathbf Y}} = {\mathbb Q} \Sigma_{{\mathbf f}{\mathbf f}} {\mathbb Q}' + \Sigma_{{\mathbf E} {\mathbf E}}
\end{equation}
with
\begin{equation} \label{Eq:covMatAll}
\Sigma_{{\mathbf Y}{\mathbf Y}} =
\left(
  \begin{array}{ccc}
    \Sigma_{{\mathbf Y}_1 {\mathbf Y}_1} & \ldots & \Sigma_{{\mathbf Y}_1 {\mathbf Y}_R} \\
    {\vdots} & \ddots & \vdots \\
    \Sigma_{{\mathbf Y}_R {\mathbf Y}_1} & \ldots & \Sigma_{{\mathbf Y}_R {\mathbf Y}_R} \\
  \end{array}
\right)
\end{equation}
where the diagonal blocks $\Sigma_{{\mathbf Y}_r {\mathbf Y}_r}$ are defined by (\ref{Eq:covROIfm}) for $r=1, \ldots, R$, and the off-diagonal block $\Sigma_{{\mathbf Y}_j {\mathbf Y}_k} = Cov[{\mathbf Y}^{\mathcal C}_j(t), {\mathbf Y}^{\mathcal C}_k(t)] = {\mathbf Q}_j \Sigma_{\mathbf{f}_j \mathbf{f}_k} {\mathbf Q}_k'$ for $j \neq k$ is $n_j \times n_k$ cross-covariance matrix between the node time series ${\mathbf Y}^{\mathcal C}_j(t)$ and ${\mathbf Y}^{\mathcal C}_k(t)$ of cluster $j$ and cluster $k$. The factor decomposition of the high-dimensional node-wise covariance matrices of ${\mathbf Y}^{\mathcal C}_r(t)$ and ${\mathbf Y}(t)$ in (\ref{Eq:covROIfm}) and (\ref{Eq:covModelAll}) allows for massive dimension-reduction. 
It provides an efficient way to compute the large whole-network dependency matrix $\Sigma_{{\mathbf Y}{\mathbf Y}}$ in (\ref{Eq:covMatAll}), by reconstruction from smaller pair-wise between-cluster dependence blocks $\Sigma_{{\mathbf Y}_j {\mathbf Y}_k}$, which can be further approximated by lower-dimensional matrix $\Sigma_{\mathbf{f}_j \mathbf{f}_k}$.
Moreover, the approximation using a low-rank matrix plus a diagonal matrix can produce better-conditioned estimates for the large covariance structure at the both levels. In the following, we make use of the low-dimensional factor covariance $\Sigma_{{\mathbf f} {\mathbf f}}$ together with the RV coefficient to derive a single-valued measure to summarize the node-wise connectivity blocks across clusters and networks at the global level.

\textit{3) Global (between-cluster) dependency.} The factor covariance matrix $\Sigma_{{\mathbf f} {\mathbf f}}$ is a block matrix that model instantaneous (lag zero) dependency structure between clusters 
\[
\Sigma_{{\mathbf f}{\mathbf f}} =
\left(
  \begin{array}{ccc}
    \Sigma_{\mathbf{f}_1 \mathbf{f}_1} & \ldots & \Sigma_{\mathbf{f}_1 \mathbf{f}_R}  \\
    \vdots & \ddots & \vdots \\
    \Sigma_{\mathbf{f}_R \mathbf{f}_1} & \ldots & \Sigma_{\mathbf{f}_R \mathbf{f}_R} \\
  \end{array}
\right).
\]
Each diagonal block $\Sigma_{\mathbf{f}_r \mathbf{f}_r}$ is a $m_r \times m_r$ diagonal covariance matrix that captures the total variance of factors within each cluster. While the factors within a cluster are uncorrelated, factors between different clusters may be correlated. The off-diagonal blocks $\Sigma_{\mathbf{f}_j \mathbf{f}_k} = Cov[\mathbf{f}_j(t), \mathbf{f}_k(t)]$ for $j \neq k$ are $m_j \times m_k$ cross-covariance matrices between factors $\mathbf{f}_j(t)$ and $\mathbf{f}_k(t)$, satisfying $\Sigma_{\mathbf{f}_j \mathbf{f}_k} = \Sigma_{\mathbf{f}_k \mathbf{f}_j}'$, that can summarize cross-dependence between clusters $j$ and $k$.

\textit{4) Global (between-network) dependency:} The measure of dependency between the sub-networks of clusters can be conveniently derived based on the covariances between the factor time series from the clusters in different networks. Let ${\mathbb F}_{s}(t) = [\mathbf{f}_{s 1}^{'}(t), \ldots, \mathbf{f}_{s d_s}^{'}(t)]'$ denotes collectively the corresponding factors of all the clusters in sub-network $s$, with a total dimension $L_s = \sum_{r \in \left\{ 1, \ldots, d_s \right\}} m_r$, which summarizes ${\mathbf Y}^{\mathcal W}_s(t)$. The dependence within a network $s$, $\Sigma_{{\mathbf Y}_s {\mathbf Y}_s}$ can be characterized by the ${L_s} \times {L_s}$ covariance matrix $\Sigma_{{\mathbb F}_s {\mathbb F}_s} = Cov[{\mathbb F}_s(t)]$. The dependence between network $p$ and network $q$ is captured by the ${L_p} \times {L_q}$ cross-covariance matrix between ${\mathbb F}_{p}(t)$ and ${\mathbb F}_{q}(t)$, denoted as $\Sigma_{{\mathbb F}_p {\mathbb F}_q} = Cov[{\mathbb F}_p(t), {\mathbb F}_q(t)]$.

\textit{5) RV coefficients:} The between-cluster and between-network connectivity above are represented by block covariance matrices between multiple factor time series (possibly of different dimensions) across clusters and networks. We propose to use the RV coefficient \citep{Escoufieri1973} as a single-valued measure for the linear dependence between factors of different clusters and networks. It is a multivariate generalization of the squared correlation coefficient which measures normalized dependence between two univariate time series. The RV coefficient between factors in clusters $j$ and $k$ is defined by
\[
RV^C_{jk} = \frac{\tr(\mathbf{C}_{\mathbf{f}_j \mathbf{f}_k} \mathbf{C}_{\mathbf{f}_k \mathbf{f}_j})}{\sqrt{\tr(\mathbf{C}_{\mathbf{f}_j \mathbf{f}_j} \mathbf{C}_{\mathbf{f}_j \mathbf{f}_j}) \tr(\mathbf{C}_{\mathbf{f}_k \mathbf{f}_k} \mathbf{C}_{\mathbf{f}_k \mathbf{f}_k})}}
\]
and between the networks $p$ and $q$ by
\[
RV^S_{pq} = \frac{\tr(\mathbf{C}_{{\mathbb F}_p {\mathbb F}_q} \mathbf{C}_{{\mathbb F}_q {\mathbb F}_p})}{\sqrt{\tr(\mathbf{C}_{{\mathbb F}_p {\mathbb F}_p} \mathbf{C}_{{\mathbb F}_p {\mathbb F}_p}) \tr(\mathbf{C}_{{\mathbb F}_q {\mathbb F}_q} \mathbf{C}_{{\mathbb F}_q {\mathbb F}_q})}}
\]
where $\mathbf{C}_{\mathbf{f}_j \mathbf{f}_k} = (\Sigma_{\mathbf{f}_j \mathbf{f}_j})^{-1/2} \Sigma_{\mathbf{f}_j \mathbf{f}_k} (\Sigma_{\mathbf{f}_k \mathbf{f}_k})^{-1/2}$ \ and $\mathbf{C}_{\mathbb F_p \mathbb F_q} = (\Sigma_{\mathbb F_p \mathbb F_p})^{-1/2} \Sigma_{\mathbb F_p \mathbb F_q} (\Sigma_{\mathbb F_q \mathbb F_q})^{-1/2}$ are the correlation matrices. The correlations and RV coefficients provide results that are more easily interpretable than the covariances when measuring the strength of connectivity, as both $\mathbf{C} = [\rho_{pq}]$, $\rho^2_{pq} \in [0,1]$ and $RV_{pq} \in [0,1]$ are constrained to take values only in the unit interval $[0,1]$. A value of $RV_{pq}$ close to one indicates strong connection between network $p$ and $q$, whereas a value of zero indicates there is no connection.

\vspace{-0.1in}

\subsection{Model Identifiability}
We now discuss the identifiability issue of the covariance of the common component, ${\mathbb Q} {\mathbf f}(t)$ in the global model (\ref{Eq:ModelAll}). One key feature is that the mixing matrix ${\mathbb Q}$ is block diagonal. This guarantees that the cross-dependence between clusters will be captured only by the covariance matrix $\Sigma_{{\mathbf f} {\mathbf f}}$ and not by ${\mathbb Q}$. The dependence between the pair of clusters $j$ and $k$ is directly contained in the cross-covariance matrix $\Sigma_{{\mathbf f}_j{\mathbf f}_k}$.
Similar to the local FA model, ${\mathbb Q}$ and ${\mathbf f}(t)$ are not separately identifiable, since ${\mathbb Q} \mathbf{f}(t) = {\mathbb Q} {\mathbb U} {\mathbb U}^{-1} \mathbf{f}(t) = {\mathbb Q}^* \mathbf{f}^*(t)$ for any invertible matrix ${\mathbb U}$ such that ${\mathbb U} {\mathbb U}^{-1} = {\mathbb I}$, where ${\mathbb Q}^* = {\mathbb Q} {\mathbb U}$ and $\mathbf{f}^*(t) = {\mathbb U}^{-1} \mathbf{f}(t)$. However, the covariance matrix of the common component ${\mathbb Q} {\mathbf f}(t)$ is identifiable as follows
\begin{equation} \notag
Cov[{\mathbb Q} {\mathbf f}(t)] =  {\mathbb Q}\Sigma_{{\mathbf f} {\mathbf f}}{\mathbb Q}' = {\mathbb Q} {\mathbb U} {\mathbb U}^{-1} \Sigma_{{\mathbf f} {\mathbf f}} ({\mathbb U}')^{-1} {\mathbb U}' {\mathbb Q}' = {\mathbb Q}^* \Sigma_{{\mathbf f} {\mathbf f}}^{*} {\mathbb Q}^{*'}
\end{equation}
where ${\mathbb Q}^* = {\mathbb Q} {\mathbb U}$ and $\Sigma_{{\mathbf f} {\mathbf f}}^{*} = {\mathbb U}' \Sigma_{{\mathbf f} {\mathbf f}} ({\mathbb U}')^{-1}$ admit a non-unique factorization.
The key question now is whether the new mixing matrix ${\mathbb Q}^*$ is also block diagonal as required by (\ref{Eq:ModelAll}). To address this important question, we use an example of only two clusters for ease of exposition. Let ${\mathbb Q} = {\mbox{diag}}({\mathbf Q}_1, {\mathbf Q}_2)$,
\[
\Sigma_{{\mathbf f} {\mathbf f}} =
\left(
\begin{array}{cc}
\Sigma_{{\mathbf f}_1 {\mathbf f}_1 } & \Sigma_{{\mathbf f}_1 {\mathbf f}_2 } \\
\Sigma_{{\mathbf f}_2 {\mathbf f}_1 } & \Sigma_{{\mathbf f}_2 {\mathbf f}_2 } \\
\end{array}
\right) \ {\mbox{and}} \ \
{\mathbb U} =
\left(
  \begin{array}{cc}
    {\mathbf U}_{11} & {\mathbf U}_{12} \\
    {\mathbf U}_{21} & {\mathbf U}_{22} \\
  \end{array}
\right)
\]
By expanding on ${\mathbb Q}^*$, we have
\begin{eqnarray*}
{\mathbb Q}^* & = &
\left(
  \begin{array}{cc}
     {\mathbf Q}_1 {\mathbf U}_{11} & {\mathbf Q}_1 {\mathbf U}_{12}  \\
     {\mathbf Q}_2 {\mathbf U}_{21}  & {\mathbf Q}_2 {\mathbf U}_{22}  \\
  \end{array}
\right)
\end{eqnarray*}
For ${\mathbb Q}^*$ to be block diagonal, it is sufficient to set ${\mathbf U}_{12} = {\bf 0}$ and ${\mathbf U}_{21} = {\bf 0}$. However, since ${\mathbb U} {\mathbb U}' = {\mathbb I}$, it follows that ${\mathbf U}_{11} {\mathbf U}_{11}^{'} = {\mathbf I}$ and ${\mathbf U}_{22} {\mathbf U}_{22}^{'} = {\mathbf I}$. Thus, the factor loading matrix will be identifiable up to orthonormal transformations only within each cluster.

\vspace{-0.1in}

\section{Estimation and Inference}

Inferring dependence in a network between a large number of nodes involves estimating the high-dimensional covariance matrix $\Sigma_{{\mathbf Y} {\mathbf Y}}$. The traditional sample covariance matrix is no longer consistent when $N$ is large and is not invertible when $N > T$. Our primary objectives are to estimate the dependence quantities: (1.) $\Sigma_{{\mathbf Y}{\mathbf Y}}$ which models the dependence across all nodes in the entire network; (2.) $\Sigma_{{\mathbf Y}_r {\mathbf Y}_r}$ the dependence across nodes within each cluster; (3.) $\Sigma_{{\mathbf f}_j {\mathbf f}_k}$ and $\Sigma_{{\mathbb F}_p {\mathbb F}_q}$ the dependence between any pairs of clusters and sub-networks of clusters.

\vspace{-0.1in}

\subsection{PCA Estimation}

In this section, we develop a two-step procedure to estimate the high-dimensional dependence based on the proposed MSFA model. The estimation of the whole-network covariance matrix is reduced to the estimation of sub-matrices of much smaller dimensions. The estimation procedure is summarized in Algorithm 1. In Step 1, we apply the method of PCA to estimate the parameters ${\mathbf Q}_r$ and $\mathbf{f}_r(t)$ in the local-level model (\ref{Eq:ROIfm}) to construct the covariance estimates within each cluster. In Step 2, we integrate these local estimators to derive the estimators of the global-level dependence quantities in (\ref{Eq:ModelAll}), i.e. the between-cluster and between-network factor covariance ($\Sigma_{{\mathbf f}_j {\mathbf f}_k}$ and $\Sigma_{{\mathbb F}_p {\mathbb F}_q}$) and RV coefficients ($RV^C_{jk}$ and $RV^S_{pq}$). In the local estimation, the PC estimates of ${\mathbf{f}}_r(t)$ and ${\mathbf Q}_r$ can be computed conveniently via eigenvalue-eigenvector analysis of the sample covariance matrix, for the $n_r \times n_r$ cross-sectional covariance $\Sigma_{{\mathbf Y}_r {\mathbf Y}_r} = E[{\mathbf Y}^{\mathcal C}_r(t) {\mathbf Y}'^{\mathcal C}_r(t)]$ when $T \geq n_r$ [Step 1.1(a)], and on the $T \times T$ temporal covariance $\Sigma^{'}_{{\mathbf Y}_r {\mathbf Y}_r} = E[{\mathbf Y}'^{\mathcal C}_r(t) {\mathbf Y}^{\mathcal C}_r(t)]$ when $T < n_r$ [Step 1.1(b)]. The estimator of the factor loading matrix $\widehat{\mathbf Q}_r$ are eigenvectors corresponding to the $m_r$ principal eigenvalues of the sample covariance, the factors are then estimated as $\widehat{\mathbf f}_r(t) = \widehat{{\mathbf Q}}'_r {\mathbf Y}^{\mathcal C}_r(t)$. The noise covariance $\Sigma_{{\mathbf E}_r {\mathbf E}_r}$ can be estimated based on the residuals [Step 1.2]. We then have a simple substitution estimator for the within-cluster dependence matrix in (\ref{Eq:covROIfm}) using estimates from Step 1.1 - 1.2 [Step 1.3]. In the global estimation, we use the estimated factor signals $\widehat{\mathbf f}_r(t)$ to compute estimators for the pair-wise covariance sub-matrices between clusters and between networks [Step 2.1], to generate the RV coefficient estimates [Step 2.2]. In the final steps [Step 2.3 and 2.4], the parameters of the global model (\ref{Eq:ModelAll}) and the whole-network dependence (\ref{Eq:covMatAll}) can be constructed from the component estimators in the previous steps.


The PCA extracts latent factors $\widehat{\mathbf{f}}_r(t)$ that best represent the region-specific dynamics. It estimates an ordered sequence of factor series that account for most variability of signals across all nodes within a cluster, which might not sufficiently captured by a single mean signal. One of the dominant factors would possibly be the mean. However, instead of making this imposition as in most analyses, our procedure is data-driven where these factors are learned from data according to their significance. Besides, the PC estimators of factors $\widehat{\mathbf f}_r(t)$ and factor loadings $\widehat{\mathbf Q}_r$ are consistent under general framework of large $N$ and large $T$, and in the presence of correlated noise in the signals \citep{StockWatson2002,Bai2003}. The FA model-based estimator of large covariance matrix and its inverse are shown to produce lower estimation errors and attain improved convergence rates under various norms, compared to the sample covariance \citep{Fan2011}. This can provide reliable estimation of large dependency networks.


The number of factors $m_r$ for a cluster can be objectively selected based on some threshold of the amount of variance of the signals within each cluster. The criterion is computed using the eigenvalues of the sample covariance matrix $\Sigma_{{\mathbf Y}_r {\mathbf Y}_r}$ or $\Sigma^{'}_{{\mathbf Y}_r {\mathbf Y}_r}$, which measure the estimated variances of the individual factors. Precisely, $m_r$ can be estimated by
\begin{equation}\label{Eq:var-r}
\widehat{m}_r = \argmax_{\ell\in\{1,2, \ldots, L_r\}} \ \frac{\sum_{i=1}^{\ell} \widehat{\lambda}_{r i}}{\sum_{i=1}^{\kappa} \widehat{\lambda}_{r i}} \ \leq \tau
\end{equation}
where $\kappa = \min(n_r, T)$, $\ell\in\{1,2, \ldots, L_r\}$ is an evaluated candidate value of $m_r$, with $L_r$ a bounded integer such that $m_r \leq L_r \leq \kappa$ and $\tau = [0 ,\ 1]$ is a global threshold for all clusters.
The proportion of variance explained by the first $\ell$ components is equal to the ratio of the sum of $\ell$ largest sample eigenvalues to the sum of all eigenvalues. Naturally, the proportion is subjectively selected by the users via the threshold.
However, once it is specified, PCA can objectively extract a number of optimal latent components. 

\begin{algorithm}[H]\footnotesize
\caption{ PCA for MSFA Model Estimation}
	\begin{algorithmic}[1]
	 \State {\bf Step 1:} \underline{Local Estimation}: Input: $T \times n_r$ data matrix ${\mathbf Y}_r = [{\mathbf Y}^{\mathcal C}_r(1), \ldots, {\mathbf Y}^{\mathcal C}_r(T)]'$, $T \times m_r$ matrix of factor signals ${\mathbf f}_r = [{\mathbf f}_r(1), \ldots, {\mathbf f}_r(T)]'$ at cluster $r$, and number of clusters $R$
		\vspace{-0.05in}
		\begin{enumerate}
			\item[1.1] \textbf{for} $r=1$ to $R$, apply PCA to sample covariance matrix to compute $\widehat{{\mathbf f}}_r$ and $\widehat{{\mathbf Q}}_{r}$.
			\vspace{-0.1in}
			\begin{enumerate}[label=(\alph*),leftmargin=-0.01in]

			\item \textbf{if} $T$ $ \geq $ $n_r$ \textbf{then}

			\begin{itemize}

			\item Compute orthonormal eigenvectors ${\mathbf V}_{r 1}$, $\ldots$, ${\mathbf V}_{r n_r}$ and associated eigenvalues $\widehat{\lambda}_{r 1} \geq \ldots \geq \widehat{\lambda}_{r n_r} >0$ of cross-sectional sample covariance matrix ${\mathbf Y}_r' {\mathbf Y}_r / T$

			\item Compute $\widehat{m}_r$ according to (\ref{Eq:var-r}) or (\ref{Eq:BIC})

			\item Define $\widehat{\mathbf{Q}}_{r} = [ \mathbf{V}_{r 1}, \ldots, \mathbf{V}_{r m_r}] \in {\mathbb{R}}^{n_r \times m_r}$ eigenvectors corresponding to the $m_r$ largest eigenvalues $\widehat{\lambda}_{r 1}, \ldots, \widehat{\lambda}_{r m_r}$, and $\widehat{\Sigma}_{{\mathbf f}_r {\mathbf f}_r} = {\mbox{diag}}(\widehat{\lambda}_{r 1}, \ldots, \widehat{\lambda}_{r m_r})$

			\item Compute $\widehat{\mathbf f}_r = {\mathbf Y}_r	\widehat{{\mathbf Q}}_r$ and $\widehat{\Sigma}_{{\mathbf f}_r {\mathbf f}_r} = \widehat{\mathbf f}'_r \widehat{\mathbf f}_r / T$

			\end{itemize}

			\item \textbf{if} $T$ $ < $ $n_r$ \textbf{then}

			\begin{itemize}

			\item Compute orthonormal eigenvectors ${\mathbf V}_{r 1}$, $\ldots$, ${\mathbf V}_{r T}$ and associated eigenvalues $\widehat{\lambda}_{r 1} \geq \ldots \geq \widehat{\lambda}_{r T} >0$ of temporal sample covariance matrix ${\mathbf Y}_r {\mathbf Y}_r' / T$

			\item Compute $\widehat{m}_r$ according to (\ref{Eq:var-r}) or (\ref{Eq:BIC})

			\item Define $\mathbf V = [ \mathbf{V}_{r 1}, \ldots, \mathbf{V}_{r m_r}] \in {\mathbb{R}}^{T \times m_r}$ eigenvectors corresponding to the $m_r$ largest  eigenvalues, and $\mathbf{D} = {\mbox{diag}}(\widehat{\lambda}_{r 1}, \ldots, \widehat{\lambda}_{r m_r})$
			
			\item Compute $\widehat{\mathbf{f}}_{r}= \sqrt{T} \mathbf V  \mathbf{D}^{1/2}$ and $\widehat{\Sigma}_{{\mathbf f}_r {\mathbf f}_r} = \mathbf{D}$
			
			\item Compute $\widehat{{\mathbf Q}}_r = {\mathbf Y}'_r \widehat{\mathbf f}_r \mathbf{D}^{-1} / T$
			
			\end{itemize}

			\end{enumerate}

			\item[1.2] Compute noise covariance from residuals $\widehat{\Sigma}_{{\mathbf E}_r {\mathbf E}_r} = {\mbox{diag}}(\widehat{\sigma}_{e_{r1}}^2, \ldots, \widehat{\sigma}_{e_{r n_r}}^2)$, $\widehat{\sigma}_{e_{ri}}^2 = \frac{1}{T} \sum_{t=1}^T \widehat{e}_{ri}^2(t)$ with $\widehat{\mathbf E}_r(t) = {\mathbf Y}_{r}(t) - \widehat{\mathbf{Q}}_{r} \widehat{\mathbf{f}}_r(t)$

			\vspace{-0.05in}

			\item[1.3] Estimate within-cluster dependency $\widehat{\Sigma}_{{\mathbf Y}_r {\mathbf Y}_r} = \widehat{\mathbf Q}_r \widehat{\Sigma}_{\mathbf{f}_r \mathbf{f}_r} \widehat{{\mathbf Q}}_r' + \widehat{\Sigma}_{{\mathbf E}_r {\mathbf E}_r}$
			
		\end{enumerate}
		
	\State {\bf Step 2:} \underline{Global Estimation}:

		\begin{enumerate}

		\item[2.1] Estimate between-cluster and between-network dependency, $\widehat{\Sigma}_{{\mathbf f}_j {\mathbf f}_k} = \frac{1}{T} \sum_{t=1}^T \widehat{\mathbf f}_j(t) \widehat{\mathbf f}'_k(t)$ and $\widehat{\Sigma}_{{\mathbb F}_p {\mathbb F}_q} = \frac{1}{T} \sum_{t=1}^T \widehat{\mathbb F}_p(t) \widehat{\mathbb F}'_q(t)$
		for all $j \neq k$ and $p \neq q$. $\widehat{\mathbb F}_{p}(t) = [\widehat{\mathbf{f}}_{p1}^{'}(t), \ldots, \widehat{\mathbf{f}}_{p d_p}^{'}(t)]'$ and $\widehat{\mathbb F}_{q}(t) = [\widehat{\mathbf{f}}_{q1}^{'}(t), \ldots, \widehat{\mathbf{f}}_{q d_q}^{'}(t)]'$ are estimated cluster- and network-specific factor signals.

		\item[2.2] Estimate RV-based between-cluster and between-network dependency $\widehat{RV}^C_{jk}$ and $\widehat{RV}^S_{pq}$ by substitution using $\widehat{\mathbf C}_{{\mathbf f}_j {\mathbf f}_k} = (\widehat{\Sigma}_{\mathbf{f}_j \mathbf{f}_j})^{-1/2} \widehat{\Sigma}_{\mathbf{f}_j \mathbf{f}_k} (\widehat{\Sigma}_{\mathbf{f}_k \mathbf{f}_k})^{-1/2}$ and $\widehat{\mathbf C}_{{\mathbb F}_p {\mathbb F}_q} = (\widehat{\Sigma}_{\mathbb F_p \mathbb F_p})^{-1/2} \widehat{\Sigma}_{\mathbb F_p \mathbb F_q} (\widehat{\Sigma}_{\mathbb F_q \mathbb F_q})^{-1/2}$.

		\item[2.3] Estimate global mixing matrix by $\widehat{\mathbb Q} = {\mbox{diag}}(\widehat{\mathbf Q}_1, \ldots, \widehat{\mathbf Q}_R)$, noise covariance by $\widehat{\Sigma}_{{\mathbf E} {\mathbf E}} = {\mbox{diag}}(\widehat{\Sigma}_{{\mathbf E}_1 {\mathbf E}_1}, \ldots, \widehat{\Sigma}_{{\mathbf E}_R {\mathbf E}_R})$ and global factor covariance matrix $\widehat{\Sigma}_{{\mathbf f} {\mathbf f}}$ by substituting estimated blocks $\widehat{\Sigma}_{{\mathbf f}_r {\mathbf f}_r}$ on the diagonal elements and $\widehat{\Sigma}_{{\mathbf f}_j {\mathbf f}_k}$, $j \neq k$ for the off-diagonals.

		\item[2.4] Estimate whole-network dependency $\widehat{\Sigma}_{{\mathbf Y} {\mathbf Y}} = \widehat{{\mathbb Q}} \widehat{\Sigma}_{\mathbf{f} \mathbf{f}} \widehat{{\mathbb Q}}' + \widehat{\Sigma}_{{\mathbf E} {\mathbf E}}$ or by substituting into (\ref{Eq:covMatAll}) the estimated elementary blocks $\widehat{\Sigma}_{{\mathbf Y}_r {\mathbf Y}_r}$, $r=1, \ldots, R$ and $\widehat{\Sigma}_{{\mathbf Y}_j {\mathbf Y}_k} = \widehat{{\mathbf Q}}_j \widehat{\Sigma}_{\mathbf{f}_j \mathbf{f}_k} \widehat{{\mathbf Q}}_k'$, $j \neq k$.

		\end{enumerate}

	\end{algorithmic}
\end{algorithm}

Alternative method is via model selection using BIC \citep{Bai2003}
\begin{equation}\label{Eq:BIC}\begin{split}
\widehat{m}_r = \argmax_{\{1, \ldots, L_r\}} \left\{ \ln \left(\frac{1}{n_r T} \sum_{t=1}^{T} \|\widehat{\bf E}_t(r)\|_2^2\right) \right. \\ 
\left. + r \left(\frac{n_r+T}{n_r T}\right) \ln\left(\frac{n_r T}{n_r+T}\right) \right\}
\end{split}
\end{equation}
where $\|{\bf x}\|_2$ denotes the Euclidean norm of a vector ${\bf x}$

\vspace{-0.1in}
\subsection{Asymptotic Properties of the Estimator}

We first specify some regularity conditions (Assumptions 1-4 in Appendix 7.1). In following Proposition, we present the limiting distributions for the PCA estimates $\widehat{{\mathbf f}}_r(t)$ and $\widehat{{\mathbf Q}}_r$ for each cluster $r$, as defined in Step 1.1 (a) of Algorithm 1. It follows from results derived in \citep{Bai2003} (Theorem 1 and 2) and \citep{bai2013} (Theorem 1). The consistency of $\widehat{{\mathbf f}}_r(t)$ and $\widehat{{\mathbf Q}}_r$ can also be established based on results in \citep{StockWatson2002}.

\textit{Proposition 1 (Asymptotic normality of $\widehat{{\mathbf f}}_r(t)$ and $\widehat{{\mathbf Q}}_r$):} Suppose that Assumptions 1-4 and additional Assumptions E-G in \citep{Bai2003} hold. Let $\widehat{\bf q}_{ri}$ be the $i$-th row vector of $\widehat{\mathbf Q}_r = [\widehat{\bf q}_{r1}, \ldots, \widehat{\bf q}_{r n_r}]'$. Then as $n_r$, $T \rightarrow \infty$ with $\sqrt{T}/n_r \rightarrow 0$, we have for each $i$
\begin{equation}\label{asym-Q}
\sqrt{T}(\widehat{\bf q}_{ri} - {\bf q}_{ri}) \stackrel{d}{\rightarrow} N(\bf{0}, (\Sigma'_{\mathbf{f}_r \mathbf{f}_r})^{-1} {\mathbf \Phi}_{\textit{i}} \Sigma^{-1}_{\mathbf{f}_r \mathbf{f}_r})
\end{equation}
Furthermore, if $\sqrt{n_r}/T \rightarrow 0$, for each $t$
\begin{equation}\label{asym-f}
\sqrt{n_r}(\widehat{{\mathbf f}}_r(t) - {\mathbf f}_r(t)) \stackrel{d}{\rightarrow} N(\bf{0}, {\mathbf \Gamma}_{\textit{t}}).
\end{equation}
Our proposed local (within-cluster) factor-based covariance estimator is a special case of the principal orthogonal complement thresholding (POET) estimator of Fan \citep{Fan2013}. The POET estimates a sparse error covariance matrix for the approximate factor model (correlated noise) by adaptive thresholding of principal orthogonal complements (i.e. remaining components of the sample covariance after taking out the first $m_r$ PCs), as computed by $\widehat{\Sigma}_{{\mathbf E}_r {\mathbf E}_r} = 1/T \sum_{t=1}^{T} {\mathbf E}_r(t){\mathbf E}'_r(t)$. The diagonal matrix computed in Step 1.2 is an extreme case of this sparse covariance estimate by choosing a threshold of correlation elements equal to one. Thus, we can derive the rates of convergence for our estimator $\widehat{\Sigma}_{{\mathbf Y}_r {\mathbf Y}_r}$ for a strict factor model (uncorrelated noise) based on results in \citep{Fan2013} for the POET large covariance estimator under various norms. We consider, for example, the weighted quadratic norm ${\left\| {\mathbf A} \right\|}_{\Sigma} = {n_r}^{-1/2} \left\| {\Sigma}^{-1/2}{\mathbf A}{\Sigma}^{-1/2}  \right\|_F$, as in following Proposition.

\textit{Proposition 2 (Rate of convergence for $\widehat{\Sigma}_{{\mathbf Y}_r {\mathbf Y}_r}$):} Under Assumptions 1-4, the within-cluster covariance estimator as defined in Step 1.3 of Algorithm 1 satisfies
\begin{equation}\label{rate-cov}
{\left\| \widehat{\Sigma}_{{\mathbf Y}_r {\mathbf Y}_r} - \Sigma_{{\mathbf Y}_r {\mathbf Y}_r} \right\|}_{\Sigma} = O_p(\frac{\sqrt{n_r}\text{log} (n_r)}{T} + m_n \omega_T^{1-\delta})
\end{equation}
where
\begin{equation*}
\omega_T = \sqrt{\frac{\text{log} (n_r)}{T}} + \frac{1}{\sqrt{n_r}}
\end{equation*}
and $m_n = \max_{i \leq n_r} {\sigma}_{e_{ii}}^{2\delta}$ for some $\delta \in [0, \ 1]$ is the measure of sparsity condition on $\Sigma_{{\mathbf E}_r {\mathbf E}_r}$.

The consistency of the constructed global covariance estimator (defined in Step 2.5) for high dimensions can be implied by the consistency of estimator for each of individual component blocks at the local level. In this paper, the improved consistency is shown by simulation in Table IV (Section IV.C), as indicated by the lower estimation standard errors compared to other high-dimensional covariance estimators. The complete proof of the consistency and convergence rates for our novel estimator will be developed in future work.

\vspace{-0.1in}
\subsection{Statistical Inference}

To test for the statistical significance of the between-cluster and between-network
dependence, as measured by the RV coefficient, we propose a formal inferential
procedure for testing the null versus alternative hypotheses
\vspace{-0.05in}
\[
H_0: \ \ RV_{jk} = 0  \ \ {\mbox{vs.}} \ \ H_1: RV_{jk} > 0
\]
which denotes the absence or presence of a significant connectivity between the clusters (or networks) $j$ and $k$. 

A large value of the sample RV coefficient may not necessarily imply statistical significance in connectivity because it needs to be compared to some reference null distribution. 
One way of approximating null distribution of RV coefficients is by random permutation of the temporal order of factor series ${\mathbf f}_r(1), \ldots, {\mathbf f}_r(T)$ and computes the RVs. However, this is computationally expensive as there are $T!$ possible permutations to repeat.
We follow \citep{Josse2008} to approximate the exact null distribution. Standardized RV (or z-score) is used as the test statistics
\vspace{-0.02in}
\[
t_{RV} = \frac{\widehat{RV}_{jk} - E_{H_0}(\widehat{RV}_{jk})}{\sqrt{{Var}_{H_0}(\widehat{RV}_{jk})}}
\]
where $E_{H_0}\widehat{RV}$ and ${Var}_{H_0}\widehat{RV}$ are the estimates of first and
second moments of null distribution of the permuted RVs \citep{Kazi-Aoual1995}
\begin{equation}\label{E-RV}
E_{H_0}(\widehat{RV}) = \frac{\sqrt{\beta_j \times \beta_k}}{T - 1} \ \ \text{with} \ \ \beta_i = \frac{(\tr(\Sigma_{\mathbf{f}_i \mathbf{f}_i}))^2}{\tr(\Sigma_{\mathbf{f}_i \mathbf{f}_i} \Sigma_{\mathbf{f}_i \mathbf{f}_i})}
\end{equation}
\vspace{-0.12in}
\begin{equation}\label{Var-RV}
{Var}_{H_0}(\widehat{RV}_{jk}) = \frac{2(T - 1 - \beta_j)(T - 1 - \beta_k)}{(T + 1)(T - 1)^2(T - 2)} \left(1 + \frac{T - 3}{2 T (T-1)} \tau_j \tau_k\right)
\end{equation}
\vspace{-0.2in}
where
\begin{equation} \notag
\tau_j = \frac{T - 1}{(T - 3)(T - 1 -\beta_j)} \left( T(T + 1) \frac{\sum_{i} [\widehat{\mathbf f}_j' \widehat{\mathbf f}_j]_{ii}^2}{\tr((\widehat{\mathbf f}_j' \widehat{\mathbf f}_j)(\widehat{\mathbf f}_j' \widehat{\mathbf f}_j))} - (T - 1)(\beta_j +2) \right).
\end{equation}
$[{\mathbf H}]_{ii}$ denotes the $i$-th diagonal element of matrix ${\mathbf H}$. The test statistics $t_{RV}$ has an asymptotic standard normal distribution $N(0,1)$ under the null hypothesis where the true RV coefficient is zero.
In practice, $\beta_j$ and $\beta_k$ are not known and hence are estimated by replacing
$\Sigma_{\mathbf{f}_j}$ and $\Sigma_{\mathbf{f}_k}$ by their corresponding estimators
$\widehat{\Sigma}_{\mathbf{f}_j}$ and $\widehat{\Sigma}_{\mathbf{f}_k}$ respectively.
A connection is considered statistically significant if the absolute value of $t_{RV}$ for the estimated RV coefficient is greater than a threshold at the $p$-th percentile of $N(0,1)$, which is set here as $p = 100\times(1-\alpha/(2D))$ with $\alpha$ the significance level and $D$ the number of coefficients to be tested. The significance level from testing multiple connectivity entries are adjusted using the Bonferroni method. Note that in (\ref{E-RV}), the sampling distribution of sample RV coefficients even under null, depends on the sample size $T$ and the complexity of the covariance matrix for each pair of clusters, as encoded by $\beta_i$. The RVs take high values when $T$ is small and $\Sigma_{\mathbf{f}_i \mathbf{f}_i}$ is very high-dimensional. Modified versions of RV coefficients \citep{jose2014} can be used in future works to solve this limitation.

\vspace{-0.05in}


\section{Simulations}

We evaluate the proposed MSFA model for estimating large-scale, community structure of connectivity networks at the node-cluster level, using simulated network data.
We focus on comparing the performance of (1.) the MSFA model-based estimator with the sample covariance matrix and other large-dimensional covariance estimators for the whole-network (between-node) connectivity and (2.) the RV-based coefficients using the cluster-specific factor time-series with that using mean time-series for the between-cluster connectivity. The connectivity between time series from distinct network nodes typically exhibits high level of correlations (both auto- and cross-) and modularity. To emulate both these dependence structure, we used a covariance-stationary structured vector autoregressive (VAR) model of order one, to generate the node wise time-series for $R$ clusters
\begin{equation}\label{Eq:StrucVAR}
{\mathbf Y}(t) = \mathbf\Phi {\mathbf Y}(t-1) \ + \ {\mathbf W}(t)
\end{equation}
where $\mathbf\Phi$ is a $N \times N$ global VAR coefficient matrix that measures effective (directed) connectivity of the whole network, and with a block structure to represent the network modules.
The diagonal block ${\mathbf\Phi}_{rr}$ is the $n_r \times n_r$ coefficient matrix which quantifies the dependence of ${\mathbf Y}^{\mathcal C}_r(t)$ on ${\mathbf Y}^{\mathcal C}_r(t-1)$ and measures the directed connectivity between nodes within the cluster $r$. A non-zero $(i, j)-$element of ${\mathbf\Phi}_{rr}$ indicates the presence of directed influence in a Granger-causality sense from node $j$ to node $i$ within cluster $r$. The off-diagonal block ${\mathbf\Phi}_{jk}$, $j \neq k$ measures the directed influence from cluster $k$ to cluster $j$. ${\mathbf W}(t) = [{\mathbf W}_{1}^{'}(t), \ldots, {\mathbf W}_{R}^{'}(t)]'$ is a $N \times 1$ Gaussian white noise with $E[{\mathbf W}(t)] = {\bf 0}$ and $Cov[{\mathbf W}(t)] = \Sigma_{{\mathbf W} {\mathbf W}} = \sigma^2_{\mathbf W} \mathbf{I}_N$.

Under model (\ref{Eq:StrucVAR}), the functional connectivity in ${\mathbf Y}(t)$ is related
to its effective counterpart by
\begin{equation} \label{Eq:covStrucVAR}
\Sigma_{{\mathbf Y} {\mathbf Y}} = {\mathbf \Phi} \Sigma_{{\mathbf Y} {\mathbf Y}} {\mathbf \Phi}' \ + \ \Sigma_{{\mathbf W} {\mathbf W}}
\end{equation}
where the process covariance matrix is invariant over time i.e. $\Sigma_{{\mathbf Y} {\mathbf Y}} = Cov[{\mathbf Y}(t)] = Cov[{\mathbf Y}(t-1)]$, which can be re-written in a vectorized form \
$vec( \Sigma_{{\mathbf Y} {\mathbf Y}} ) = {(\mathbf{I}_{N^2} - \mathbf A)}^{-1} vec( \Sigma_{{\mathbf W} {\mathbf W}} )
$ \
where $\mathbf A$ = $({\mathbf \Phi}\otimes{\mathbf \Phi})$ is a $N^2 \times N^2$ matrix. All absolute eigenvalues of ${\mathbf \Phi}$ are assumed to be less than one to ensure stationarity of the process and invertibility of $\mathbf A$.

We construct a synthetic network of 5 clusters ${\mathcal C}_1, \ldots, {\mathcal C}_5$ each with 25 nodes. We assume following structure for the ground-truth connectivity matrix for generating process (\ref{Eq:StrucVAR})
\[
\widetilde{\mathbf\Phi} =
\left(
  \begin{array}{ccccc}
    \widetilde{\mathbf\Phi}_{11} & \widetilde{\mathbf\Phi}_{12} & {\bf 0} & {\bf 0} & {\bf 0} \\
    {\bf 0} & \widetilde{\mathbf\Phi}_{22} & {\bf 0} & {\bf 0} & \widetilde{\mathbf\Phi}_{25} \\
    {\bf 0} & {\bf 0} & \widetilde{\mathbf\Phi}_{33} & {\bf 0} & {\bf 0} \\
		\widetilde{\mathbf\Phi}_{41} & {\bf 0} & \widetilde{\mathbf\Phi}_{43} & \widetilde{\mathbf\Phi}_{44} & \widetilde{\mathbf\Phi}_{45} \\
		{\bf 0} & {\bf 0} & {\bf 0} & {\bf 0} & \widetilde{\mathbf\Phi}_{55} \\
  \end{array}
\right).
\]
The synthetic network is modular, allowing strong connections between nodes within each cluster but weak connections across different clusters. The nodes within clusters are allowed to be highly inter-connected but scarcely connected to nodes in other clusters, i.e. all diagonal-blocks $\widetilde{\mathbf\Phi}_{rr}$ have full entries and the non-zero off-diagonal blocks $\widetilde{\mathbf\Phi}_{jk}$ are sparse. All other entries of $\widetilde{\mathbf\Phi}$ are zero. The values of the non-zero entries were randomly selected. The noise variance is set $\sigma^2_{\mathbf W} = 0.1$.

We used one realization of the synthetic VAR coefficient matrix $\widetilde{\mathbf\Phi}$ and the implied covariance matrix $\widetilde\Sigma_{{\mathbf Y} {\mathbf Y}}$ by (\ref{Eq:covStrucVAR}) to generate synthetic network time series data. We computed the corresponding correlation matrix $\widetilde{\mathbf C}_{{\mathbf Y} {\mathbf Y}} = (\widetilde{\Sigma}_{{\mathbf Y} {\mathbf Y}})^{-1/2} \widetilde{\Sigma}_{{\mathbf Y} {\mathbf Y}} (\widetilde{\Sigma}_{{\mathbf Y} {\mathbf Y}})^{-1/2}$ and RV coefficient matrix $\widetilde{RV}_{{\mathbf Y} {\mathbf Y}} = [\widetilde{RV}_{jk}]$ from $\widetilde\Sigma_{{\mathbf Y} {\mathbf Y}}$, as ground-truth for evaluation for the whole-network between-node and between-cluster connectivity. We compared the MSFA model-based estimator $\widehat{\mathbf C}_{{\mathbf Y} {\mathbf Y}} = (\widehat{\Sigma}_{{\mathbf Y} {\mathbf Y}})^{-1/2} \widehat{\Sigma}_{{\mathbf Y} {\mathbf Y}} (\widehat{\Sigma}_{{\mathbf Y} {\mathbf Y}})^{-1/2}$ (as computed in Step 2.4 of Algorithm 1) with the sample correlation matrix, and the estimator $\widehat{RV}_{{\mathbf Y} {\mathbf Y}}$ computed by using the mean with using the factor series of each cluster.
To investigate the performance of the estimators under different scenarios of dimensionality $T < N$, $T \approx N$ and $T > N$, we varied the number of time points from $T = 50$ to $T = 250$ with an increment of 25. The time series dimension is fixed as $N = 125$ with $n_r = 25$ per cluster. 100 simulations were repeated for each $T$. To measure the estimation performance, we evaluated the total squared errors over all entries between the true and the estimated dependency matrices, defined for the whole-network between-node and between-cluster connectivity as $\|\widehat{\mathbf C}_{{\mathbf Y} {\mathbf Y}} - \widetilde{\mathbf C}_{{\mathbf Y} {\mathbf Y}}\|_F^2$ and $\|\widehat{RV}_{{\mathbf Y} {\mathbf Y}} - \widetilde{RV}_{{\mathbf Y} {\mathbf Y}}\|_F^2$, where $\|{\bf H}\|_F = {\tr({\bf H}'{\bf H})}^{1/2}$ denotes Frobenius norm of matrix ${\bf H}$.

\vspace{-0.05in}
\subsection{Results for Fixed $m_r$}

\begin{figure}[!ht]
\hspace{-0.55 cm}
\centering
	\begin{minipage}[t]{0.45\linewidth}
		\centering
		\subfloat[${\mathbf C}_{{\mathbf Y} {\mathbf Y}}$, Fixed $m_r$]{\includegraphics[width=1\linewidth,keepaspectratio]{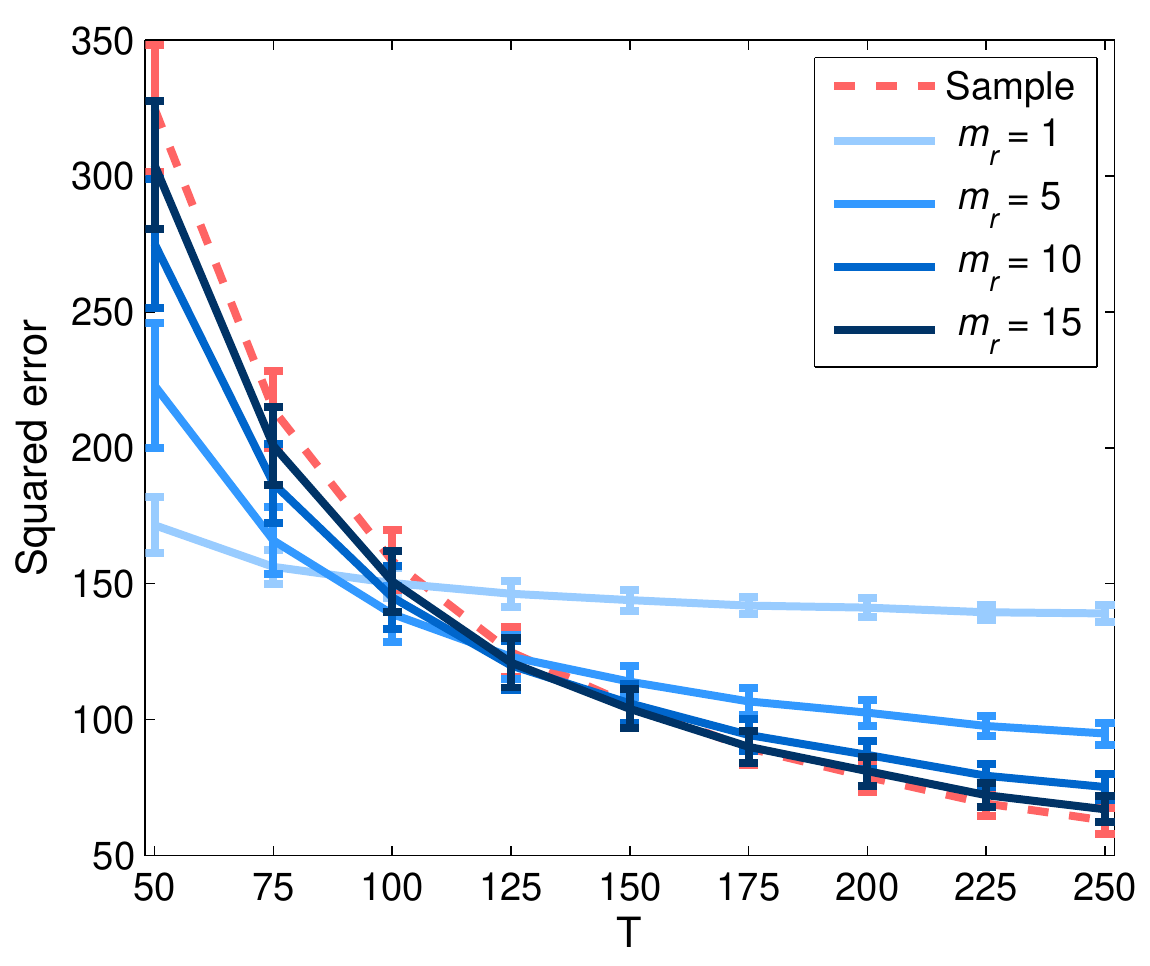}}
	\end{minipage}
	\begin{minipage}[t]{0.45\linewidth}
		\centering
		\subfloat[${RV}_{{\mathbf Y} {\mathbf Y}}$, Fixed $m_r$]{\includegraphics[width=1\linewidth,keepaspectratio]{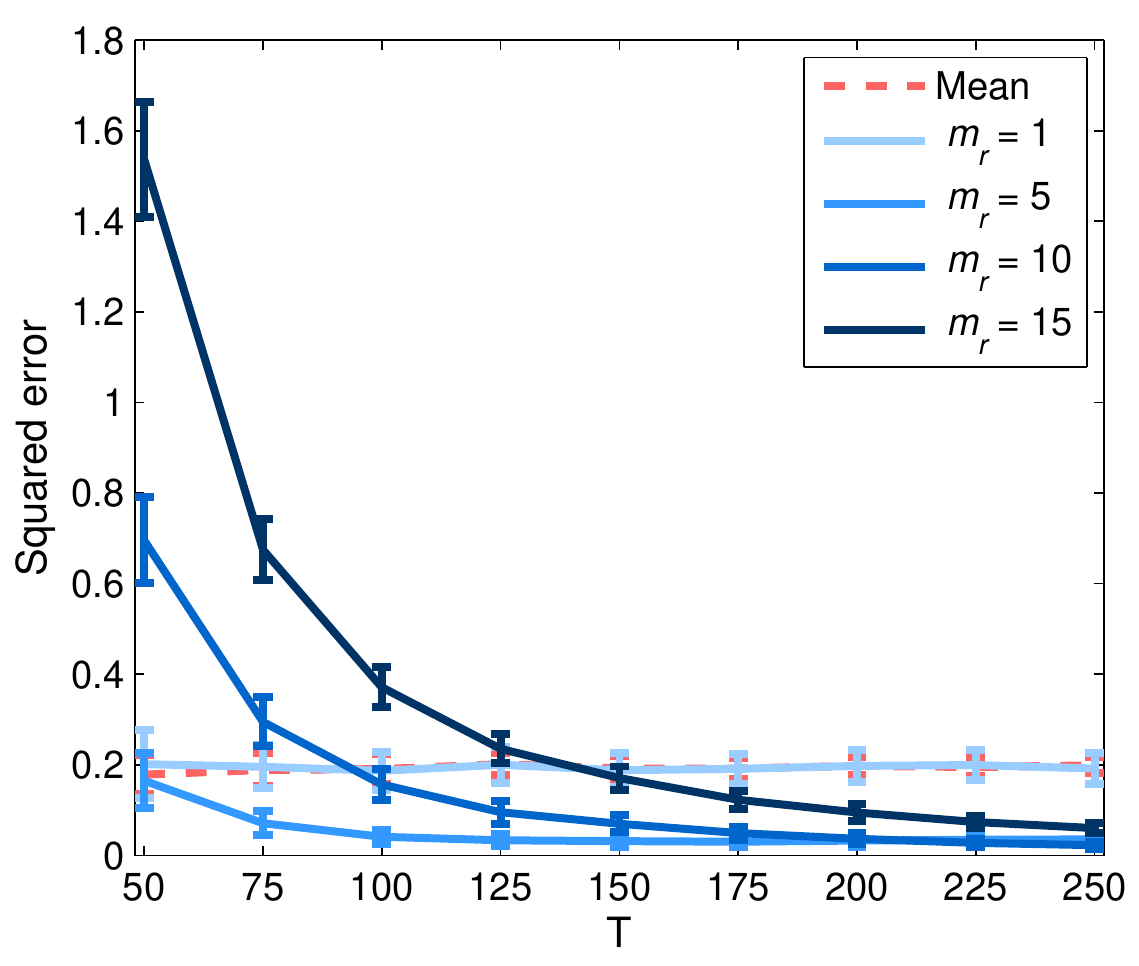}}
	\end{minipage} \\ \vspace{0.2 cm}
	\hspace{-0.55 cm}
		\begin{minipage}[t]{0.45\linewidth}
		\centering
		\subfloat[${\mathbf C}_{{\mathbf Y} {\mathbf Y}}$, Adaptive $m_r$]{\includegraphics[width=1\linewidth,keepaspectratio]{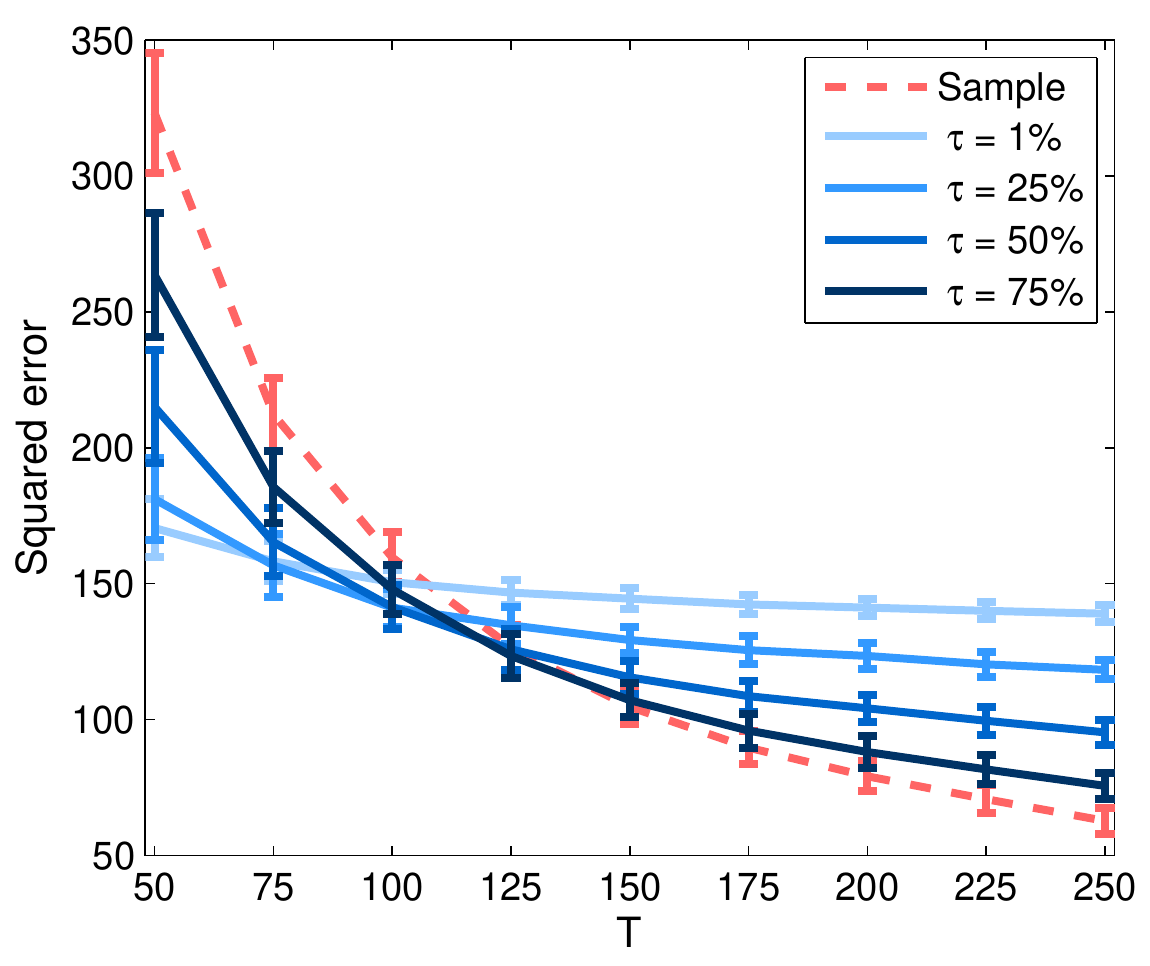}}
	\end{minipage}
	\begin{minipage}[t]{0.45\linewidth}
		\centering
		\subfloat[${RV}_{{\mathbf Y} {\mathbf Y}}$, Adaptive $m_r$]{\includegraphics[width=1\linewidth,keepaspectratio]{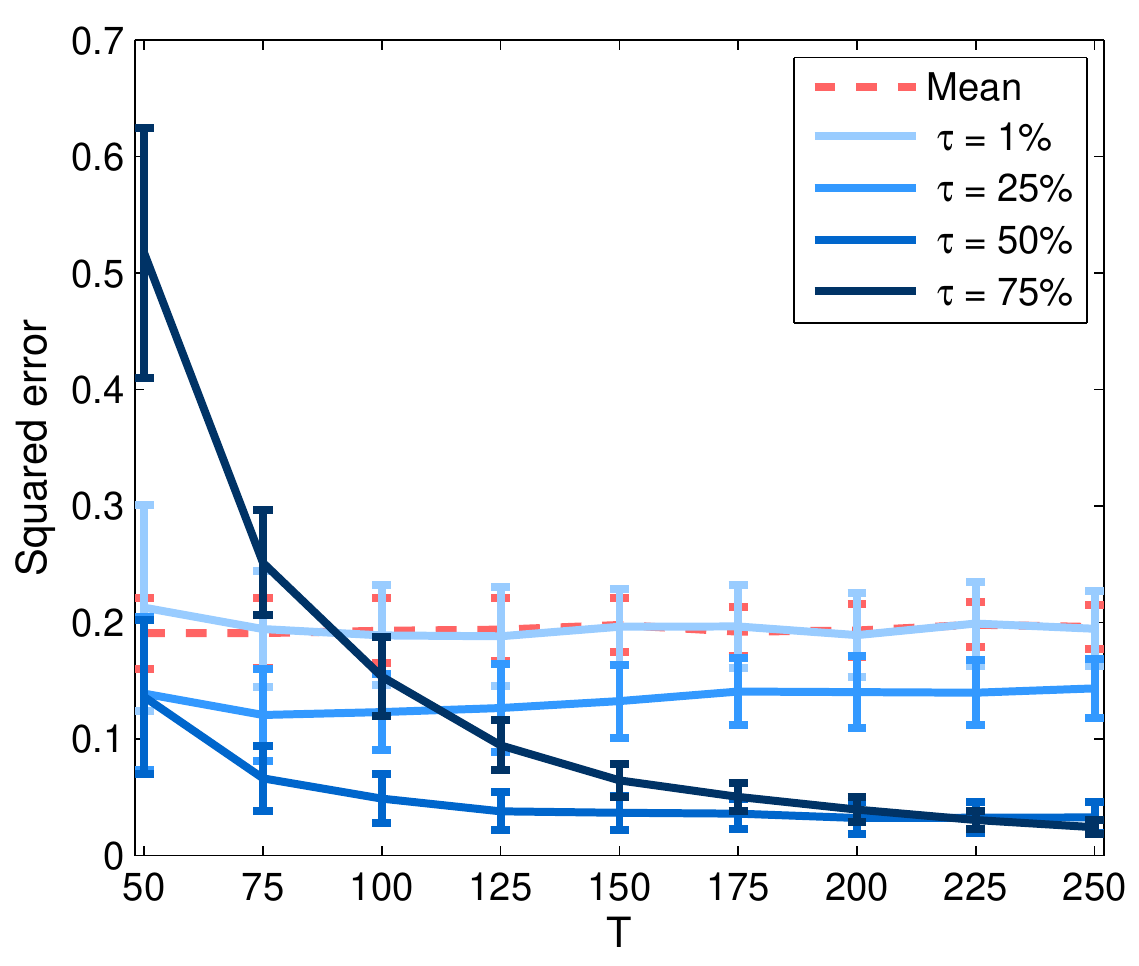}}
	\end{minipage}
\caption{Estimation errors in Frobenious norm as a function of sample size $T$. (a),(c) Correlation matrix of whole-network between-node connectivity $\widehat{\mathbf C}_{{\mathbf Y} {\mathbf Y}}$ using sample correlation matrix and MSFA estimator. (b),(d) RV coefficients of between-cluster connectivity $\widehat{RV}_{{\mathbf Y} {\mathbf Y}}$ using mean and factor time series. Number of factors for MSFA estimators: (Top) Fixed for each cluster $m_r =$ 1, 5, 10 and 15. (Bottom) Selected for each cluster based on percentage of data variance explained with thresholds $\tau = 1\%, 25\%, 50\%$ and $75\%$. Lines and error bars represent averages and standard deviations over 100 realizations of a covariance-stationary VAR model for simulated network time series data.}
\label{Fig:SimMSE}
\end{figure}

Fig.~\ref{Fig:SimMSE} plots the means and standard deviations of estimation errors under Frobenius norm for various estimators over 100 replications of simulations, as a function of sample size $T$. In Fig.~\ref{Fig:SimMSE}(top), we evaluate the factor-based estimates for different number of factors $m_r =$ 1, 5, 10 and 15 fixed for each cluster. 

For the node-wise connectivity (Fig.~\ref{Fig:SimMSE} (a)), when $T$ is small relative to dimension $N$, the MSFA estimator of $\widehat{\mathbf C}_{{\mathbf Y} {\mathbf Y}}$ clearly outperforms the sample correlation matrix, with substantially lower estimation errors especially when $T < N$. This suggests the robustness of the MSFA estimator in small-sample settings, probably due to construction of the large covariance matrix from lower-dimensional factor-based sub-matrices, which are reliably estimated by the PCA based on larger amount of data as $T > n_r$. As expected, the sample covariance with small samples produces poor estimate, which is ill-conditioned when $T \approx N$ and becomes singular when $T < N$. Using smaller number of factors $m_r$ tends to perform better than large $m_r$ when $T$ is small. This may be because the number of reliably estimated principal components is limited by the sample size $T$ and inclusion of more components when $T$ is small will probably induce noisy estimates. Another reason is that the factor-based covariance estimator using more components will converge to the sample covariance matrix as a limit.

As expected, estimation errors of both methods drop as $T$ increases. When $T$ is large, $T > N$ where more data is available for estimation, the sample covariance, however, performs better than factor-based estimator which basically relies on subspace approximation of the full covariance matrix. In contrast to small $T$, factor-based estimates perform better with increase of $m_r$ under large $T$, because additional components can better explain connectivity structure in the data.


\begin{figure}[ht!]
	\hspace{0.5 cm}
	\begin{minipage}[t]{0.27\linewidth}
		\centering
		\subfigure[]{\includegraphics[width=1.1\linewidth,keepaspectratio]{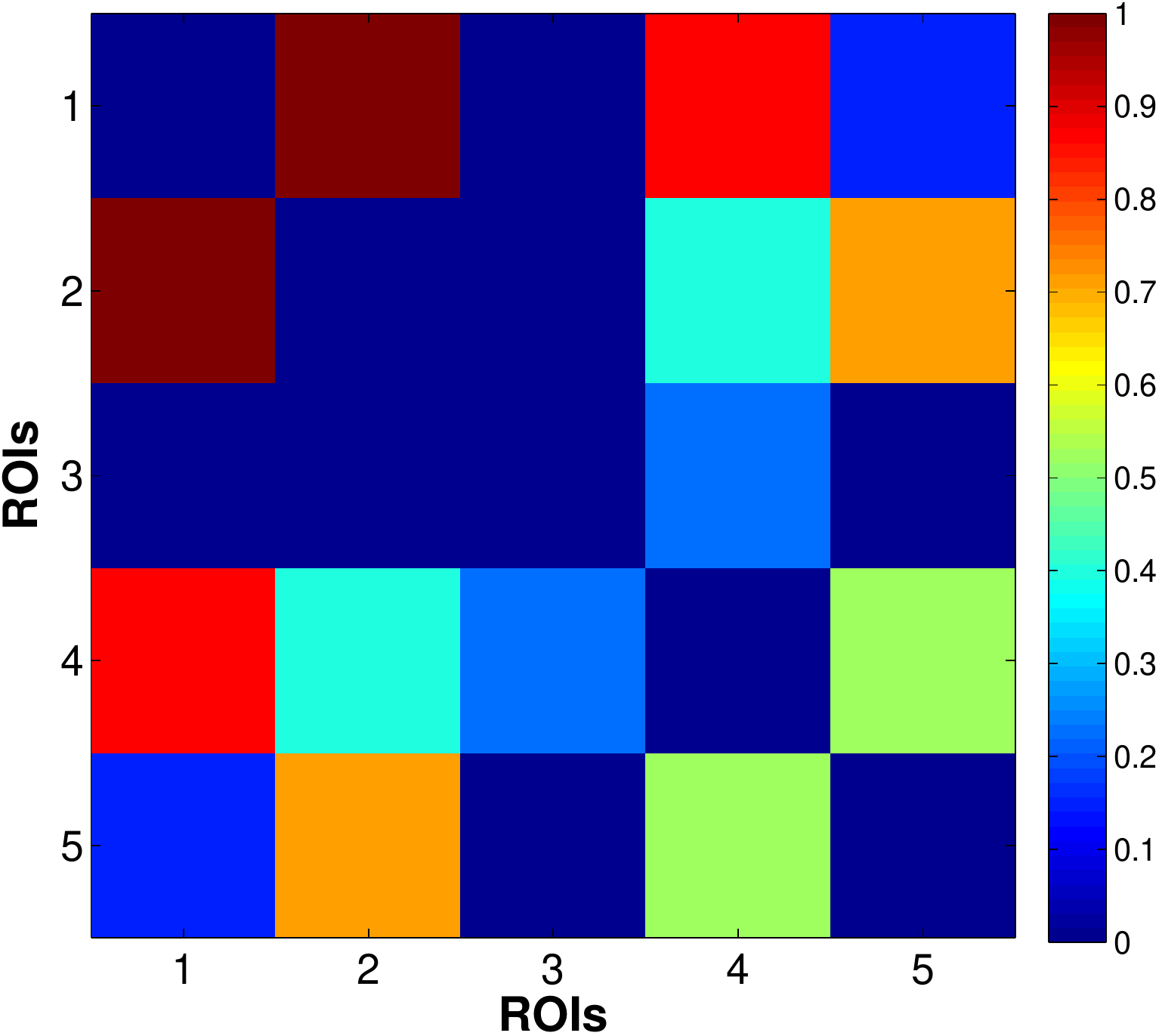}}
	\end{minipage}
	\hspace{0.5 cm}
	\begin{minipage}[t]{0.27\linewidth}
		\centering
		\subfigure[]{\includegraphics[width=1.1\linewidth,keepaspectratio]{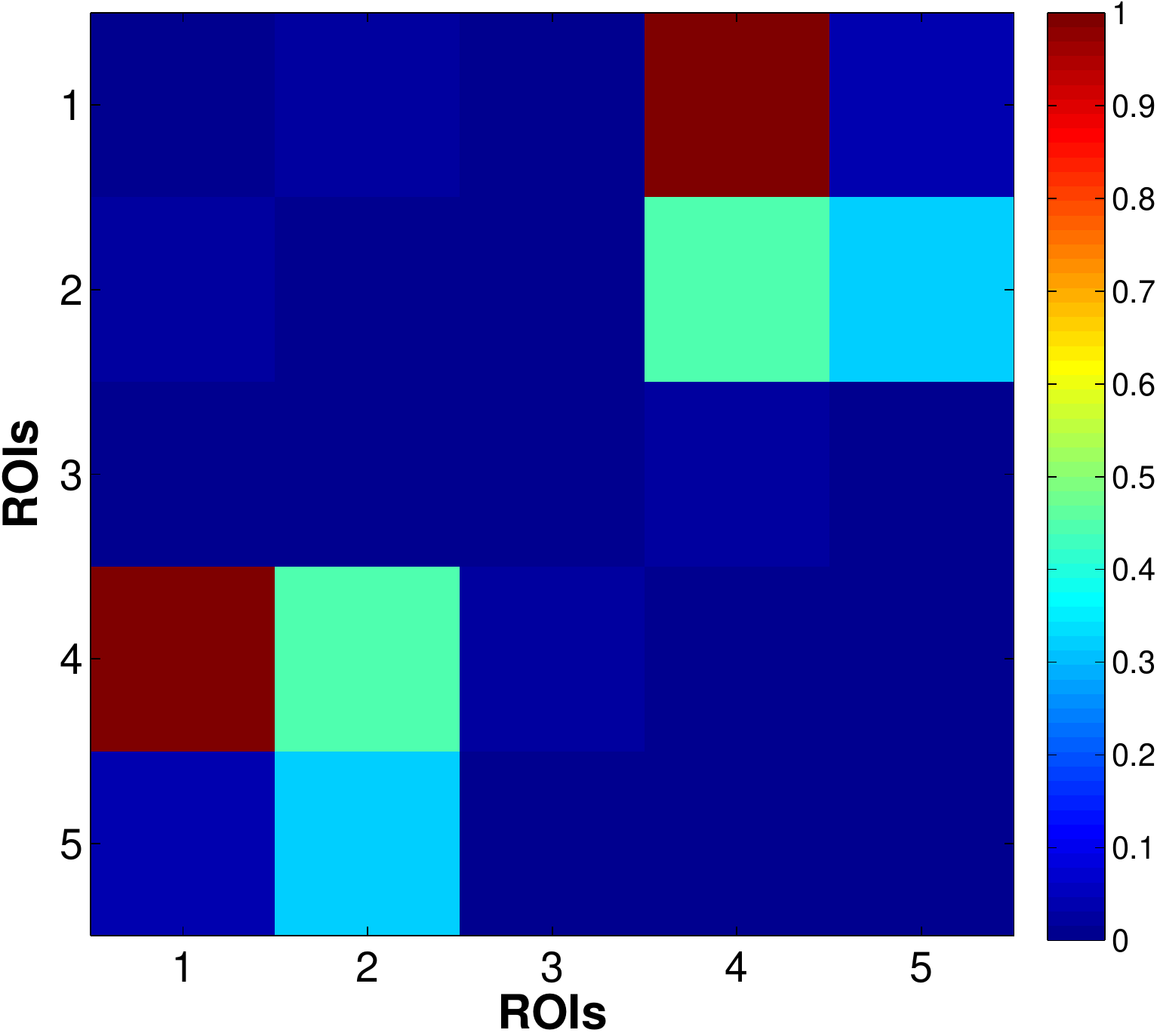}}
	\end{minipage}
	\hspace{0.5 cm}
	\begin{minipage}[t]{0.27\linewidth}
		\centering
		\subfigure[]{\includegraphics[width=1.1\linewidth,keepaspectratio]{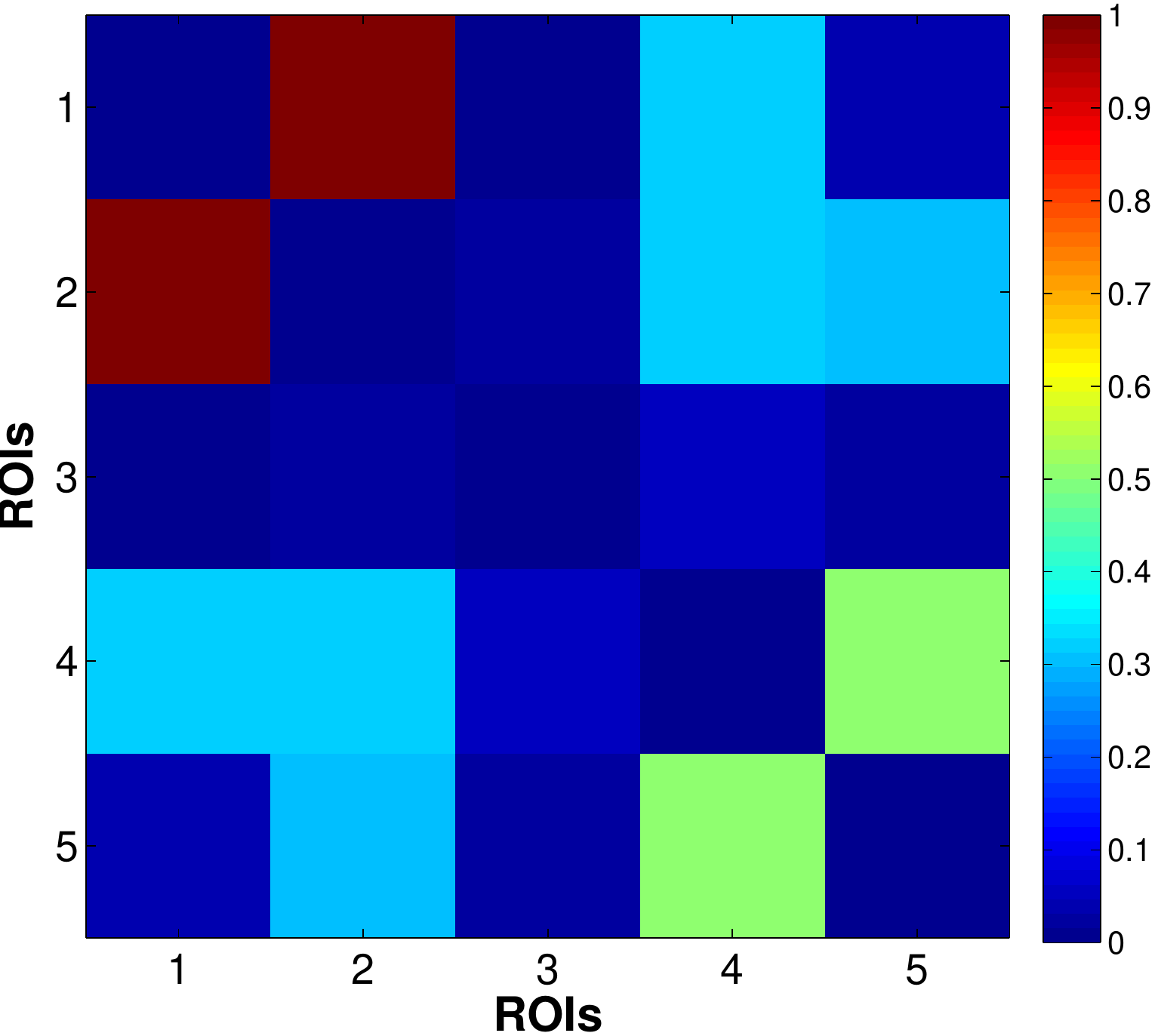}}
	\end{minipage}
\caption{(a) The true RV-based between-cluster connectivity matrix $\widetilde{RV}_{{\mathbf Y} {\mathbf Y}}$. (a) Estimates $\widehat{RV}_{{\mathbf Y} {\mathbf Y}}$ based on mean time-series. (c) Estimates factor time-series with $m_r = 5$. Sample size is $T=150$. The RV coefficient estimates are computed from averaged mean and factor time-series over the 100 realizations.}
\label{Fig:RV-Comp-Sim}
\end{figure}

For between-cluster connectivity (Fig.~\ref{Fig:SimMSE}(b)), estimator $\widehat{RV}_{{\mathbf Y} {\mathbf Y}}$ based on the factors generally gives lower errors than the mean time series, especially for $m_r = 5$, suggesting that the factors can better characterize the correlations within the clusters. The first factor $m_r = 1$ behaves similarly to the mean, as evident from the same errors across $T$. Fig.\ref{Fig:RV-Comp-Sim} shows the true and estimated RV-based connectivity matrices computed from the averaged correlation matrix of the mean and factor time series (with $m_r = 5$) over 100 replications. We can see that the factor-based estimates more closely resemble the true connectivity pattern, e.g. accurately identifying the connections between clusters (${\mathcal C}_1$-${\mathcal C}_2$) and (${\mathcal C}_4$ and ${\mathcal C}_5$) which are mis-detected by estimates based on average time series.

\vspace{-0.05in}

\subsection{Results for Adaptive $m_r$}

We also evaluated the performance of MSFA estimator with $m_r$ adaptively selected for each cluster according to criteria in (\ref{Eq:var-r}) and (\ref{Eq:BIC}). Table~\ref{Table:Est-mr} shows the values of $m_r$ selected using BIC and thresholds $\tau = 1\%, 25\%, 50\%$ and $75\%$ of variance. The results are averages over 100 realizations. As expected, more factors are selected to explain greater variability of the data. Use of $\tau = 1\%$ and BIC select the first principal component. BIC tends to suggest low number of factors.
To cover $25\%$ of the total variance, the first two principal components are sufficient consistently for all clusters. These most likely capture the dominant information that is shared across all the clusters. The number of additional components selected by using the higher percentage $50\%$ and $75\%$ varies for 
different clusters. These capture the detailed variability distinctive to individual clusters. Moreover, 
fewer factors are needed for clusters that are strongly and densely connected with other clusters, e.g. ${\mathcal C}_4$ and ${\mathcal C}_1$ as shown in Fig.~\ref{Fig:RV-Comp-Sim}(a), compared to clusters with few and weak connections e.g. ${\mathcal C}_3$. This is because highly correlated time series contains many redundant information and thus can be explained by only few common factors. Therefore, the optimal number of factors, 
which varies according to dependency structure of each cluster, should be selected adaptively rather than 
held fixed for the entire brain as in most PCA analyses of fMRI.

\begin{table}[!t]
\renewcommand{\arraystretch}{0.7}
\caption{Number of factors selected for each cluster using different thresholds $\tau$ for percentage of data variance.}
\vspace{0.3 cm}
\label{Table:Est-mr}
\centering
\begin{tabular}{c|ccccc}
  \hline \hline
\multirow{2}{*}{Cluster, ${\mathcal C}_r$} & \multirow{2}{*}{BIC} &
\multicolumn{4}{c}{Percentage of variance explained} \\
\cline{3-6}
& & $\tau = 1\%$ & $\tau = 25\%$ & $\tau = 50\%$ & $\tau = 75\%$\\
\hline
1 & 1 & 1 & 2 & 4 & 9 \\
2 & 1 & 1 & 2 & 5 & 11 \\
3 & 1 & 1 & 2 & 6 & 12 \\
4 & 1 & 1 & 2 & 5 & 10 \\
5 & 1 & 1 & 2 & 5 & 11  \\
\cline{1-6}
Average & 1 & 1 & 2 & 5 & 11  \\ \hline \hline
\end{tabular}
\end{table}


Fig.~\ref{Fig:SimMSE}(bottom) plots the estimation errors obtained using $m_r$ obtained 
adaptively for the different thresholds of variance. Similar to the results for the fixed $m_r$ in Fig.\ref{Fig:SimMSE}(top), the MSFA estimators with adaptive $m_r$ show improved performance over the sample correlation matrix for between-node connectivity (in Fig.~\ref{Fig:SimMSE}(c)) and RV coefficient based on average time series for between-cluster connectivity (in Fig.~\ref{Fig:SimMSE}(d)). To compare performance of the adaptive and the fixed $m_r$ fairly, we contrast the results for $\tau = 50\%$ and $75\%$ (with respective average number of selected factors over all clusters $m_r$ = 5 and 11 as in Table~\ref{Table:Est-mr}) with that for $m_r$ = 5 and 10 in Fig.~\ref{Fig:SimMSE}(top). The adaptive-$m_r$ approach generally outperforms the fixed-$m_r$ approach for both scales of connectivity, particularly when $T$ is small compared to $N$. This suggests that use of adaptive $m_r$ which is able to capture the cluster-specific dependency structure, can improve the connectivity estimates. Based on these simulation results, we will use the MSFA estimator with adaptive $m_r$ to analyze real fMRI data.

\vspace{-0.05in}

\subsection{Comparison with Other Covariance Estimators}

In addition to the factor-based estimators, various regularization methods have been proposed in recent years for estimating a large covariance matrix and its inverse (or precision matrix). The first class of estimators are based on the shrinkage of sample covariance eigenvalues \citep{ledoit2004}. The second includes regularizing the covariance matrix by banding, tapering and thresholding \citep{Bickel2008,Cai2012}. 
The third imposes sparsity on the precision matrix in graphical models with $\ell_1$ penalization \citep{Yuan2007,Cai2011}, extended to a low-rank plus sparse estimation by combining latent variable and graphical modeling \citep{Chandra2011}. However, the main aim of this paper is to develop a covariance modeling approach for the purpose of analyzing large dependence in networks with multi-scale structure, but not to compete with other advanced large covariance estimators. Therefore, for evaluation purposes, we compare the performance of our proposed MSFA model-based estimator only with two well-known high-dimensional covariance estimators as benchmarks: the shrinkage estimator of Ledoit and Wolf (LW) \citep{ledoit2004}, and the graphical lasso (glasso) regularized estimator of Friedman et al. \citep{friedman2008}. We focus on assessing the estimation of a single-block, high-dimensional covariance matrix i.e. the whole-network node-wise connectivity. Note that our method offers additional advantage of multi-scale covariance analysis over the above-mentioned methods which mostly estimate a single-block covariance matrix, and our framework can potentially be extended to accommodate the shrinkage and sparsity.


We performed statistical comparisons between the estimators with repeated ANOVA tests via pairwise confidence intervals, using a linear mixed effects model with the squared error as response variable. There is no significant difference in performance between a pair of estimators, if the computed confidence interval for the difference in their squared errors contains zero. The confidence intervals are adjusted using the Bonferroni's method for multiple comparisons at a global confidence level of $95\%$. The estimation errors of ${\mathbf C}_{{\mathbf Y} {\mathbf Y}}$ over 100 replications by various covariance estimators for different $T$ are reported in Table~\ref{Table:Comp-covs}. For the glasso, penalty parameter $\rho = 0.5$ is used, and the covariance estimate was obtained from the estimated inverse. As expected, both LW and glasso significantly outperform the sample covariance when $T < N$, for $T = 50$, but fail to deliver any advantages or even perform worse when $T$ is large. Interestingly, the MSFA estimator is shown to improve substantially over the both large covariance estimators with significantly lower estimation errors for all cases of dimensionality, and only slightly underperformed relative to the sample covariance for large $T$. This suggests that the MSFA provides a more robust and better-conditioned covariance estimator. There is also significant improvement by using $\tau = 75\%$ as compared to $\tau = 50\%$ for all settings.

\begin{table}[!t]
\renewcommand{\arraystretch}{0.8}
\caption{Averages and standard deviations over 100 replications of errors under Frobenius norm for various covariance estimators of whole-network connectivity ${\mathbf C}_{{\mathbf Y} {\mathbf Y}}$  for different sample size $T$ and fixed dimension $N = 125$: Sample covariance, Ledoit-Wolf, graphical lasso, and MSFA ($\tau = 50\%$, $\tau = 75\%$).}
\vspace{0.3 cm}
\label{Table:Comp-covs}
\resizebox{1\textwidth}{!}{\begin{minipage}{\textwidth}
\centering
\begin{tabular}{c|ccccc}
\hline \hline
\multirow{2}{*}{\normalsize{$T$}} & \normalsize{Sample} & \normalsize{Ledoit} & \normalsize{Graphical} & \multicolumn{2}{c}{\normalsize{MSFA}} \\
& \normalsize{Cov.} & \normalsize{Wolf} & \normalsize{Lasso} & \normalsize{$\tau = 50\%$} & \normalsize{$\tau = 75\%$} \\
\hline
50 & 323.62 (22.35) & 241.45 (10.80) & 285.85 (15.85) & 162.53 (9.37) & 156.20 (8.43) \\
100 & 160.20 (10.86) & 209.18 (6.98) & 217.28 (7.86) & 129.67 (5.17) & 120.64 (4.24) \\ 				
150 & 104.33 (7.53) & 189.97 (4.61) & 197.02 (4.92) & 117.01 (3.32) & 107.75 (2.66) \\ 				
200 & 79.03 (6.10) & 179.19 (4.31) & 189.64 (4.90) & 110.94 (2.64) & 101.31 (2.12) \\ 				
250 & 62.65 (4.46) & 170.62 (3.48) & 184.73 (4.58) & 107.30 (1.96) & 97.46 (1.45) \\ \hline \hline 				
\end{tabular}
\end{minipage} }
\end{table}

\vspace{-0.05in}

\section{Application to Connectivity in fMRI Data}

In this section, we analyze a high-dimensional real resting-state fMRI data using the proposed MSFA approach. Spontaneuous fluctuations of the blood-oxygen-level-dependent (BOLD) fMRI signals during rest, are temporally correlated across distinct brain regions, revealing large-scale coherent spatial patterns called resting-state networks (RSNs) \citep{HeuveHulshoff2010}. We investigated three hierarchical levels of nested modularity in the resting-state brain functional networks, namely, (1.) voxels--ROIs; \ (2.) ROIs--functional systems; \ and (3.) systems--whole brain, in terms of within-module and between-module functional connectivity.
We first partition the whole-brain network into a set of anatomically-parcellated ROIs and extract a few latent factors by PCA to summarize the massive voxel data within each ROI. The high-dimensional voxel-wise connectivity within ROIs is characterized by the low-dimensional factors.
A similar approach \citep{Sato2010} uses PCs from each ROI to analyze between-ROI connectivity, however, neglected the fine-scale between-voxel connectivity. 

\vspace{-0.1in}
\subsection{Resting-State fMRI Data}

\textit{1) Data acquisition and preprocessing:} We studied resting-state fMRI data of 10 subjects from a data set available at NITRC (https://www.nitrc.org/projects/nyu\_trt/). A time series of $T = 197$ BOLD functional images were acquired on a Siemens Allegra 3.0-Tesla scanner using a T2-weighted gradient-echo planar imaging (EPI) sequence (TR/TE = 2000/25 ms; voxel size = $3\times3\times3$ mm$^3$; matrix = $64\times64$; 39 slices). Subjects were asked to relax during scans. The data were preprocessed with motion correction, normalization and spatial smoothing.

\textit{2) Parcellation:} We used the AAL atlas to obtain an anatomical parcellation of the whole brain into 90 ROIs. In this study, the ROIs were grouped into six pre-defined resting-state system networks of similar anatomical and functional properties, based on the templates in \citep{Allen2011,Allen2012,Li2011}. The considered RSNs include sub-cortical (SCN), auditory (AN), sensorimotor (SMN), visual (VN), attentional (ATN) and default mode network (DMN) 
The ROIs and their mapping to corresponding RSNs with overlapping are given in Appendix 7.2. We followed the ROI abbreviations in \citep{Salvador2005}.

\vspace{-0.05in}

\subsection{Analysis of Voxel-wise Connectivity}

Voxel-wise analysis is challenging due to low signal-to-noise ratio at individual voxels. The standard approach is to compute the average time series over each
ROI. In contrast, the  MSFA approach achieves dimension-reduction which leads to reliable and computationally efficient connectivity estimates. Moreover, by using PCA, the MSFA approach retains only the dominant information (components with largest eigenvalues). Those with smaller eigenvalues are considered to correspond to either measurement noise, machine noise and the weak physiological signals of non-interest. We fitted a local FA model on the voxel time series for each ROI using PCA estimation and then constructed the covariance matrix between voxels as quantified by $\widehat{\Sigma}_{{\mathbf Y}_r {\mathbf Y}_r}$. The number of factors, $m_r$, was selected data-adaptively for each ROI based on threshold $\tau = 20\%$ or maximum $1\%$ of the number of voxels. The threshold was chosen such that the number of factors are sufficient to capture the variability in the data and hence the fine dependency structure particularly at the voxel scale, while excluding fluctuations that are probably noises which could induce spurious connectivity estimates. For this data, a total of 145 factors were selected to represent brain activity from 183,348 voxels of the entire brain volume. Thus, this represents a massive reduction of the dimension of $99.0002\%$ (since total number of factors
$145$ is only $0.08\%$ of the total number of voxels). The number of voxel time series and the selected $m_r$ are given in Appendix Table 1.

\begin{figure}[!t]
\captionsetup[subfigure]{labelformat=empty}
	\vspace{-2 cm}
	\hspace{2.5 cm}
    \begin{minipage}[t]{.35\linewidth}
		\centering
    \subfigure[]{\includegraphics[width=1.00\linewidth,keepaspectratio]{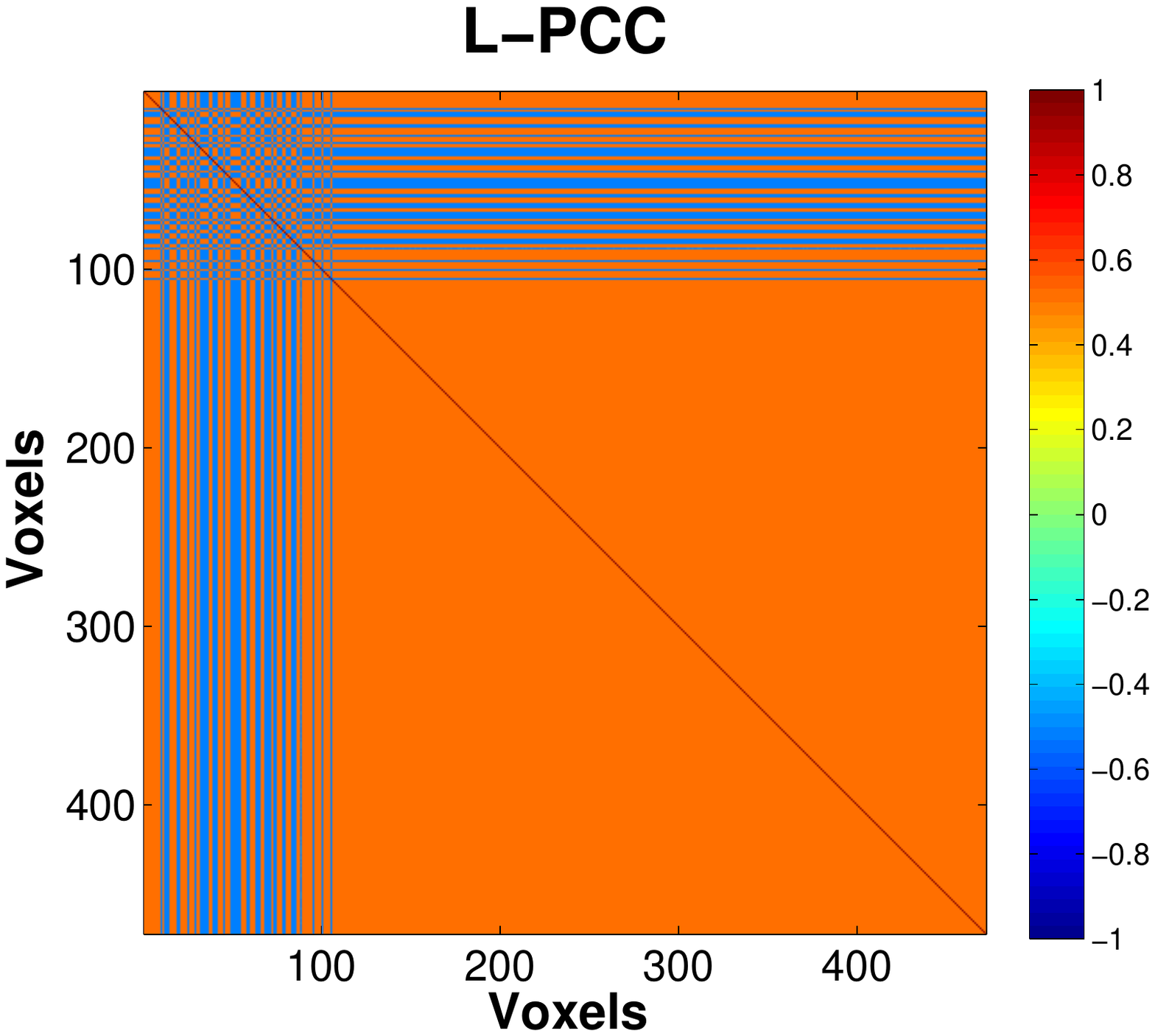}}
  \end{minipage} \hspace{-0.5 cm}
  \begin{minipage}[t]{.35\linewidth}
		\centering
    \subfigure[]{\includegraphics[width=1.00\linewidth,keepaspectratio]{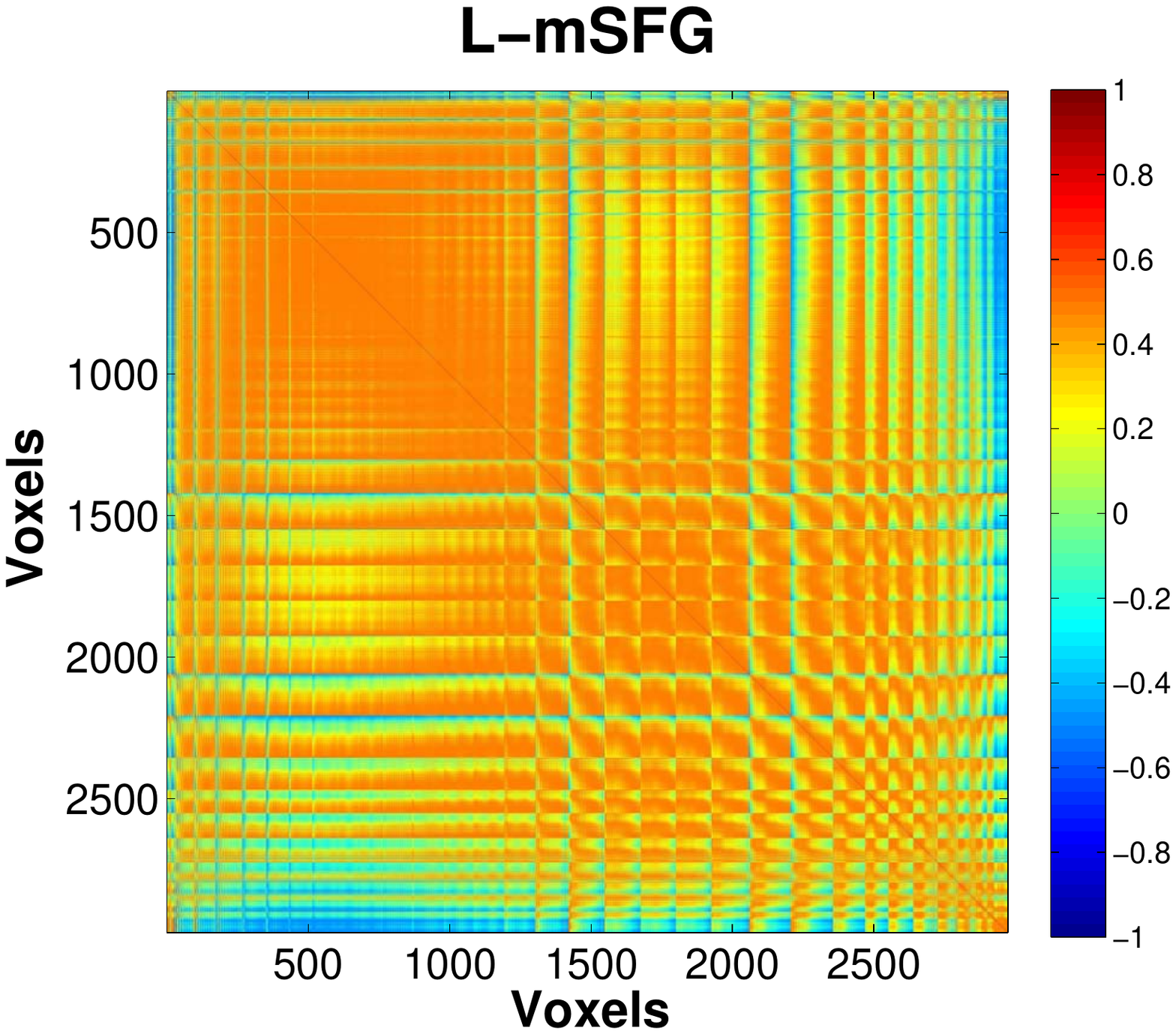}}
  \end{minipage}
	
	\vspace{-4.5 cm}
	\hspace{2.5 cm}
  \begin{minipage}[t]{.35\linewidth}
		\centering
    \subfigure[]{\includegraphics[width=1.00\linewidth,keepaspectratio]{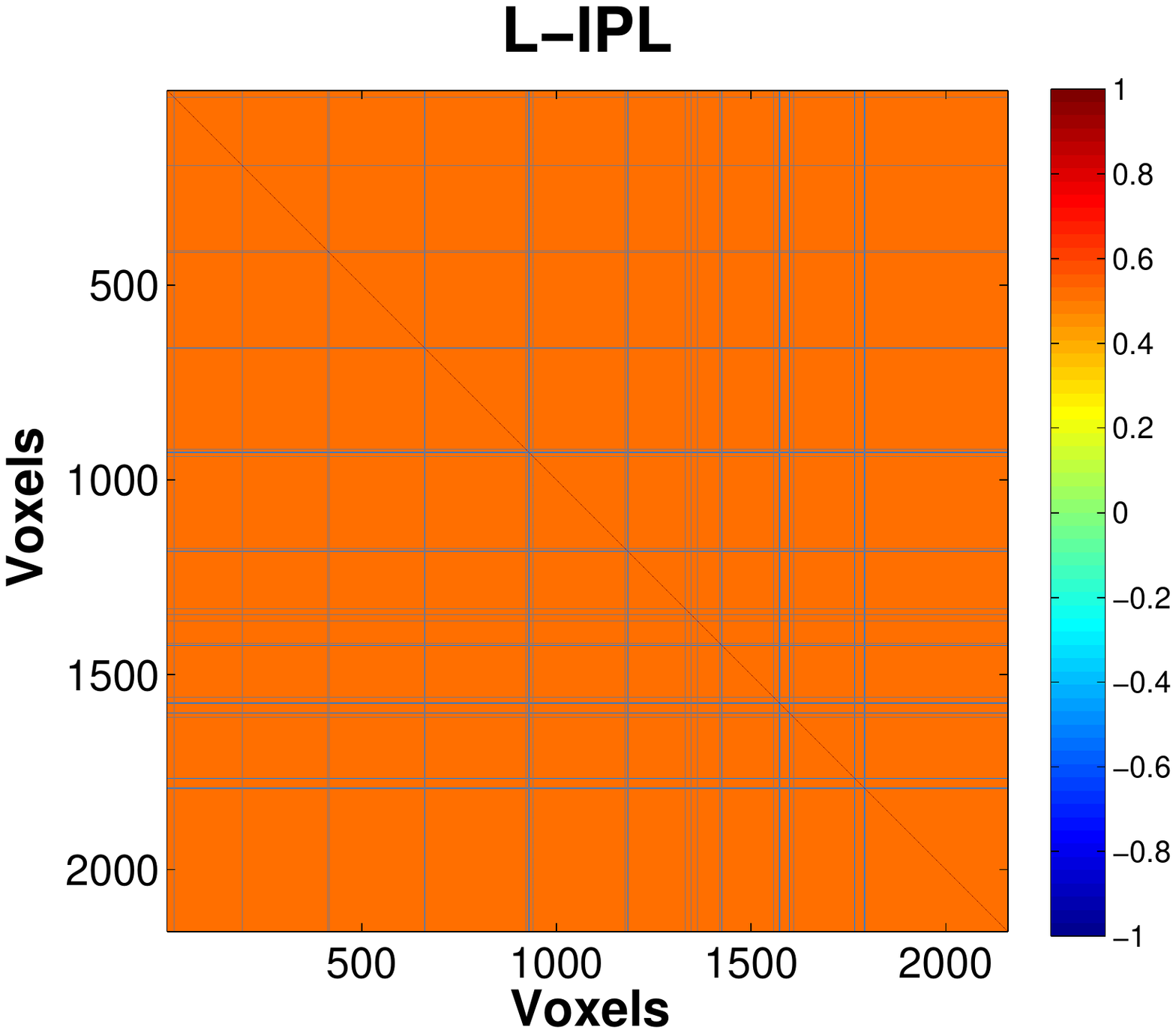}}
  \end{minipage} \hspace{-0.5 cm}
  \begin{minipage}[t]{.35\linewidth}
		\centering
    \subfigure[]{\includegraphics[width=1.00\linewidth,keepaspectratio]{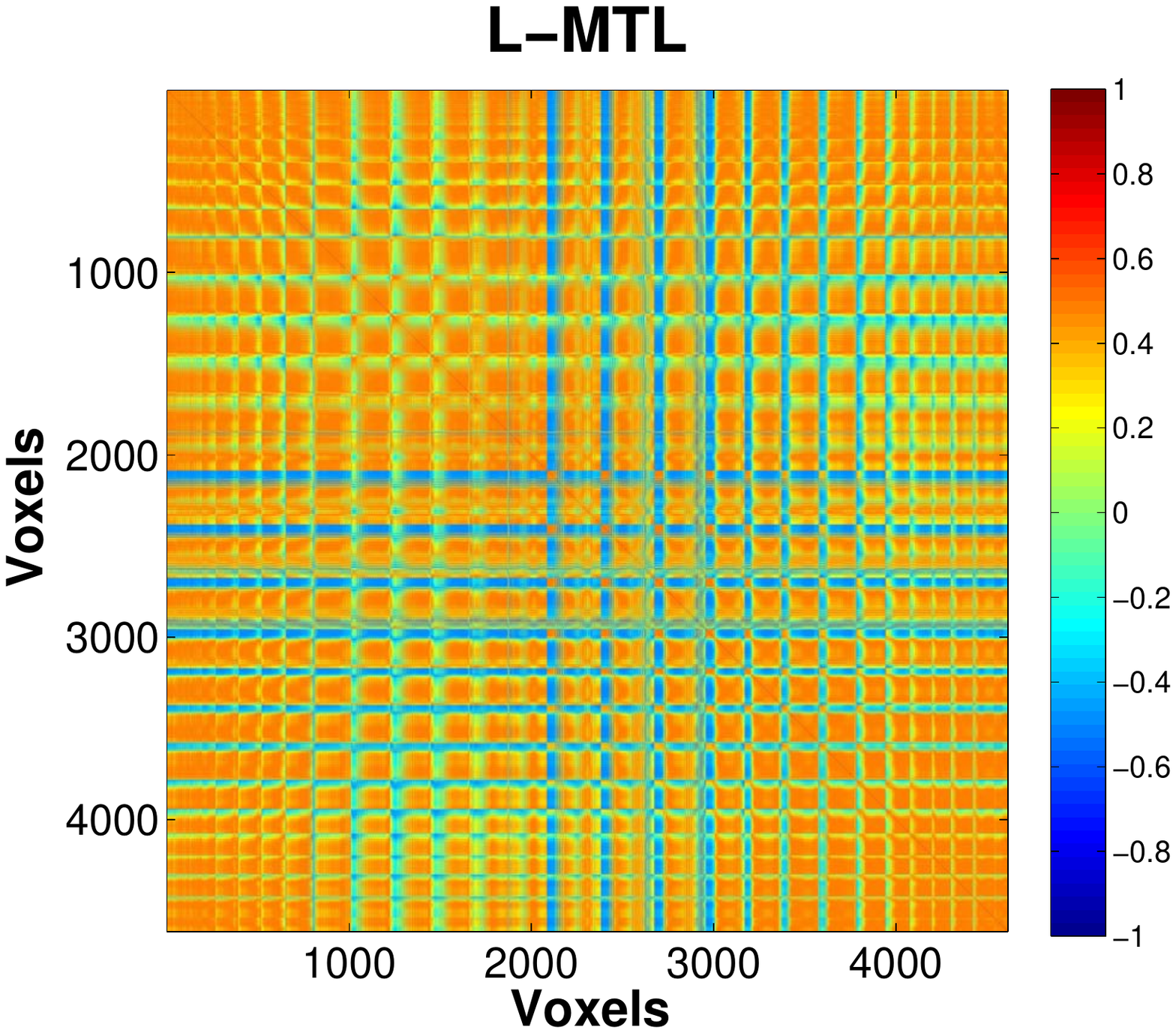}}
  \end{minipage}
	\vspace{-2.5 cm}
	\caption{Estimated within-ROI voxel-wise functional connectivity matrices from resting-state fMRI data of a subject, shown for four selected brain regions (from the left cerebral hemispheres) which belong to the default mode network. PCC: posterior cingulate cortex, mSFG: medial superior frontal gyrus, IPL: inferior parietal lobe, MTG: middle temporal gyrus.}
	\label{Fig:Within-ROI}
	\vspace{-0.05in}
\end{figure}

Fig.~\ref{Fig:Within-ROI} shows the estimated correlation matrices of voxels in four key ROIs that belongs to the DMN a for single subject. The regions of DMN, a well-known RSN, has been reported to exhibit increased activation and correlation in neuronal activities during rest compared to goal-oriented tasks, suggesting it as an important idling mode of the brain \citep{RaichleSnyder2007}.
To our knowledge, our study is probably among the few reporting the voxel-level connectivity in the DMN regions. The estimates using only a small number of factors is able to reveal the existence of complex, large amount of interactions between massive voxels, within a small brain region, even during resting state.
Within these regions, there appears to be a pervasively strong connectivity between many pairs of voxels. We can see that the PCC, a major hub that is strongly inter-connected with many other brain regions, also exhibits the strongest intra-connectivity within the region itself. This is followed by IPL which is another active region of the DMN. 

\vspace{-0.05in}

\subsection{Analysis of Between-ROI Connectivity}

\begin{figure}[!t]
	\begin{minipage}[t]{\linewidth}
		\centering
		\includegraphics[width=0.45\linewidth,keepaspectratio]{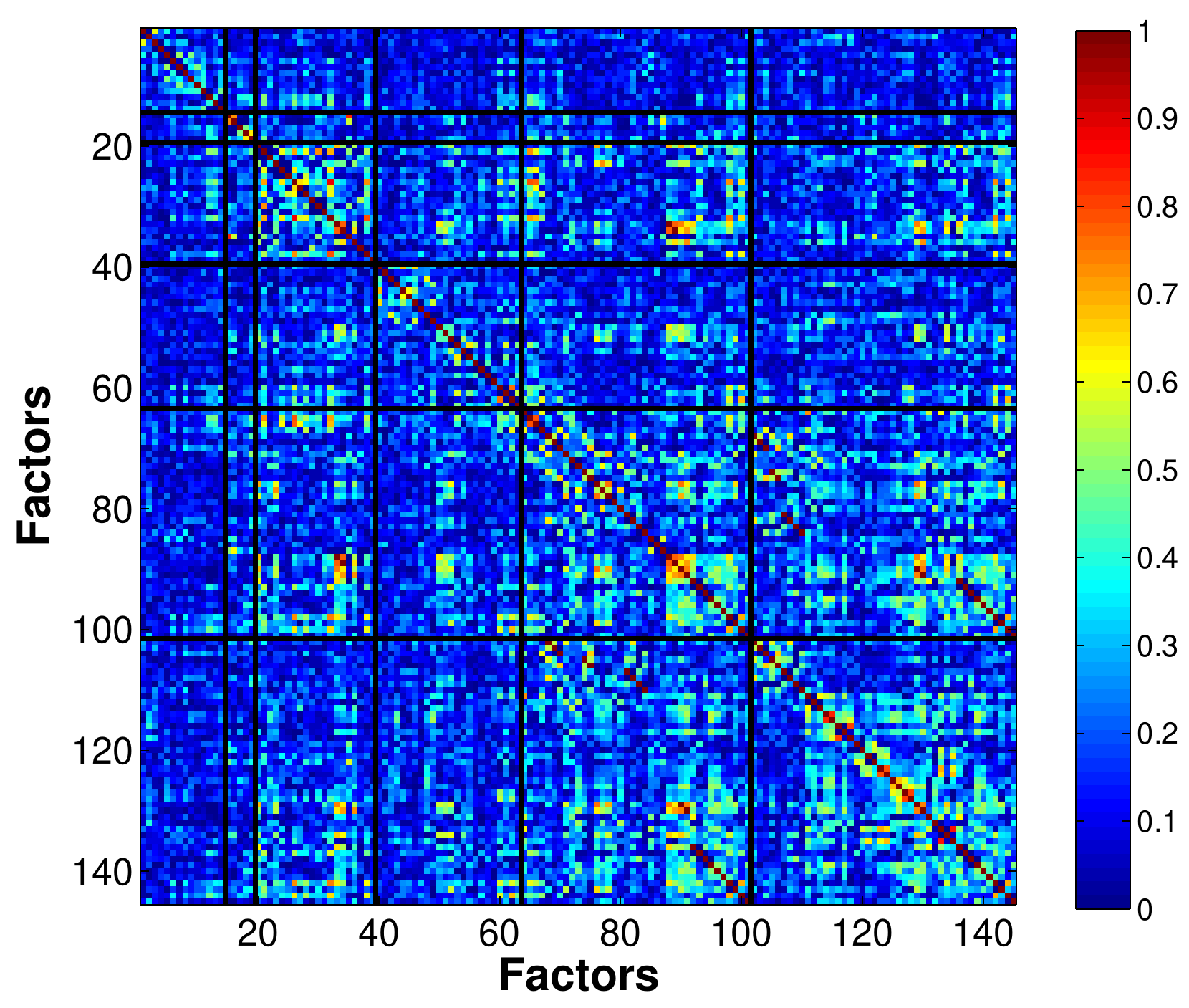}
	\end{minipage}
	\vspace{-0.2in}
	\caption{Correlation matrix of global factor concatenating factors for all ROIs.}
\label{Fig:Corr-GlobalFac}
\end{figure}

In this section and the next, we applied the global factor model to analyze the higher scales of functional connectivity between ROIs (and between functional system sub-networks), via the correlations between factors associated with each region. The analysis is illustrated using a single subject's data. We further computed the RV coefficient as a single measure to summarize the strength of connectivity between the pairs of ROIs (or networks). Fig.~\ref{Fig:Corr-GlobalFac} shows the correlation matrix, $\Sigma_{{\mathbf f}{\mathbf f}}$ (absolute-valued for comparison with the non-negative RVs) of the factor time series over the entire brain, constructed form sub-blocks of correlation matrices of factors between every pair of ROIs. Note that the factors are highly correlated across different regions despite being independent within a region. This suggests that, while the intra-regional connectivity is captured mainly by the mixing matrix, the inter-regional connectivity is  quantified through the dependence between factors across regions. Our method extends the roles of the factor series beyond merely explaining the variance of the data, as in the conventional factor-based connectivity analysis which is limited by the independent component assumption.

\begin{figure}[!ht]
\hspace{0.6 cm}
	\begin{minipage}[t]{0.45\linewidth}
		\centering
		\subfigure[]{\includegraphics[width=0.8\linewidth,keepaspectratio]{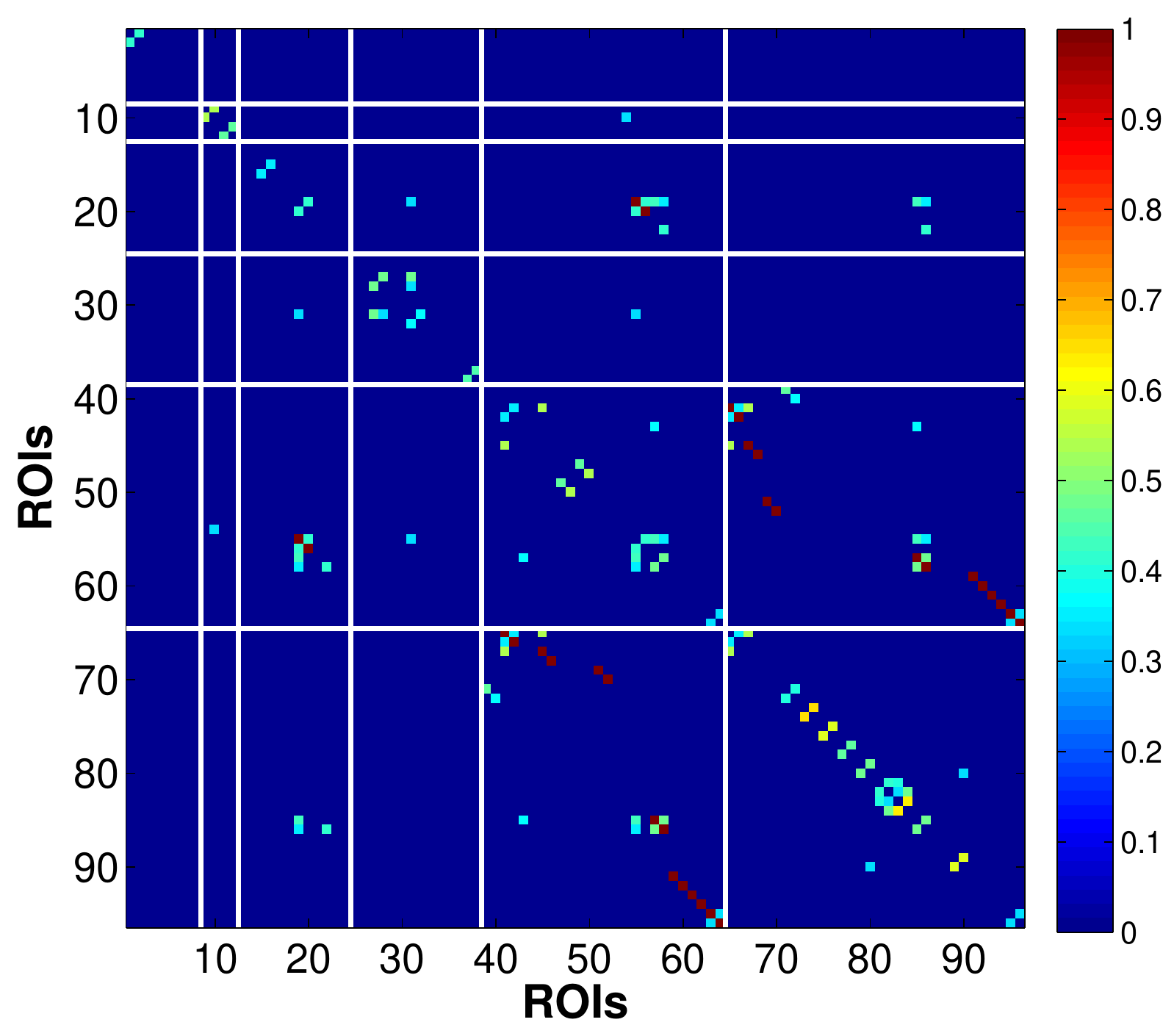}}
	\end{minipage}
	\hspace{-0.18 cm}
	\begin{minipage}[t]{0.45\linewidth}
		\centering
		\subfigure[]{\includegraphics[width=0.8\linewidth,keepaspectratio]{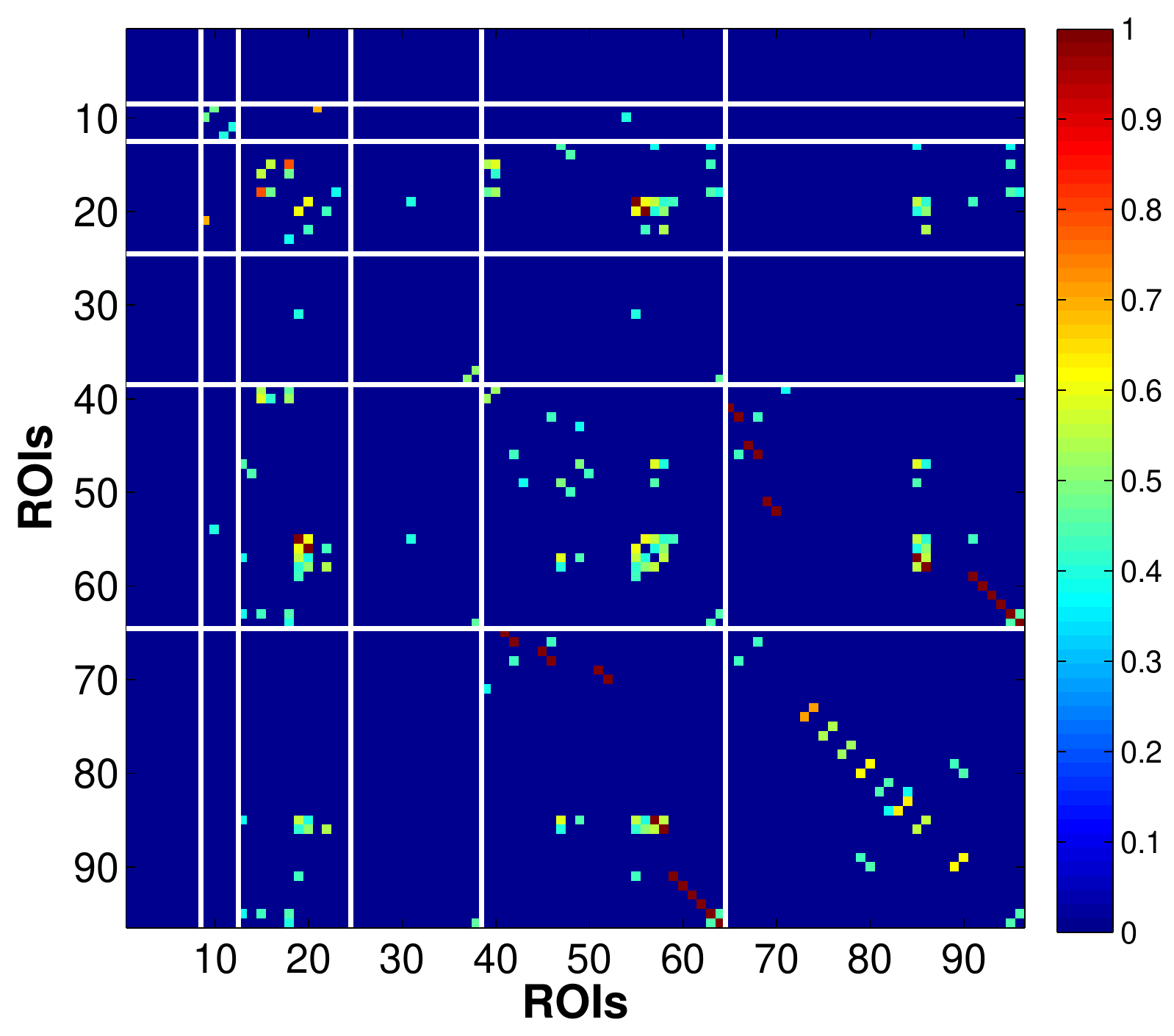}}
	\end{minipage}
	\vspace{-0.1 cm}
\caption{Comparison of between-ROI functional connectivity measured by RV coefficients based on mean time series (a) and common factors (b) from each ROI. The connections shown are significantly different from zero, with absolute values of the standardized RVs greater than the upper-bound of $(1-\alpha/D)\times100\%$ ($\alpha = 0.05$, $D = 96\times96 = 9216$) Bonferroni-adjusted confidence interval under the null hypothesis of no connections between ROIs. The null-distributions of the RV coefficients are assumed normal.}
\label{Fig:Between-ROI}
\vspace{-0.02in}
\end{figure}

Fig.~\ref{Fig:Between-ROI} shows the between-ROI RV-coefficient-based connectivity $RV^C_{jk}$, computed from the correlations of mean and factor time series between ROIs. The results from our method have identified a markedly modular structure of the brain functional networks during rest, as reported in previous studies \citep{Ferrarini2009}. The ROIs within a resting-state network with similar functional relevance are more densely connected with each others, as evident particularly for the SMN, ATN and DMN. Whereas, the ROIs from different networks are sparsely connected especially between SCN, AN and VN with the other networks. However, relatively denser connections between ROIs were found across the SMN, ATN and DMN, with the strongest strength of connectivity between the ATN and DMN. Compared to our proposed method, the usual mean-time series approach gives a sparser between-ROI connectivity. A natural question here is whether the mean-time series approach has an inflated false negative (low power in detecting connectivity) or the proposed method has an inflated false positive. We believe that the former is more likely based on our simulation results.

\begin{figure}[!t]
\captionsetup[subfigure]{labelformat=empty}
\captionsetup{position=top}
\captionsetup{font=large}
    \begin{minipage}[t]{.28\linewidth}
		\centering
    \subfigure[(a) Sensorimotor Net]{\includegraphics[width=0.68\linewidth,keepaspectratio]{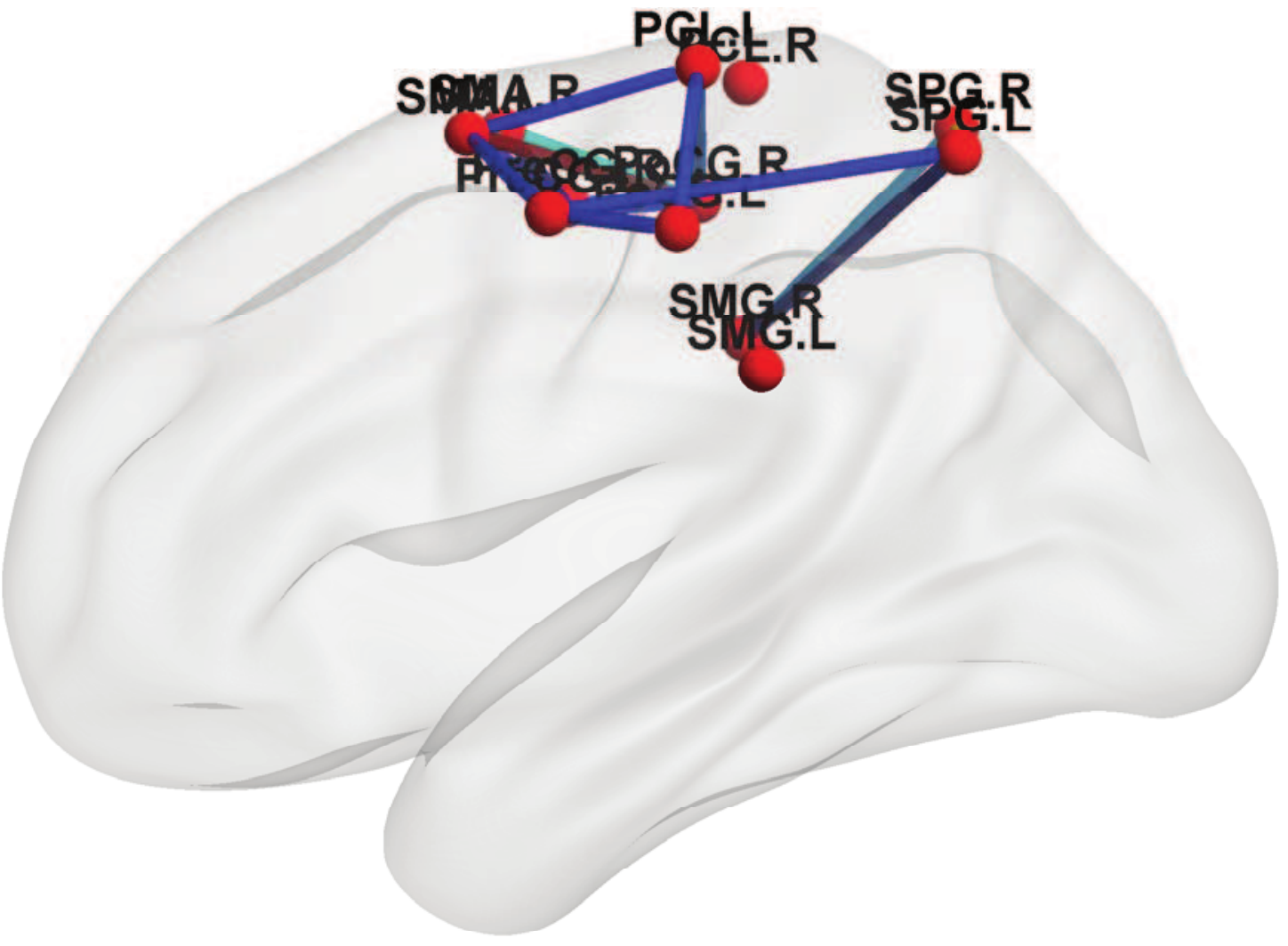}}
  \end{minipage} \hspace{-1.5 cm}
  \begin{minipage}[t]{.28\linewidth}
		\centering
    \subfigure{\includegraphics[width=0.7\linewidth,keepaspectratio]{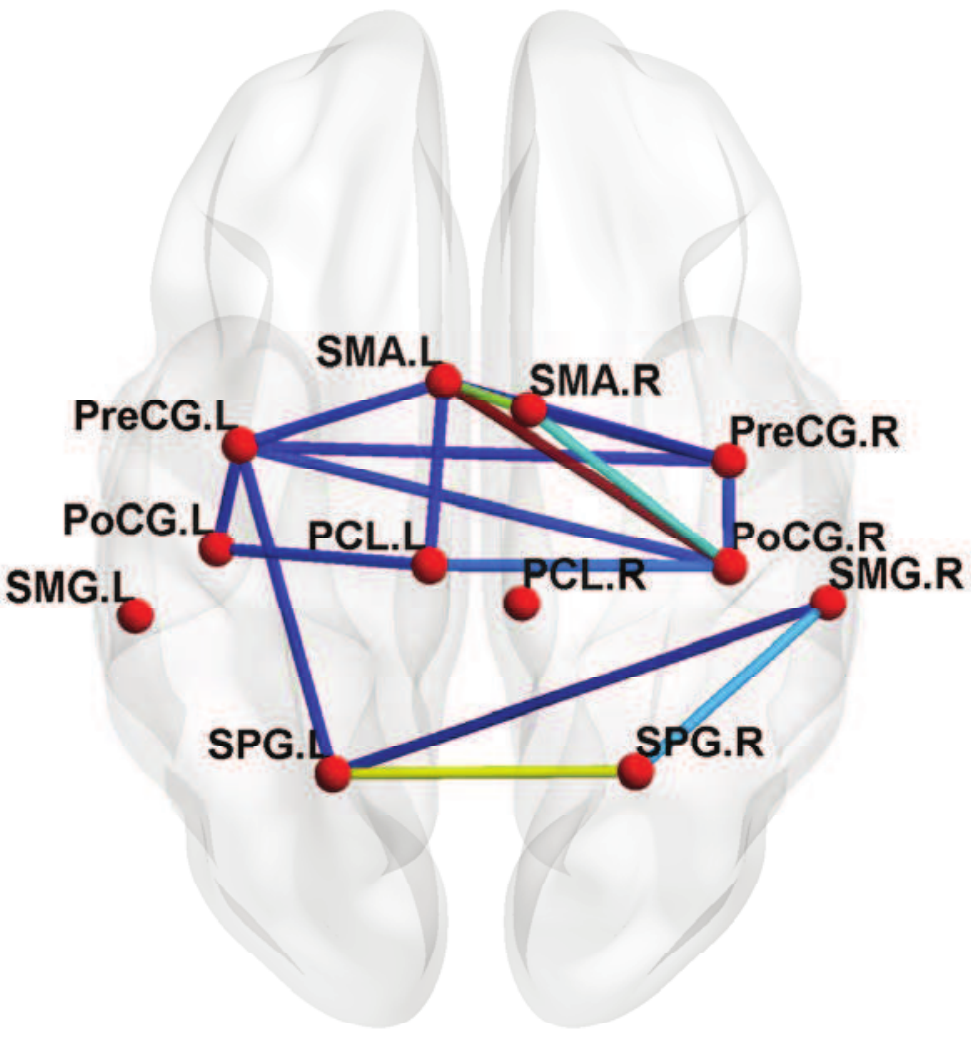}}
  \end{minipage} \hspace{-0.5 cm}
  \begin{minipage}[t]{.28\linewidth}
		\centering
    \subfigure[(b) Visual Net]{\includegraphics[width=0.7\linewidth,keepaspectratio]{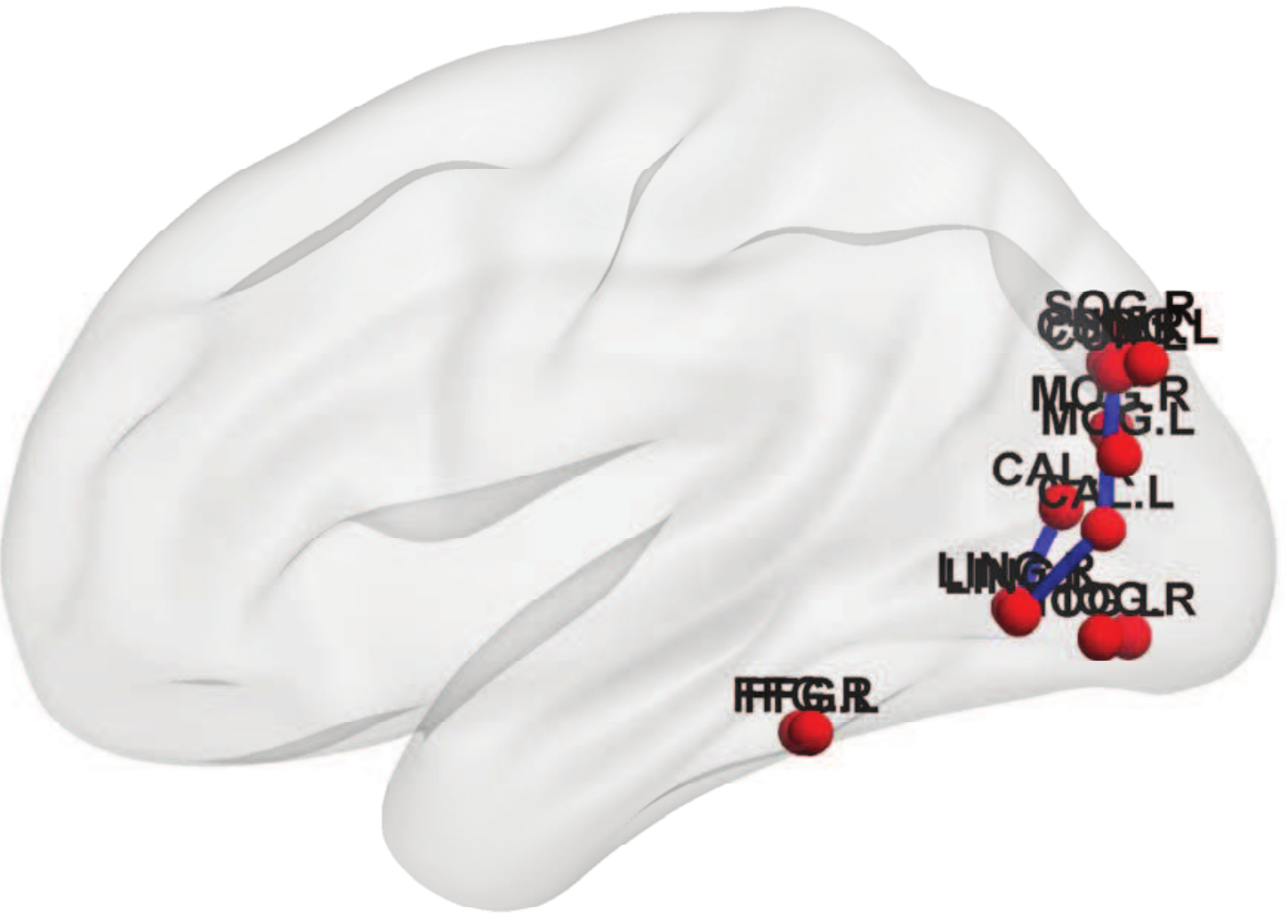}}
  \end{minipage} \hspace{-1.5 cm}
  \begin{minipage}[t]{.28\linewidth}
		\centering
    \subfigure{\includegraphics[width=0.55\linewidth,keepaspectratio]{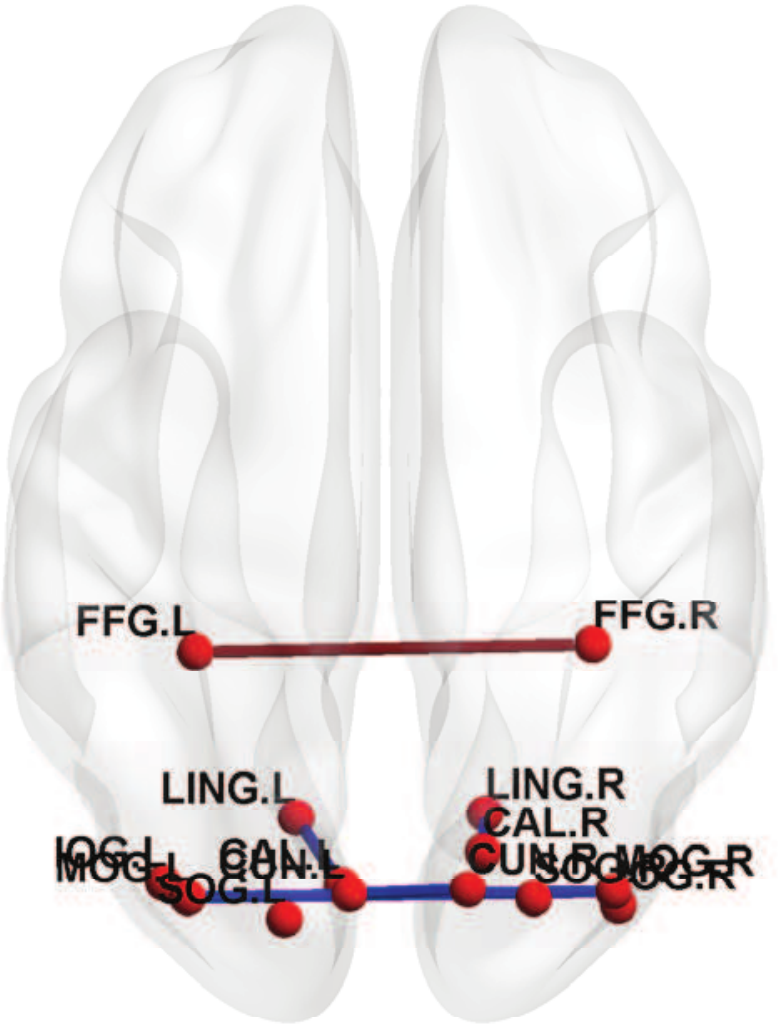}}
  \end{minipage}
	
	\vspace{0.1 cm}
  \begin{minipage}[t]{.28\linewidth}
		\centering
    \subfigure[(c) Attentional Net]{\includegraphics[width=0.7\linewidth,keepaspectratio]{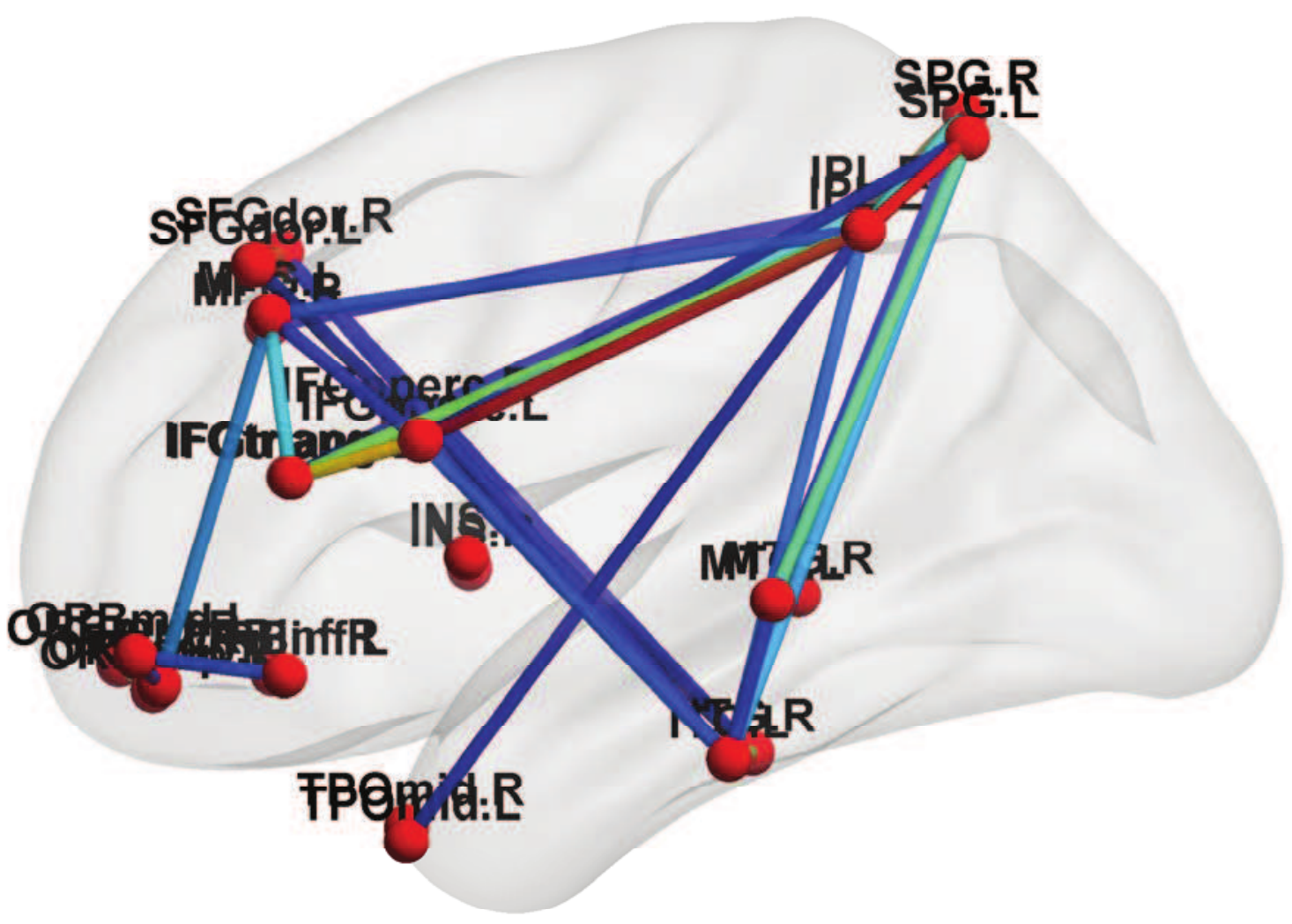}}
  \end{minipage} \hspace{-1.5 cm}
  \begin{minipage}[t]{.28\linewidth}
		\centering
    \subfigure{\includegraphics[width=0.75\linewidth,keepaspectratio]{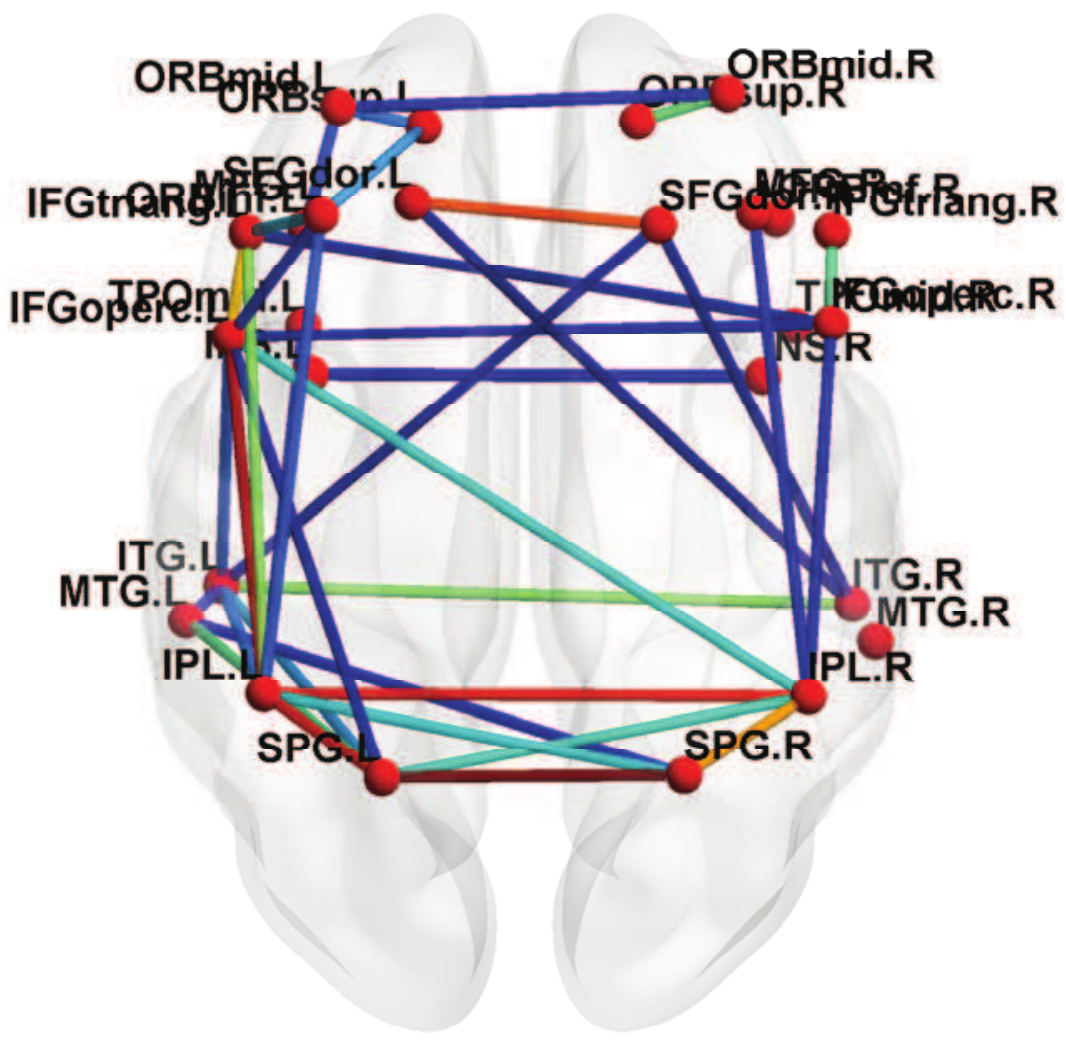}}
  \end{minipage} \hspace{-0.5 cm}
  \begin{minipage}[t]{.28\linewidth}
		\centering
    \subfigure[(d) Default Mode Net]{\includegraphics[width=0.72\linewidth,keepaspectratio]{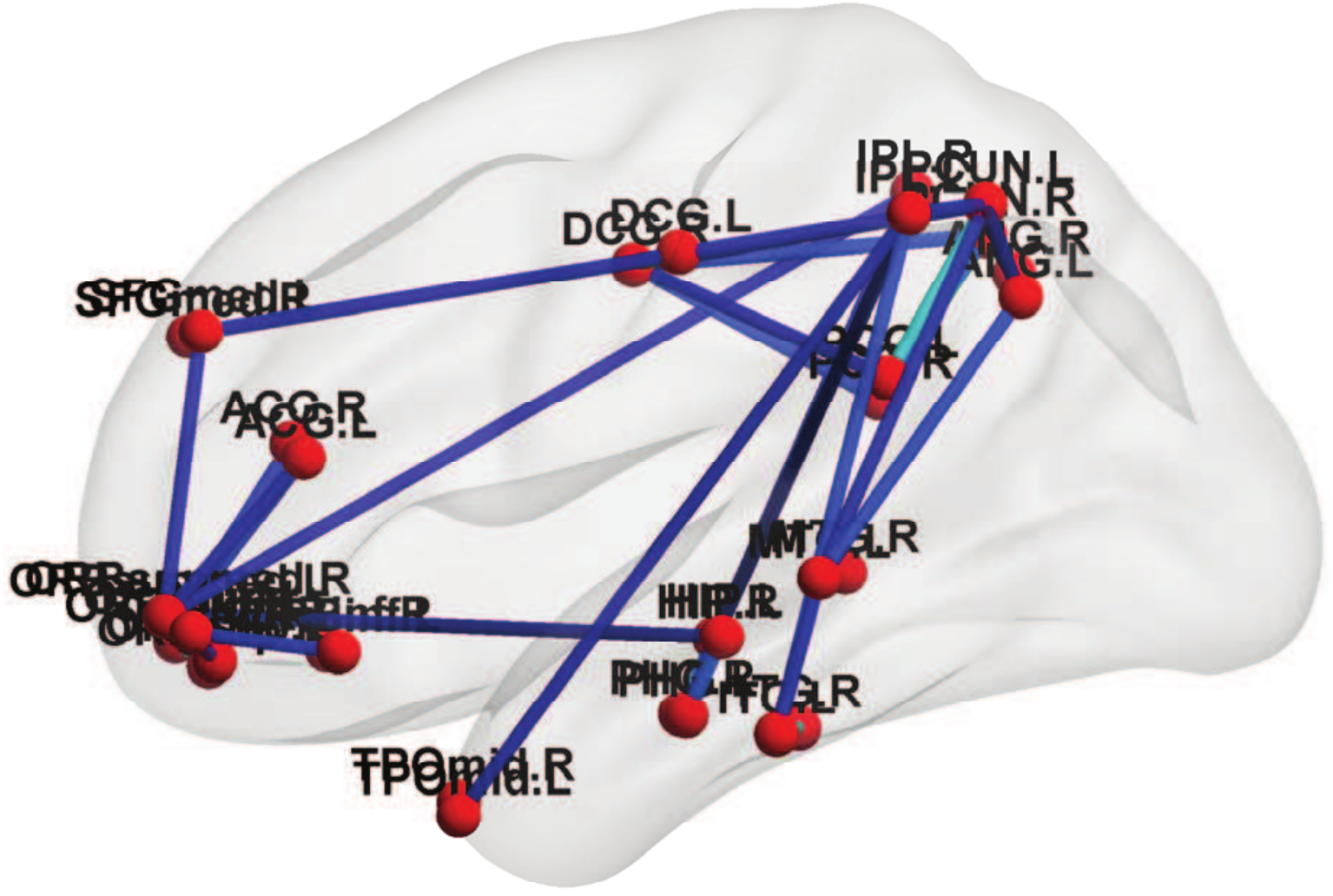}}
  \end{minipage} \hspace{-1.5 cm}
  \begin{minipage}[t]{.28\linewidth}
		\centering
    \subfigure{\includegraphics[width=0.7\linewidth,keepaspectratio]{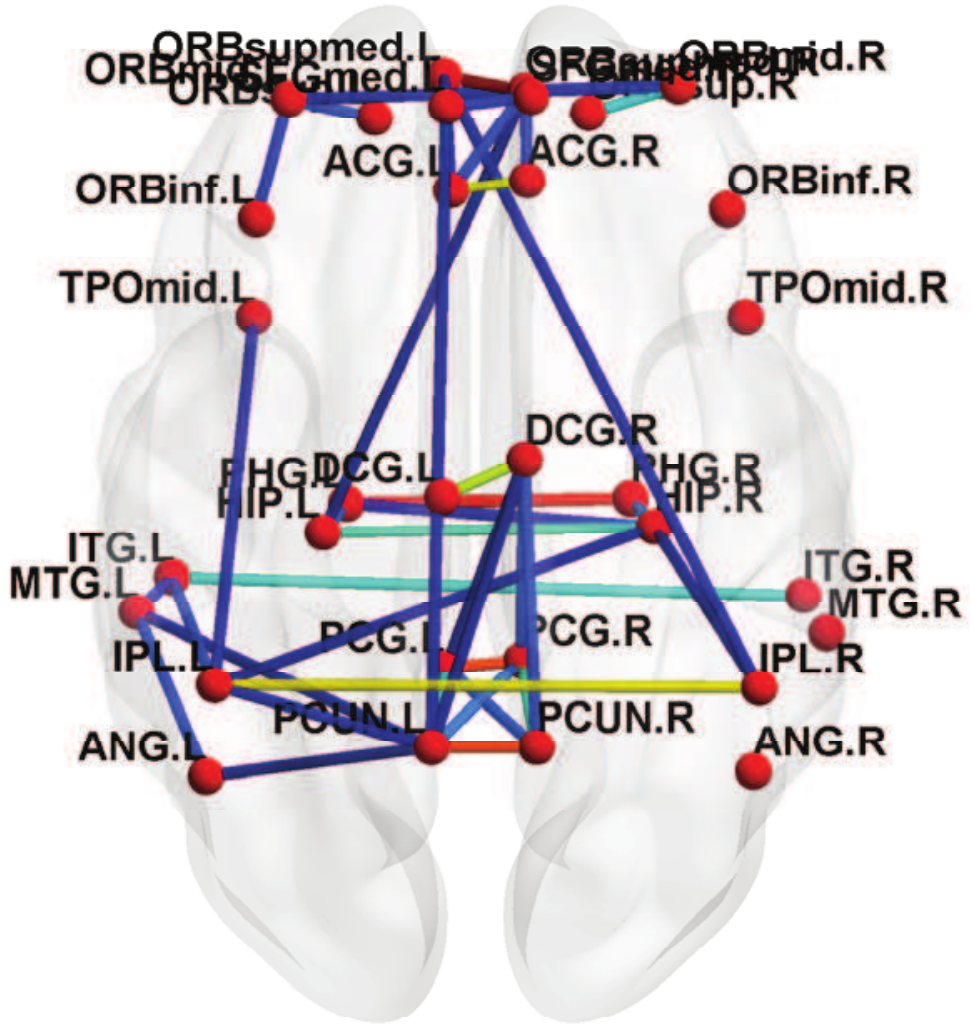}}
  \end{minipage}
	\vspace{-0.2 cm}
	\caption{Topological representations of between-ROI connectivity within each resting-state networks. Links represents significant RV connections with absolute values of standardized coefficients greater than a threshold value of 3 at the 99.97-th percentile of $N(0,1)$ under null hypothesis of no connections.}
	\label{Fig:Within-Net}
	\vspace{-0.05in}
\end{figure}

Fig.~\ref{Fig:Within-Net} shows the topological maps of the ROI-wise connectivity within four RSNs, inferred by the estimated factor-based RV coefficients. Only significant connections with standardized coefficients greater than a threshold value of 3 are shown. The estimates by our approach shows that the ROIs are inter-connected within common functional and anatomical domains, revealing distinct spatial patterns, as identified using the spatial ICA in \citep{Allen2011,Li2011,Allen2012}.
The sensorimotor network, centered at central sulcus, covers primary somatosensory, primary motor and supplementary motor cortex as reported by \citep{Biswal1995}, located in regions e.g. precentral gyrus, postcentral gyrus and supplementary motor area. 
The visual network involves regions in the occipital lobes \citep{GrillMalach2004}. The ATN, involved in attentional processing and monitoring, consists of few sub-networks including the dorsal and ventral systems \citep{Vossel2014}.
The resting-state DMN consists of highly inter-correlated ROIs related to posterior cingulate cortex (PCC)/precuneus, medial prefrontal cortex and the left and right inferior parietal lobule. The PCC is correctly identified as a major hub of the DMN, strongly connected with other regions, as reported in many studies \citep{FranssonMarrelec2008}. 

\vspace{-0.05in}

\subsection{Analysis of Between-Network Connectivity}

\begin{figure}[!t]
\hspace{0.2 cm}
	\begin{minipage}[b]{0.31\linewidth}
		\centering
		\subfigure[]{\includegraphics[width=1.02\linewidth,keepaspectratio]{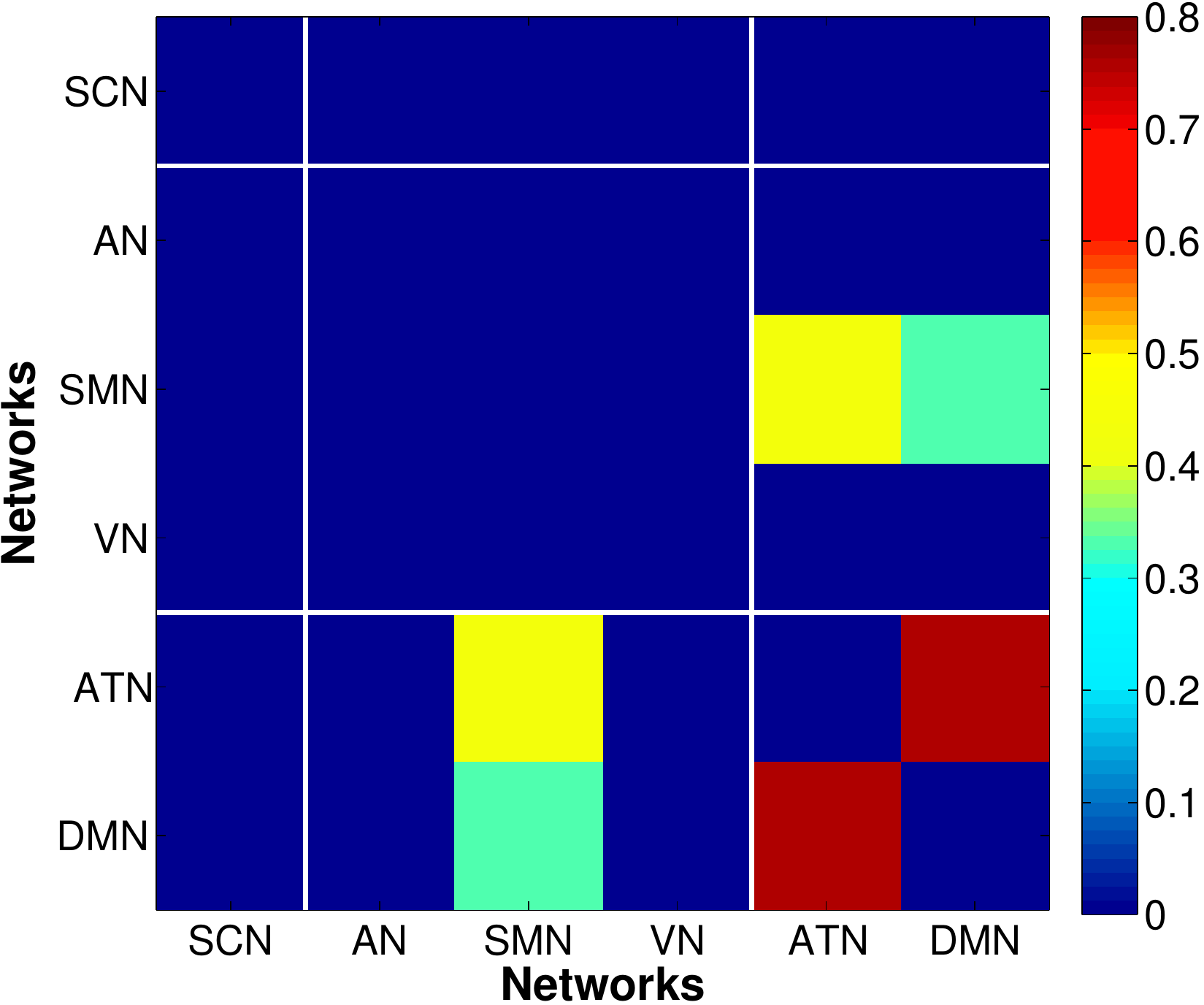}}
	\end{minipage}
	\hspace{-0.1 cm}
	\begin{minipage}[b]{0.311\linewidth}
		\centering
		\subfigure[]{\includegraphics[width=1.02\linewidth,keepaspectratio]{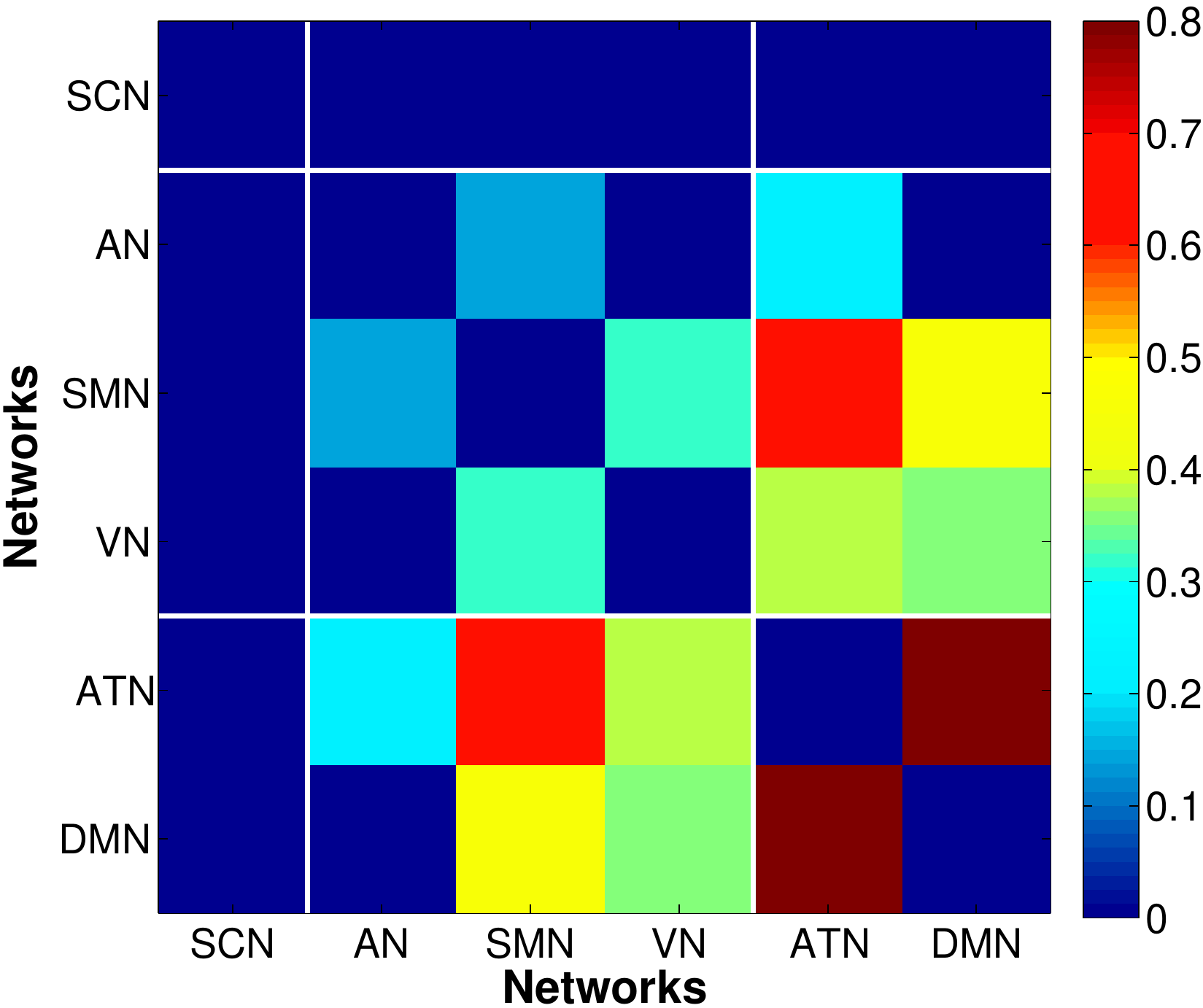}}
	\end{minipage}
	\hspace{-0.1 cm}
	\begin{minipage}[b]{0.314\linewidth}
		\centering
		\subfigure[]{\includegraphics[width=1.02\linewidth,keepaspectratio]{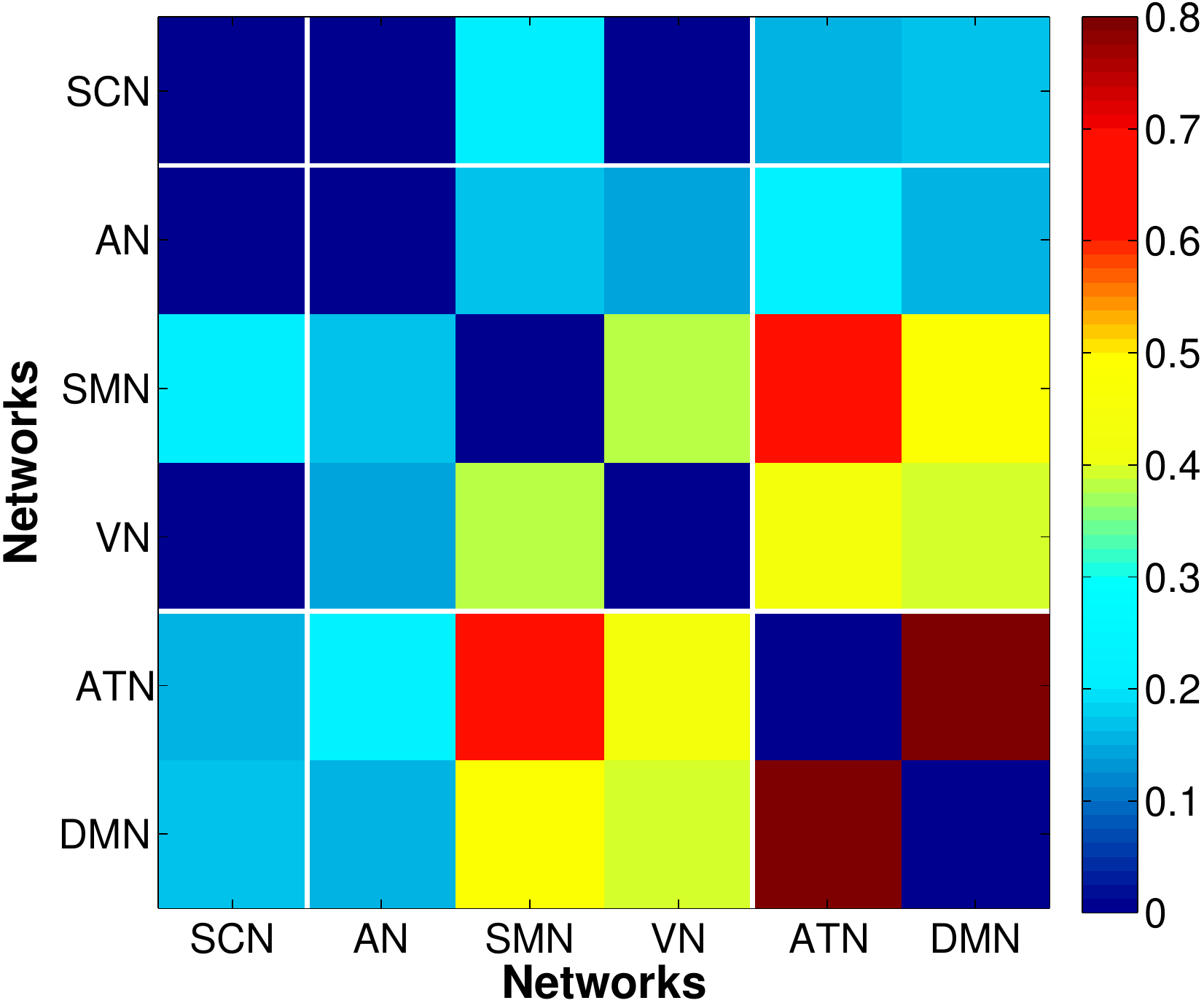}}
	\end{minipage}
\caption{Comparison of between-network functional connectivity measured by RV coefficient-based dependencies using (a): mean time series. (b): common factors that account for $1\%$ and (c): $20\%$ of the variance of the data, across six resting-state brain networks. 
The connections shown are significantly different from zero, with absolute values of the standardized RVs greater than the upper-bound of $(1-\alpha/D)\times100\%$ ($\alpha = 0.05$, $D =  6\times6 = 36$) Bonferroni-adjusted confidence interval under the null hypothesis of no connections between networks. The null-distributions are assumed normal.}
\label{Fig:Between-Net}
\vspace{-0.02in}
\end{figure}

Fig.~\ref{Fig:Between-Net} shows the between-network RV-coefficient-based connectivity $RV^S_{pq}$ computed from the correlations between mean and factor time series across six resting-state brain networks. The mean-time series approach produced the sparsest between network connectivity. The proposed method captured a more extensive connectivity where the strength and extent is greater when there are more factors per network (higher percentage of variance explained: $\tau = 20\%$ vs $\tau = 1\%$. 
It is appropriate to use small number of dominant factors because connectivity captured by the most factors could be spurious (due to random noise generated by the magnet and spread across the entire space) or an artifact (due to smoothing effect).

The VN, AN and SMN are related to the lower-level sensory processing while ATN and DMN to the higher-order cognitive functions. Within the cluster of low-level sensory-specific RSNs, we observe the presence of intra-dependency between these sensation networks which is consistent with the results in the similar study of directed connectivity across RSNs \citep{Li2011}. Using factors that explain only $1\%$ of the data variance, we found dependency of SMN with both AN and VN, with slightly stronger connection strength between the SMN and VN, and detected an additional VN-AN connection with the $20\%$ of variation, which, however, are completely missed by the estimates based on the single average time series. For the high-level cognitive RSNs, the strong dependency between the ATN and DMN, in fact the strongest among all the dependencies across RSNs, are identified consistently by both approaches, although slightly more pronounced from the factor-based estimates. This is in accordance with the recent findings of anti-correlations between the default and attentional systems in numerous studies \citep{Fox2005}. Using the $20\%$ of variance is also able to reveal the weak connections between the SCN with other networks.

The sensory-motor networks and cognitive networks are inter-dependent. The SMN exhibits strong correlation with the cognitive networks, followed by VN and AN, as detected by factor-based approach. The ATN has stronger connections with the sensory networks compared to the DMN, with the most pronounced synchrony with the SMN. The strong connection between the ATN and SMN found here is agreement with the findings using ICA \citep{Allen2011}, but was unable to be detected by Bayesian network model \citep{Li2011}. Besides, The DMN might play a pivotal roles in integrating information from all other systems, indicated by presence of connections with all networks including the subcortical network as detected using the $20\%$ of variation. In summary, the RSNs display modular organization where networks with similar functional relevance are densely connected, as the connectivity at the voxel and ROI levels. However, during the resting-state, the intra-connectivity among the cognitive networks are enhanced relative to that among the sensory-motor networks which are usually more correlated during active state when performing tasks, which require more interactions between the sensory and motor functions.

\begin{figure}[!t]
		\centering
		\subfigure[]{\includegraphics[width=0.4\linewidth,keepaspectratio]{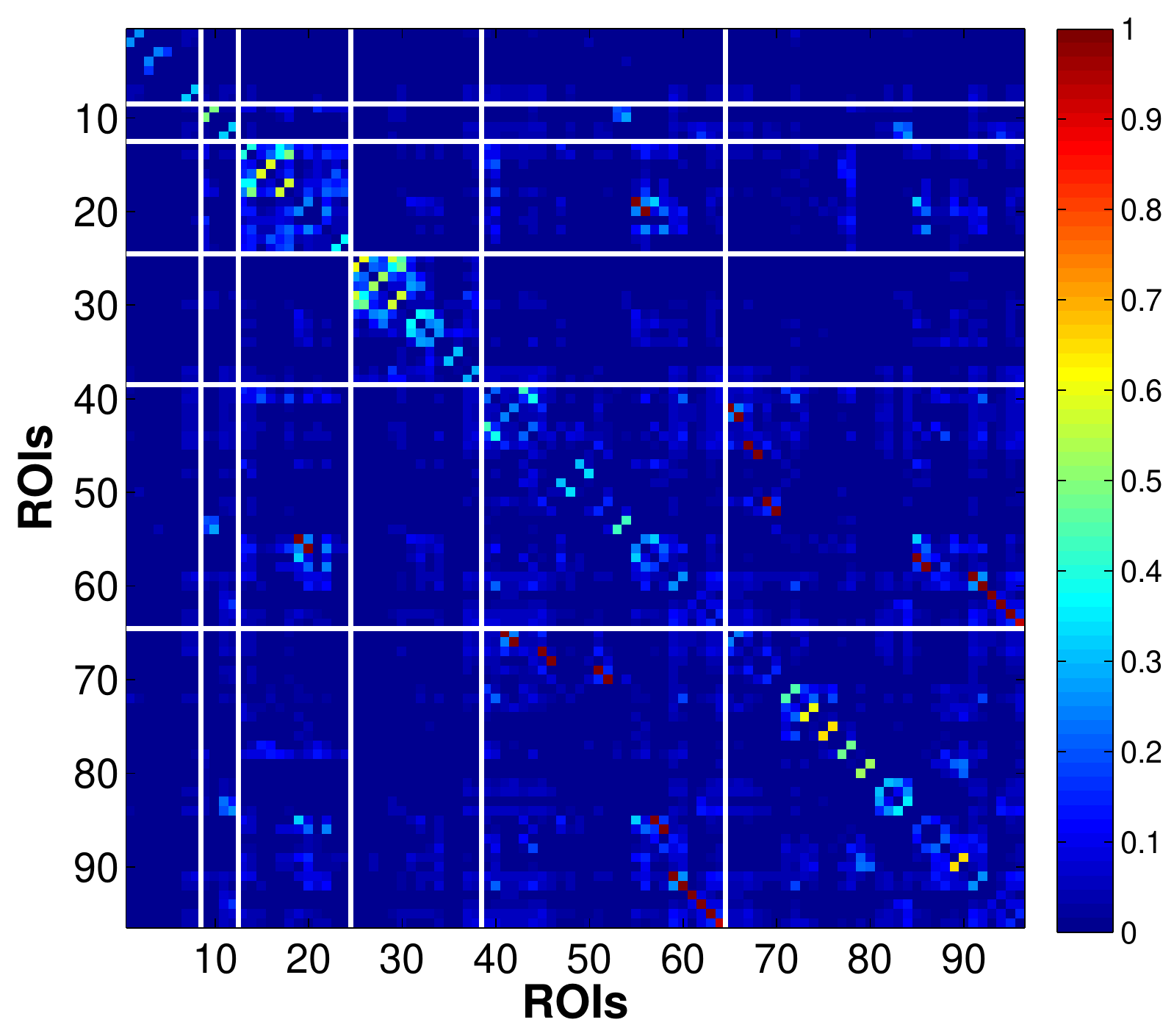}}
	\hspace{0.1 cm}
		\centering
		\subfigure[]{\includegraphics[width=0.41\linewidth,keepaspectratio]{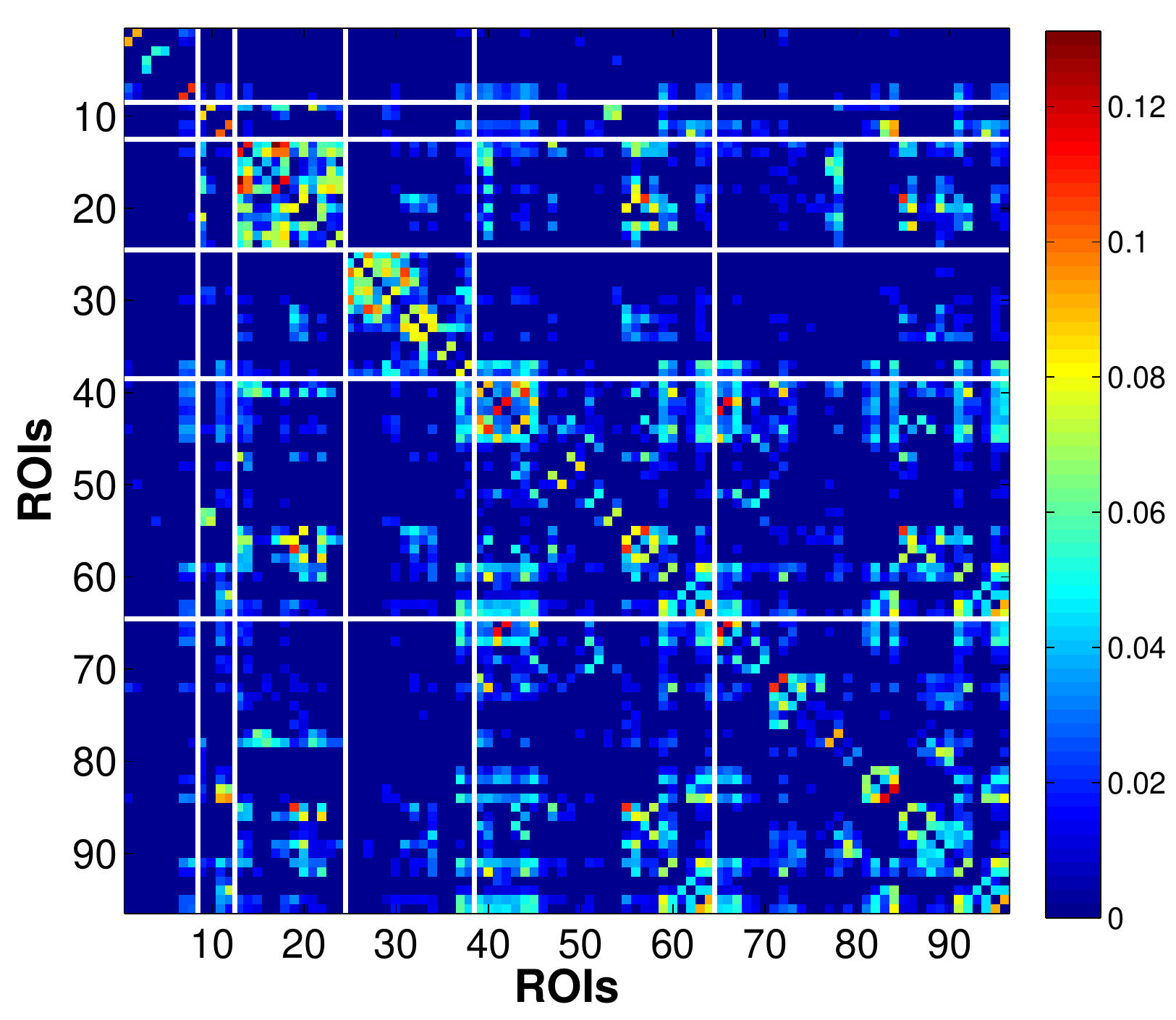}} \\
	\vspace{-0.25 cm}
		\centering
		\subfigure[]{\includegraphics[width=0.4\linewidth,keepaspectratio]{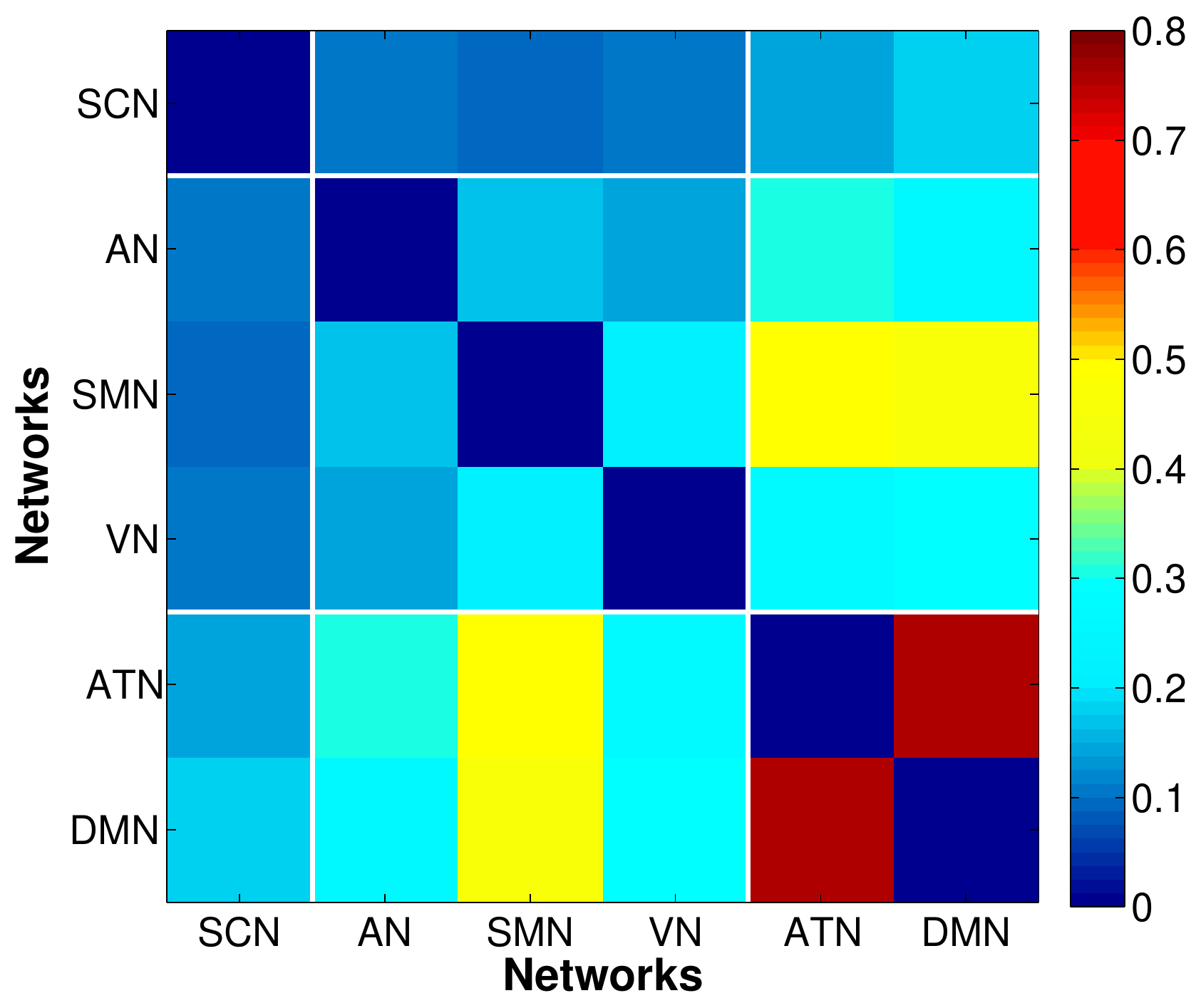}}
	\hspace{0.1 cm}
		\centering
		\subfigure[]{\includegraphics[width=0.41\linewidth,keepaspectratio]{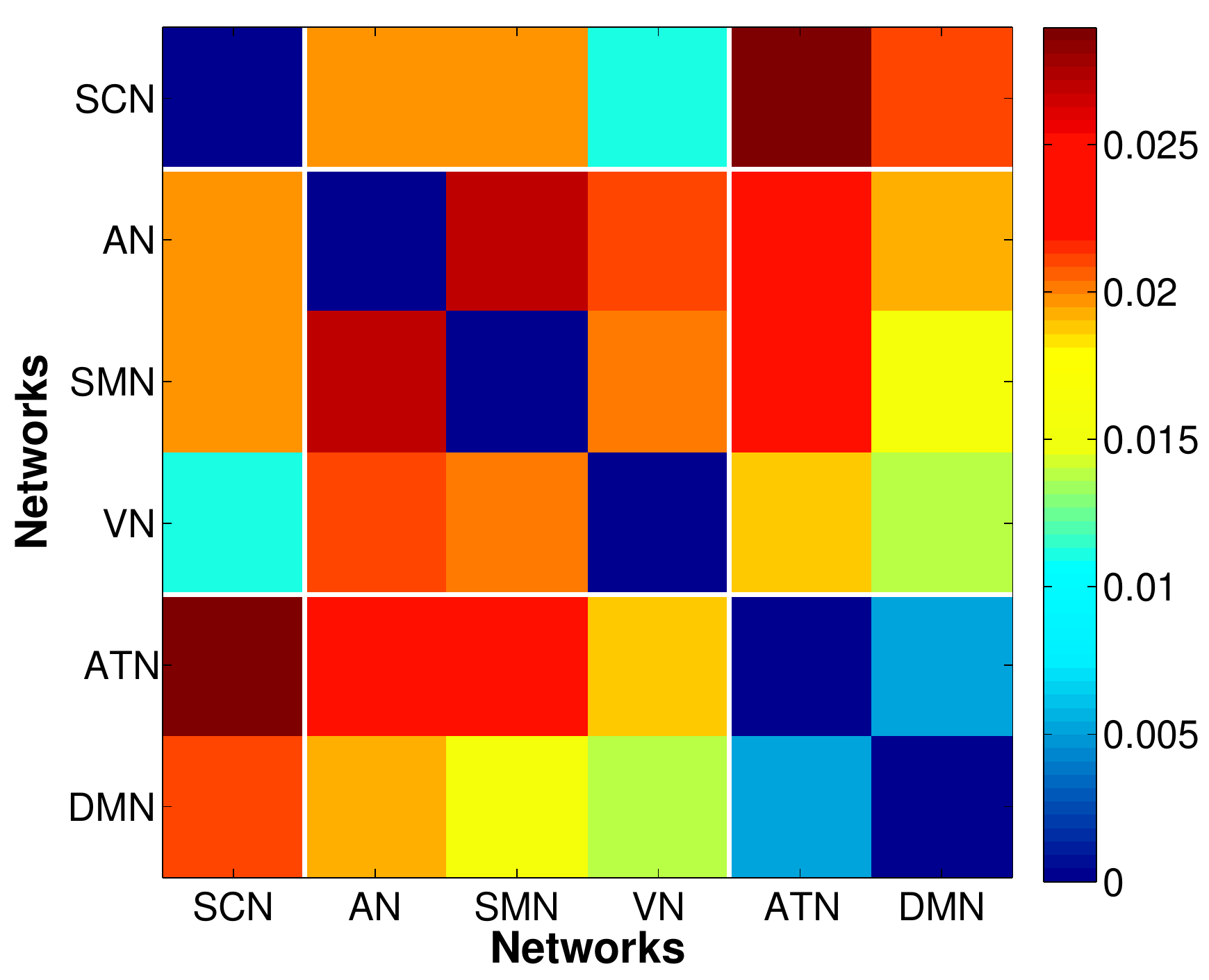}}
	\vspace{-0.02in}
\caption{Group means (left) and standard deviations (right) of the estimated resting-state functional connectivity matrices for 10 subjects, using the MSFA method with $\tau = 1\%$. (a)-(b) Between-ROIs. (c)-(d) Between-networks.}
\label{Fig:Multi-subs}
\vspace{-0.05in}
\end{figure}

\vspace{-0.01in}

\subsection{Multi-Subject Analysis}

Fig.~\ref{Fig:Multi-subs} shows the group mean (left) and variability (right) in the between-ROI and between-network functional connectivity matrices for 10 subjects. The results are averages over subject-specific estimates obtained by applying the MSFA method (with $\tau = 1\%$ of data variance) on each subject. The average connectivity matrices demonstrate that the modular structure of the between-ROI connectivity (Fig.~\ref{Fig:Multi-subs}(a)), and the strong connections between the SMN with both ATN and DMN and also between the ATN and DMN (Fig.~\ref{Fig:Multi-subs}(c)), are stable and reproducible across subjects. The variability indicates the general deviation of the subjects from the mean. From Fig.~\ref{Fig:Multi-subs}(b), the major variability in the between-ROI connectivity is found within each functional network, with the highest variation in the SMN followed by the AN. Fig.~\ref{Fig:Multi-subs}(d) reveals the most pronounced variation across subjects in the between-network connectivity is observed between the ATN and the sensory-motor networks (AN, SMN and VN), but interestingly its connection with DMN exhibits the least variability.

\vspace{-0.05in}

\section{Conclusion}

We developed a multi-scale factor analysis (MSFA) model which is a statistical approach to modeling and estimating hierarchical connectivity between nodes, clusters and sub-networks in a large-dimensional network. The MSFA provides a framework for reducing dimensionality locally (within each prescribed cluster or sub-network) by estimating the principal components series separately in each cluster. These components provide a summary of localized activity that explain the most variability within the cluster or sub-network. The proposed MSFA approach gives a good representation of multi-scale dependence and thus captures connectivity at the local (within-cluster) level and global (between clusters and between networks) level. It achieves dimension reduction in each cluster and therefore has the ability to handle massive datasets such as fMRI. The approach provides statistically reliable estimation of large-dimensional covariances to measure fine-scale functional connectivity based on the factor analysis, and summarize global-scale networks using RV dependency.

The results from our simulation studies show that the factor-based estimator outperforms the conventional sample covariance matrix for high-dimensional voxel-wise connectivity and mean-based approach for inter-regional connectivity. Applications to resting-state fMRI data demonstrate the ability of the MSFA approach in identifying the modular organization of human brain networks during rest, at the three-level hierarchy of connectivity (voxel-ROI-system network), in a unified, structural and computationally efficient way. This is in contrast with many resting-state fMRI studies that analyzed only one or two levels, often of specific networks. Our procedure is able to estimate the voxel-wise correlations in a simultaneous instead of pairwise manner, providing new insights into the resting-state connectivity at a finer scale. Moreover, our method detected connectivity of major resting-state networks, in consistency with the literature, but further reveal the global interactions across these networks, i.e. between the low-level sensory-motor and high-level cognitive functions. Future works might extend the proposed method to analyzing dynamic brain connectivity \citep{Lindquist2014,Samdin2016,Fiecas2016}. Besides, this multi-scale modeling framework can be extended to handle directed dependence based on the idea of factor-based subspace VAR analysis in \citep{Ting2014,Ting2016} by assuming the extracted factors for each cluster to follow a regional VAR model, as addressed in our recent work \citep{Wang2016}.

\section{Appendix}

\subsection{Regularity Conditions}

Let $\| {\bf H} \|_F = \sqrt{ \tr({\bf H}'{\bf H})}$ denotes Frobenius norm of vector or matrix ${\bf H}$. ${\lambda}_{\text{max}}({\bf H})$ and ${\lambda}_{\text{min}}({\bf H})$ denote the maximum and minimum eigenvalues of ${\bf H}$. We assume the following regularity conditions (used in \citep{Bai2003,Fan2013}) to guarantee consistent estimation of the factors, factor loadings and hence the constructed covariance matrix, for the regional FA models (1). We drop the index $r$ for notational simplicity. It can be shown that these conditions are applicable to the global model (2) which has the same form but an appended version of (1).

\textit{Assumption 1 (Covariance Structure)}: As $n \rightarrow \infty$,
\begin{itemize}
\item[(a)] There exists $c > 0$ such that ${\lambda}_{\text{min}}(n^{-1} {\mathbf Q}'{\mathbf Q}) > c$.
\item[(b)] There are constants $c_1, c_2 > 0$ such that $c_1 < {\lambda}_{\text{min}}(\Sigma_{{\mathbf E}{\mathbf E}}) \leq {\lambda}_{\text{max}}(\Sigma_{{\mathbf E}{\mathbf E}}) < c_2$.
\end{itemize}

\textit{Assumption 2 (Factors)}: $E{\left\| {\mathbf f}(t) \right\|}_{F}^4 \leq M < \infty$ and $T^{-1} \sum_{t=1}^T {\mathbf f}(t) {\mathbf f}'(t) \stackrel{p}{\rightarrow} {\Sigma}_{{\mathbf f} {\mathbf f}}$ as $T \rightarrow \infty$, where $\Sigma_{\mathbf{f} \mathbf{f}} = {\mbox{diag}}({\sigma}_{f_{1}}^2, \ldots, {\sigma}_{f_{n}}^2)$ with ${\sigma}_{f_{1}}^2 \geq \ldots \geq {\sigma}_{f_{n}}^2 >0$.

\textit{Assumption 3 (Factor Loadings)}: ${\left\| {\bf q}_i \right\|}_{F} \leq M < \infty$ and $n^{-1} {\mathbf Q}'{\mathbf Q} \stackrel{p}{\rightarrow} I_r$ as $n \rightarrow \infty$.

\textit{Assumption 4 (Moments of Errors)}: There exists a positive constant $M < \infty$, such that for all $i \leq n$ and $t \leq T$,
\begin{itemize}
\item[(a)] $E[e_{i}(t)] = 0$ and $E{\left| e_{i}(t) \right|}^8 \leq M$.
\item[(b)] $E[{\mathbf E}'(s){\mathbf E}(t) / n] = E[n^{-1} \sum_{i=1}^n e_{i}(s)e_{i}(t)] = {\gamma}_n(s,t)$, $\left| {\gamma}_n(s,s) \right| \leq M$ for all $s$, and $T^{-1} \sum_{s=1}^T \sum_{t=1}^T \left| {\gamma}_n(s,t) \right| \leq M$.
\item[(c)] $E[e_{i}(t)e_{j}(t)] = \tau_{ij}$ and $n^{-1} \sum_{i=1}^n \sum_{j=1}^n \left| \tau_{ij} \right| \leq M$.
\item[(d)] $E{\left| n^{-1/2} \sum_{i=1}^n \left[e_{i}(s)e_{i}(t) - E[e_{i}(s)e_{i}(t)] \right] \right|}^4 \leq M$ for every ($t,s$).
\end{itemize}

\begin{figure}[!t]
	\begin{minipage}[t]{\linewidth}
		\centering
		\includegraphics[width=0.45\linewidth,keepaspectratio]{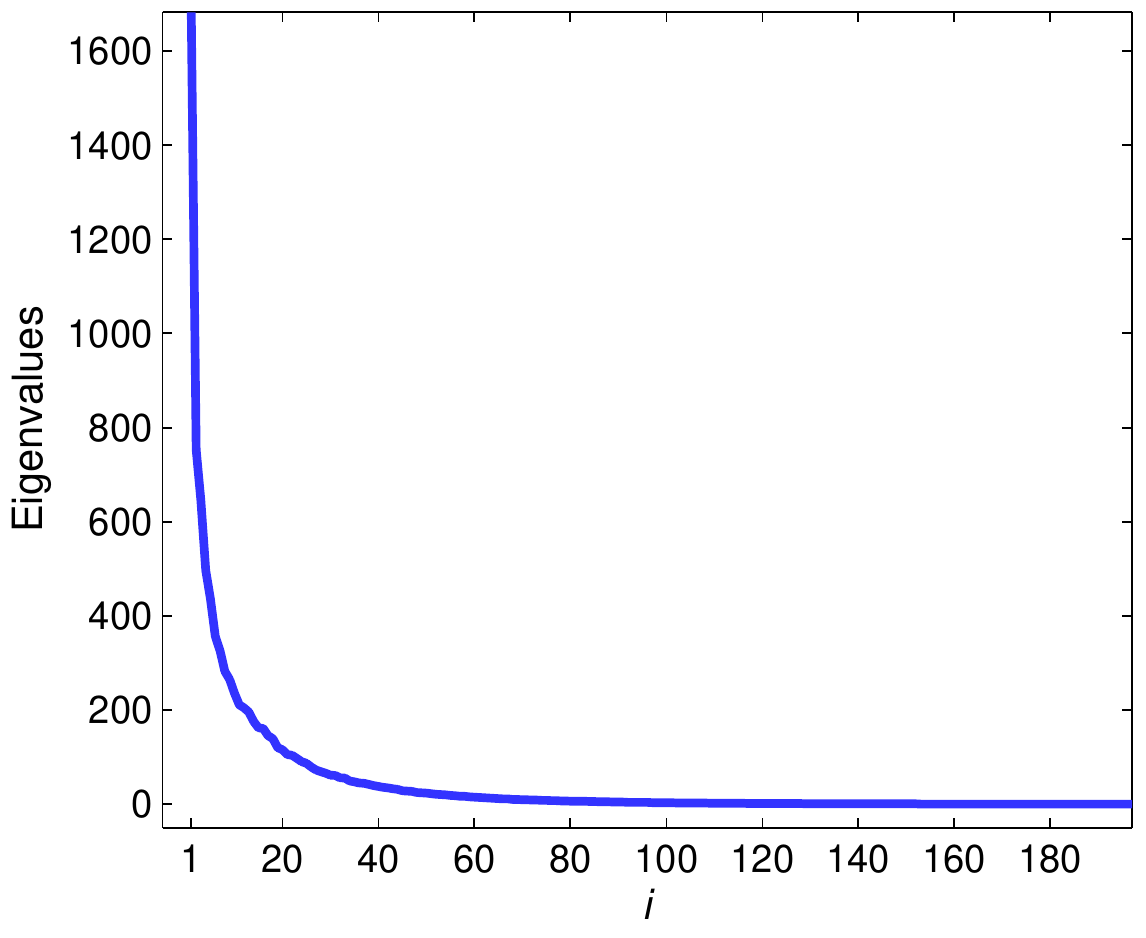}
	\end{minipage}
	\vspace{-0.6 cm}
	\caption{Eigenvalues $\widehat{\lambda}_i$ in decreasing order of the sample covariance matrix for a brain ROI with $n_r = 1017$ voxels, estimated from fMRI signals ${\mathbf Y}_r(t)$ of length $T=197$. Only the first $T$ eigenvalues are plotted.}
\label{Fig:Eigenvals-ROI}
\vspace{-0.1in}
\end{figure}

Assumption 1(a) is the pervasiveness condition which implies the first $m_r$ eigenvalues of ${\mathbf Q}_r \Sigma_{\mathbf{f}_r \mathbf{f}_r} {\mathbf Q}_r'$ are all growing to infinity with the dimensionality $n_r$. Condition 1(b) requires $\Sigma_{{\mathbf E}_r {\mathbf E}_r}$ to be well-conditioned and its eigenvalues be uniformly bounded for all large $n_r$. The decomposition (3) is then asymptotically identified as $n_r \rightarrow \infty$. These bounds also carry over to the covariance matrix where the $m_r$ largest eigenvalues of $\Sigma_{{\mathbf Y}_r {\mathbf Y}_r}$ diverge fast with $n_r$ whereas all the remaining eigenvalues are bounded as $n_r \rightarrow \infty$. The Condition 1 is sufficient to imply the existence of a $m$-factor structure in the signals ${\mathbf Y}_r(t)$. Moreover, under this condition, the eigenvectors of $\Sigma_{{\mathbf Y}_r {\mathbf Y}_r}$ corresponding to the diverging eigenvalues converge to the factor loadings, suggesting the PCA on the sample covariance is appropriate for estimating the subspace structure in a high-dimensional factor analysis. Assumption 1 is reasonable for fMRI data, as indicated by the divergence of a few eigenvalues of the sample covariance for a brain ROI with large number of nodes, with the rest being close to zero (Fig. \ref{Fig:Eigenvals-ROI}). This suggests the fMRI data has a factor structure.

\begin{figure}[!t]
	\begin{minipage}[t]{0.45\linewidth}
		\centering
		\subfigure[]{\includegraphics[width=1\linewidth,keepaspectratio]{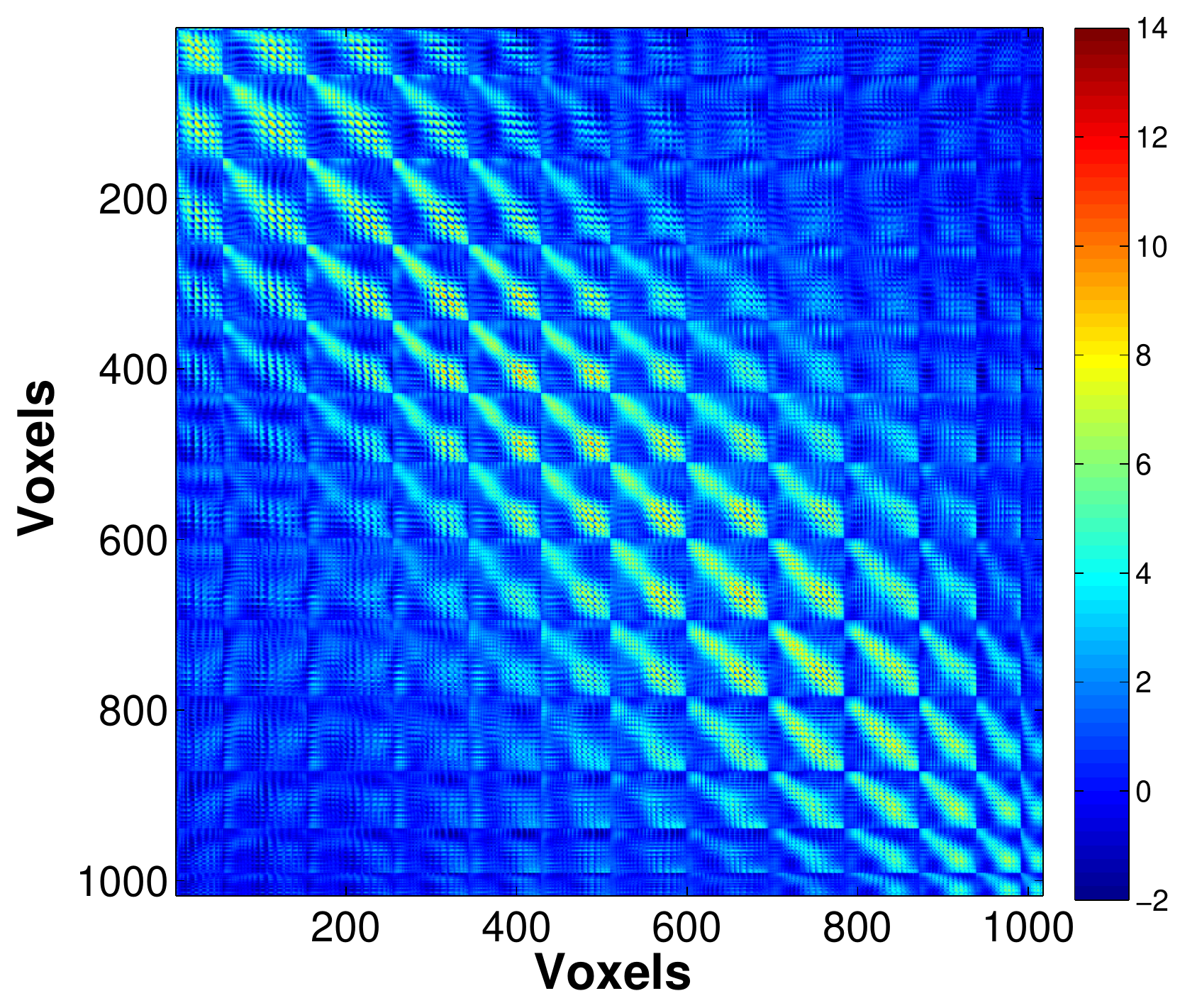}}
	\end{minipage}
	\hspace{-0.18 cm}
	\begin{minipage}[t]{0.45\linewidth}
		\centering
		\subfigure[]{\includegraphics[width=1.03\linewidth,keepaspectratio]{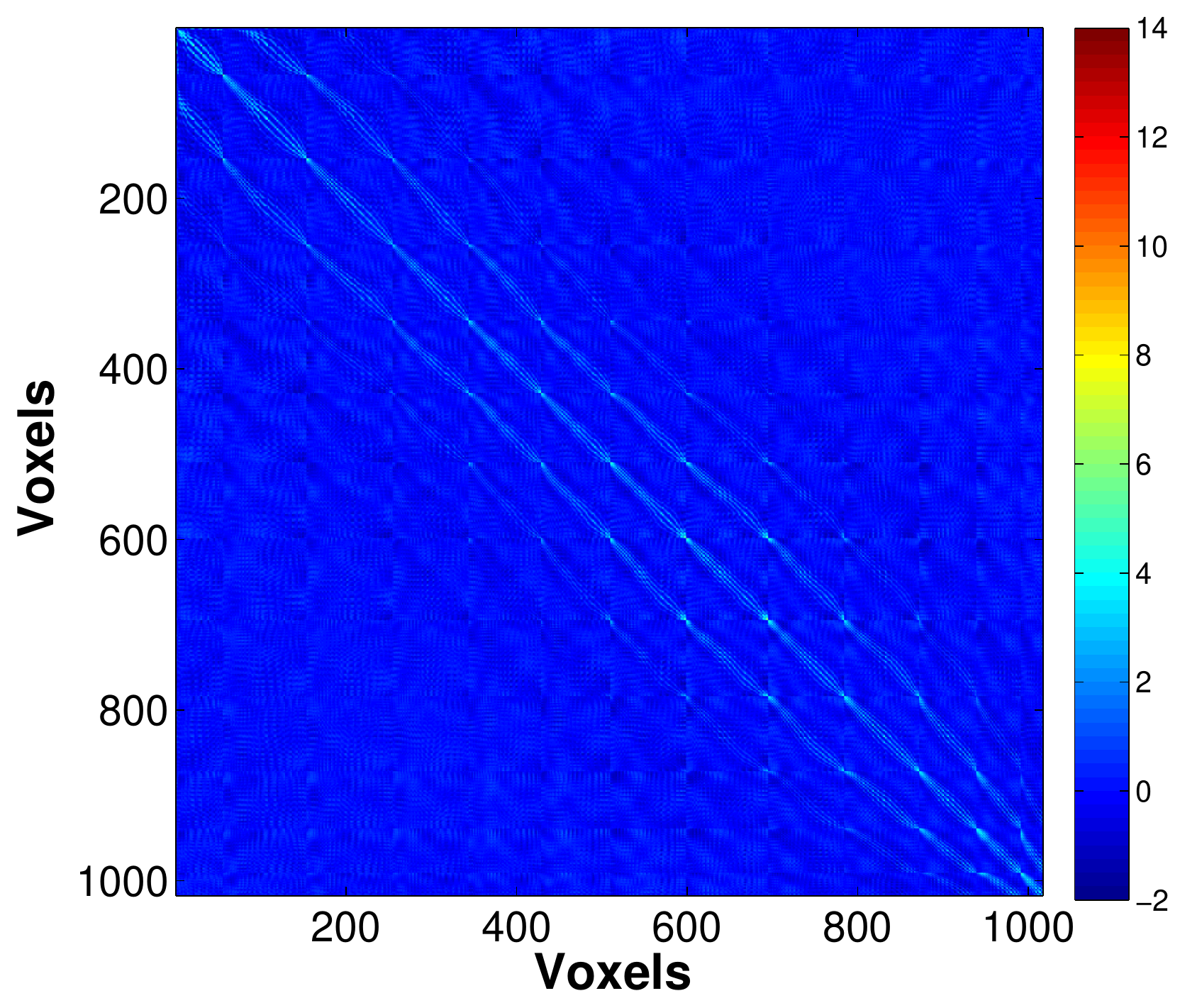}}
	\end{minipage}
	\vspace{-0.3 cm}
\caption{Decomposition of covariance matrix for a brain ROI into (a) low rank matrix $\widehat{\mathbf Q}_r \widehat{\Sigma}_{\mathbf{f}_r \mathbf{f}_r} \widehat{{\mathbf Q}}_r'$ and (b) error covariance matrix $\widehat{\Sigma}_{{\mathbf E}_r {\mathbf E}_r}$ using $m_r = 15$.}
\label{Fig:Cov-Decomp}
\vspace{-0.1in}
\end{figure}

Assumptions 2 and 3 serve as the identifiability conditions as discussed in Section II. Moreover, Assumption 2 allows $\mathbf{f}_r(t)$ to be serially correlated. Assumption 3 ensures that the factors are pervasive, i.e. having non-negligible contribution on a non-vanishing proportion of individual signal ${Y}_{ri}(t)$. Assumption 1 easily holds under this condition. Assumption 4 allows for weak serial and cross-sectional correlation in the error terms $e_{ri}(t)$, as specified in the approximate factor model \citep{Chamberlain1983}. 
In this paper, we assume the fMRI data to follow a strict factor structure \citep{BaiLi2012}, in which $e_{ri}(t)$ are independent across all $i$ and $t$ with a diagonal error covariance matrix $\Sigma_{{\mathbf E}_r {\mathbf E}_r}$. By assuming ${\sigma}_{e_{ri}}^2 \leq M < \infty$ for all $i$, with the eigenvalues of $\Sigma_{{\mathbf E}_r {\mathbf E}_r}$ simply the diagonal elements, Condition 1 holds for this special case. Moreover, given 4(a), the remaining assumptions are also satisfied under the independence of $e_{ri}(t)$. Since the approximate factor model is more general, the developed asymptotic results also apply to the strict model adopted here. Fig.\ref{Fig:Cov-Decomp} shows the factor decomposition of covariance matrix for the brain ROI, where the estimated low-rank matrix (Fig. \ref{Fig:Cov-Decomp}(a)) is dominant and accounts for most of the correlations among the voxels, whereas only a small amount of variation is picked-up by the error covariance estimated from residuals (Fig. \ref{Fig:Cov-Decomp}(b)). Besides, most of the off-diagonal elements of the residual covariance matrix $\widehat{\Sigma}_{{\mathbf E}_r {\mathbf E}_r}$ are near zero, suggesting negligible cross-correlation in the noise, and thus using strict factor model for fMRI data is not inappropriate.

\clearpage
\subsection{ROI-RSN Mapping}
\vspace{-0.2in}

\begin{table}[!th]
\renewcommand{\arraystretch}{0.9}
\caption{Brain ROIs grouped into six resting-state networks, and number of voxel fMRI time series for each ROI and selected number of factors for a subject.}
\vspace{0.2cm}
\hspace{1 cm}
\resizebox{0.58\textwidth}{!}{\begin{minipage}{\textwidth}
\begin{center}
\begin{tabular}{lcccclccc}
\hline\hline
RSN&Abbre&$n_r$&$m_r$& &RSN&Abbre&$n_r$&$m_r$ \\
\hline\hline
\multicolumn{4}{l}{\bfseries{Subcortical Network}} & & L inferior frontal gyrus, opercular & IFGoperc.L &	1007 &	1 \\
L caudate nucleus & CAU.L & 966 & 2 & & R inferior frontal gyrus, opercular & IFGoperc.R &	943 &	1 \\
R caudate nucleus & CAU.R &	984 &	2 & & L inferior frontal gyrus, triangular & IFGtriang.L &	1402 &	2 \\
L putamen & PUT.L & 1017 &	2 & & R inferior frontal gyrus, triangular & IFGtriang.R &	2237 &	1 \\
R putamen & PUT.R &	1068 & 2 & & L inferior frontal gyrus, orbital & ORBinf.L &	2167 &	2  \\
L pallidum & PAL.L & 303 & 1 & & R inferior frontal gyrus, orbital & ORBinf.R &	1671 &	2 \\
R pallidum & PAL.R &	261 &	1 & & L insula & INS.L &	1757 &	2 \\
L thalamus & THA.L &	1117 &	2 & & R insula & INS.R &	1846 &	1 \\
R thalamus & THA.R &	1022 &	2 & & L superior parietal gyrus & SPG.L &	1758 &	1 \\
\cline{1-4}
\multicolumn{4}{l}{\bfseries{Auditory Network}} & & R superior parietal gyrus & SPG.R &	1994 &	1 \\
L superior temporal gyrus & STG.L &	2161 &	1 & & L inferior parietal lobule & IPL.L &	2159 &	1 \\
R superior temporal gyrus & STG.R &	3034 &	2 & & R inferior parietal lobule & IPL.R &	1339 &	1 \\
L temporal pole: superior temporal gyrus & TPOsup.L &	1245 &	1 & & L middle temporal gyrus & MTG.L &	4614 &	2\\
R temporal pole: superior temporal gyrus & TPOsup.R &	1297 &	1 & & R middle temporal gyrus & MTG.R &	4195 &	2 \\
\cline{1-4}
\multicolumn{4}{l}{\bfseries{Sensorimotor Network}} & & L temporal pole: middle temporal gyrus & TPOsup.L &	741 &	1 \\
L precentral gyrus & PreCG.L &	3283 &	3 & & R temporal pole: middle temporal gyrus & TPOsup.R &	1174 &	1 \\
R precentral gyrus & PreCG.R &	3371 &	3 & & L inferior temporal gyrus & ITG.L &	3088 &	2 \\
L supplementary motor area & SMA.L &	2057 &	1 & & R inferior temporal gyrus & ITG.R &	3381 &	2 \\
\cline{6-9}
R supplementary motor area & SMA.R &	2237 &	2 & & \multicolumn{4}{l}{\bfseries{Default Mode Network}} \\
L postcentral gyrus & PoCG.L &	3495 &	3 & & L superior frontal gyrus, orbital & ORBsup.L &	964 &	1 \\
R postcentral gyrus & PoCG.R &	3770 &	1 & & R superior frontal gyrus, orbital & ORBsup.R &	959 &	2 \\
L superior parietal gyrus & SPG.L &	1994 &	1 & & L middle frontal gyrus, orbital & MFG.L &	899 &	1 \\
R superior parietal gyrus & SPG.R &	2183 &	1 & & R middle frontal gyrus, orbital & MFG.R &	1007 &	1 \\
L supramarginal gyrus & SMG.L &	1091 &	31 & & L inferior frontal gyrus, orbital & ORBinf.L &	1671 &	2 \\
R supramarginal gyrus & SMG.R &	1850 &	1 & & R inferior frontal gyrus, orbital & ORBinf.R &	1757 &	2 \\
L paracentral lobule & PCL.L & 1271 &	2 & & L superior frontal gyrus, medial & SFGmed.L &	2973 &	2 \\
R paracentral lobule & PCL.R & 795 &	1 & & R superior frontal gyrus, medial & SFGmed.R &	2075 &	1 \\
\cline{1-4}
\multicolumn{4}{l}{\bfseries{Visual Network}} & & L superior frontal gyrus, medial orbital & ORBsupmed.L &	728 &	1 \\
L calcarine gyrus & CAL.L &	2262 &	2 & & R superior frontal gyrus, medial orbital & ORBsupmed.R &	827 &	1 \\
R calcarine gyrus & CAL.R &	1843 &	2 & & L anterior cingulate \& paracingulate gyri & ACG.L &	1406 &	2 \\
L cuneus & PCUN.L &	1501 &	1 & & R anterior cingulate \& paracingulate gyri & ACG.R &	1274 &	1 \\
R cuneus & PCUN.R &	1405 &	1 & & L median cingulate \& paracingulate gyri & DCG.L &	1940 &	2 \\
L lingual gyrus & LING.L &	2093 &	2 & & R median cingulate \& paracingulate gyri & DCG.R &	2089 &	2 \\
R lingual gyrus & LING.G &	2329 &	2 & & L posterior cingulate gyrus & PCC.L &	472 &	1 \\
L superior occipital gyrus & SOG.L &	1330 &	1 & & R posterior cingulate gyrus & PCC.R &	312 &	1 \\
R superior occipital gyrus & SOG.R &	1403 &	1 & & L hippocampus & HIP.L &	966 &	1 \\
L middle occipital gyrus & MOG.L &	-80.73 &	2 & & R hippocampus & HIP.R &	967 &	1 \\
R middle occipital gyrus & MOG.R &	3258 &	2 & & L parahippocampal gyrus & PHG.L &	955 &	1 \\
L inferior occipital gyrus & IOG.L &	2048 & 2 & & R parahippocampal gyrus & PHG.R &	1139 &	1 \\
R inferior occipital gyrus & IOG.R &	949 &	2 & & L inferior parietal lobule & IPL.L &	2159 &	1 \\
L fusiform gyrus & FFG.L &	1005 &	2 & & R inferior parietal lobule & IPL.R &	1339 &	1 \\
R fusiform gyrus & FFG.R &	2288 &	2 & & L angular gyrus & ANG.L &	1124 &	1 \\
\cline{1-4}
\multicolumn{4}{l}{\bfseries{Attentional Network}} & & R angular gyrus & ANG.R &	1756 &	1 \\
L superior frontal gyrus, dorsolateral & SFGdor.L &	2558 &	2 & & L precuneus & PCUN.L &	3529 &	2 \\
R superior frontal gyrus, dorsolateral & SFGdor.R &	3454 &	2 & & R precuneus & PCUN.R &	3188 &	1 \\
L superior frontal gyrus, orbital & ORBsup.L &	4002 &	1 & & L middle temporal gyrus & MTG.L &	4614 &	2 \\
R superior frontal gyrus, orbital & ORBsup.R &	964 &	2 & & R middle temporal gyrus & MTG.R &	4195 &	2 \\
L middle frontal gyrus & MFG.L &	959 &	1 & & L temporal pole: middle temporal gyrus & TPOmid.L &	741 &	1 \\
R middle frontal gyrus & MFG.R &	4593 &	2 & & R temporal pole: middle temporal gyrus & TPOmid.R &	1174 &	1 \\
L middle frontal gyrus, orbital & ORBmid.L &	5009 &	1 & & L inferior temporal gyrus & ITG.L &	3088 &	2 \\
R middle frontal gyrus, orbital & ORBmid.R &	899 &	-1 & & R inferior temporal gyrus & ITG.R &	3381 &	2 \\

\hline\hline
\end{tabular}
\end{center}
\label{Table:RSN-ROI-MAP}
\end{minipage} }
\end{table}

\clearpage
\bibliographystyle{apalike}
\bibliography{Ref-JASA}

\end{document}